\documentclass[floatfix,aps,prd,abstract,twocolumn,10pt,nofootinbib,preprintnumbers ]{revtex4}

\usepackage{amsgen,amsmath,amstext,amsbsy,amsopn,amssymb}
\usepackage{graphicx}
\usepackage[usenames]{color}
\usepackage{epsfig}
\usepackage{multirow}
\usepackage{longtable}

\newcommand{\mb}{\mathbf}

\newcommand{\Ddel}{\delta_{\rm D}   }

\newcommand{\MpcOh}{ \,  \mathrm{Mpc}  \, h^{-1} }
\newcommand{\hOMpc}{ \,  \mathrm{Mpc}^{-1}  \, h  }

\newcommand{\Msun}{ \,  M_{\odot}\,h ^{-1}  }
\newcommand{\comment}[1]{}

\newcommand{\beq}{\begin{equation}}
\newcommand{\eeq}{\end{equation}}
\newcommand{\beqa}{\begin{eqnarray}}
\newcommand{\eeqa}{\end{eqnarray}}
\newcommand{\nn}{ \nonumber }

\newcommand{\cyc}{\, \mathrm{cyc.} }
\allowdisplaybreaks[1]

\begin{document}

%\preprint{Version 0.999}

%\title{  Impacts of non-Gaussianity on the covariance matrix of \\ power spectrum and bispectrum for dark matter and halos }
%\title{  The covariance of power spectrum and bispectrum for dark matter and halos }
%\title{The importance of non-Gaussian contribution to the dark matter and halo power spectrum and bispectrum covariance  } 
%\title{The importance of non-Gaussian contribution to   \\   the power spectrum and bispectrum covariance  } 
%\title{ Non-Gaussian Contribution to the Power Spectrum and Bispectrum Covariance  } 
\title{ Assessment of the Information Content of  the Power Spectrum  and Bispectrum  } 
%\title{ Non-Gaussian Contribution to the  Covariance   \\ of Power Spectrum and Bispectrum  } 

\author{Kwan Chuen Chan} \email{chan@ice.cat}
\author{Linda Blot}
\affiliation{Institute of Space Sciences, IEEC-CSIC, Campus UAB, Carrer de Can Magrans, s/n,  08193 Bellaterra, Barcelona, Spain}

\date{\today}

\begin{abstract}

  The covariance  matrix of the matter and halo power spectrum and bispectrum are studied.   Using a large suite of simulations, we find that the non-Gaussianity in the covariance  is significant already at mildly nonlinear scales.  We compute the leading disconnected non-Gaussian correction to the matter bispectrum covariance using perturbation theory, and find that the corrections result in good agreement in the mildly nonlinear regime.  The shot noise contribution to the halo power spectrum and bispectrum covariance is computed using the Poisson model, and the model yields decent agreement with simulation results. However, when the shot noise is estimated from the individual realization, which is usually done in reality, we find that the halo covariance is substantially reduced and gets close to the Gaussian covariance. This is because most of the non-Gaussianity in the covariance arises from the fluctuations in the Poisson shot noise.   We use the measured non-Gaussian covariance to access the information content of the power spectrum and bispectrum. The signal-to-noise ratio (S/N) of the matter and halo power spectrum levels off in the mildly nonlinear regime, $k \sim 0.1 - 0.2 \hOMpc $.  In the nonlinear regime the S/N of the matter and halo bispectrum increases but much slower than the Gaussian results suggest.  We find that both the S/N for power spectrum and bispectrum are overestimated by the Gaussian covariances, but the problem is much more serious for the bispectrum.  Because the bispectrum is affected strongly by nonlinearity and shot noise, inclusion of the bispectrum only adds modest amount of S/N  compared to that of the power spectrum. 

%While the Gaussian covariance suggests that the S/N of the matter bispectrum should surpass that of the matter power spectrum at  $k \sim 0.2 \hOMpc$, the full non-Gaussian results show that this only occurs at  $k \sim 1 \hOMpc $.

\end{abstract}

\maketitle

\section{ Introduction }

Current large scale structure surveys such as those aiming at measuring the baryonic acoustic oscillations are giving important insights into the physics of the universe  \cite{Percival:2001hw,Tegmark:2003ud,Eisenstein:2005su,Cole:2005sx,Tegmark:2006az,Anderson:2013zyy}. Future surveys such as Euclid \cite{EuclidRedBook} and LSST \cite{LSSTScienceBook} will measure the large scale structure over a large volume and in deep redshift ranges with unprecedented precision. To correctly interpret the data, the challenge is not only to accurately model the observables, such as the power spectrum and bispectrum, but the covariance matrix of these quantities must also be known with sufficient precision.

 The covariance is often measured by running large numbers of mocks, where the survey is modelled starting from a large scale structure simulation. In this approach the mocks not only can take into account the intrinsic correlation, but also the survey geometry and various systematic effects. To get good control of the covariance, often hundreds or even thousands of simulations are required. As it is computationally expensive to run the full $N$-body simulations, various cheap approximate methods are often used (see Ref.~\cite{Monaco:2016pys} and references therein for a review of some of the methods that have been proposed). 
 
 However,  to reach a better understanding and modelling of the covariance it is easier to start with the relatively simpler case of dark matter and halos in $N$-body simulations. This approach has been used to study the covariance of the power spectrum for dark matter \cite{Meiksin:1998mu,Scoccimarro:1999kp,Smith:2008ut,Takahashi:2009ty,Harnois-DerapsPen_2012,Blot:2014pga} and halos \cite{Smith:2008ut}. For the bispectrum PTHalos \cite{Scoccimarro:2001cj} has been used to investigate the covariance systematically \cite{Sefusatti:2006pa,Scoccimarro:2003wn}.

 Analytical methods also prove fruitful, as they give interesting insights into what the relevant contributions are for the estimation of the covariance. The dark matter power spectrum has been modelled using perturbation theory \cite{Scoccimarro:1999kp,Bertolini:2015fya,Mohammed:2016sre} and  halo model \cite{Cooray:2000ry,Mohammed:2014lja}. The Poisson model has been invoked to model the shot noise contribution to the covariance \cite{Matarrese:1997sk,Smith:2008ut}. In particular, by combining the analytical and numerical approaches, it has been realized that beat coupling or supersample covariance can be a significant contribution to the covariance in real surveys \cite{Rimes:2005dz,Hamilton:2005dx,dePutter:2012,Takada:2013bfn,Li:2014sga}.

In this paper we study the power spectrum and bispectrum covariance numerically and analytically. Taking advantage of the large suite of simulations available in the DEUS-PUR project, we  study the covariance of the dark matter and halo power spectrum and bispectrum. It is worth stressing that this is the first systematic study of the bispectrum covariance using such large number of $N$-body simulations.  We also model the covariance and compare the predictions with the numerical results. 

As the power spectrum has been well explored and the bispectrum becomes the next frontier in large scale structure, it is crucial to address how much information one can gain by going beyond the 2-point level. Previous work \cite{Sefusatti:2004xz} suggested that there is substantial information content in the high-point statistics such as the bispectrum based on the Gaussian covariance approximation. In the context of weak lensing, similar conclusion was found based on Gaussian covariance \cite{Takada:2003ef}; however, when the non-Gaussian covariance is used, the signal to noise (S/N) is substantially reduced, especially for the  bispectrum \cite{KayoTakadaJain_2013,Sato:2013mq}.  Armed with the accurate covariance measured from a large suite of simulations, we assess the information loss in the power spectrum and bispectrum due to  the correlations that arise in the nonlinear regime.

This paper is organized as follows. In Sec.~\ref{sec:Pk_darkmatter_halo}, after reviewing the basic theory of the power spectrum covariance for dark matter and halos,  we compare the model prediction against the numerical covariance. Sec.~\ref{sec:bispectrum_covariance} is devoted to the bispectrum covariance. We first lay down the theory of the dark matter and halo bispectrum covariance. We then compare the numerical bispectrum covariance with the theory predictions. The information content of the power spectrum and the bispectrum is assessed by means of the signal-to-noise ratio in Sec.~\ref{sec:signal_to_noise}. We conclude in Sec.~\ref{sec:conclusions}. We check the probability distribution of the bispectrum estimator in Appendix \ref{sec:bisp_estimaotr_distribution}.  In Appendix \ref{sec:derivation_zoo}, we show the derivation of some of formulas used in the main text. The shot noise contribution to the halo power spectrum and bispectrum covariance is derived using the Poisson model in Appendix \ref{sec:Poisson_model_PkBk}. In Appendix  \ref{sec:BinningDependence}, we check the dependence of the signal-to-noise ratio on the binning width.

%The caovarince of bispectrum is important in quantifying the information available in the bispectrum. This is important as the CMB is well studied, the next frontier is LSS. For LSS, ine has to confront with nonlinearity. The non-Gaussian covariance increases the error and hence the information contents aavailable in the LSS in the nonlinear regime.   

%% Power spectrum: numerically \cite{Meiksin:1998mu,Scoccimarro:1999kp,Takahashi:2009ty,Smith:2008ut,Blot:2014pga,Li:2014sga}.  Analytically: PT \cite{Scoccimarro:1999kp,Mohammed:2016sre} halo model \cite{Cooray:2000ry ,Mohammed:2014lja} Halo P(k): \cite{Smith:2008ut}. 

%% Beat coupling, super sample covariance \cite{Rimes:2005dz, Rimes:2005dx,Takada:2013bfn, Li:2014sga} \red{ understand the meaning of the SSC, can't see it in simulations, but comparing differetn boxes, small effects } 

%% Approximate methods PThalo for bispecturm \cite{Sefusatti:2006pa} \red{ what is the shot noise tratment there? } 

\section{ Covariance of power spectrum  }
\label{sec:Pk_darkmatter_halo}
The covariance of the dark matter power spectrum has been studied quite extensively  both numerically \cite{Meiksin:1998mu,Scoccimarro:1999kp, Takahashi:2009ty,Blot:2014pga} and theoretically \cite{Meiksin:1998mu,Scoccimarro:1999kp,Bertolini:2015fya,Mohammed:2016sre}. On the other hand the covariance of the halo power spectrum is relatively less explored, but see Refs.~\cite{Angulo:2007fw,Smith:2008ut}.  Although the focus of this section is on the  covariance of the halo power spectrum, we also present the results for dark matter for comparison. We first review the basic theory of the power spectrum covariance, which paves the way for the bispectrum covariance that we discuss later on.

\subsection{ Theory of the power spectrum covariance}

  Here we first review the theory of the covariance matrix of the matter power spectrum, and then we discuss the covariance of the halo power spectrum.

 Suppose that the Fourier modes of the density contrast $\delta $  are binned  into bands of width $\Delta k $ in Fourier space.  The power spectrum of $\delta $, $P$, is defined as
\beq
  \langle \delta( \mb{k} )   \delta( \mb{k}' ) \rangle  = P(k) \Ddel( \mb{k} + \mb{k}' )  \eeq
where  $\Ddel $ is the Dirac delta function.   From the definition, one can construct a power spectrum estimator $\hat{P}$  as (e.g.~\cite{FeldmanKaiserPeacock1994,Scoccimarro:1999kp})
\beq
\label{eq:Pk_estimator}
\hat{P}(k) =   k_{\rm F}^3  \int_{k}  \frac{ d^3 p }{ V_{\rm s}( k ) } \delta( \mb{p} ) \delta( - \mb{p} ),
\eeq 
where $ k_{\rm F} $ is the fundamental mode of the box, $2 \pi / L_{\rm box}  $ ($L_{\rm box}$ is the size of the simulation box).    Note that the integral is done over all the modes that fall into the band of width  $[ k - \Delta k /2,  k + \Delta k /2 ) $. $ V_{\rm s} $ is the volume of the spherical shell  
\beq
 V_{\rm s} (k) = \int_k d^3 p = 4 \pi k^2 \Delta k + \frac{\pi}{3} \Delta k^3. 
\eeq

The covariance  matrix of $\hat{P} $ is defined as 
\beqa
\label{eq:CP_estimator}
 C^P(k,k') 
 &\equiv & \mathrm{cov}( \hat{P}(k) , \hat{P}(k')  )    \nn \\
&=&    \langle [ \hat{P}(k) - \langle \hat{P}(k) \rangle ] [  \hat{P}(k') - \langle \hat{P}(k') \rangle    ]  \rangle  \nn \\
&=& \langle  \hat{P}(k)  \hat{P}(k') \rangle   -  \langle    \hat{P}(k) \rangle  \langle  \hat{P}(k') \rangle .  
\eeqa

\subsubsection{ Dark matter}

The covariance matrix of the matter power spectrum was first investigated using perturbation theory in Ref.~\cite{Scoccimarro:1999kp}, and it has been extended to include loop corrections in Refs.~\cite{Bertolini:2015fya,Mohammed:2016sre}.  Here we  limit ourselves to the theory laid down in Ref.~\cite{Scoccimarro:1999kp} as it is not the focus of this paper to model the matter power spectrum covariance as accurately as  possible.

Plugging Eq.~\ref{eq:Pk_estimator} into Eq.~\ref{eq:CP_estimator}, for Gaussian $\delta $, we get the Gaussian covariance of the  power spectrum estimator \cite{FeldmanKaiserPeacock1994}
\beq
\label{eq:CP_G_continuous}
C^P_{\rm G}(k,k') 
= \frac{2 k_{\rm F}^3 }{ V_{\rm s} (k) } P^2 ( k)  \delta_{  k, k' } ,
\eeq
where  $\delta_{k,k'} $ is the Kronecker delta function that ensures that the Gaussian covariance is diagonal. The Gaussian covariance is inversely proportional to the number of modes in the bin (and it is inversely proportional to the bin width $\Delta k$).   For Gaussian field, the power spectrum $P$ in Eq.~\ref{eq:CP_G_continuous} should be the linear one. As we  include non-Gaussian  corrections below, one of the power spectrum should be the 1-loop one so that it is of the same order as the non-Gaussian results. However, since the 1-loop power spectrum over-predicts the numerical power spectrum already at $k \sim 0.1 \hOMpc $ at $z=0$, we use the nonlinear power spectrum measured from simulations in place of the 1-loop result.

The non-Gaussian contribution to the power spectrum covariance comes from the trispectrum \cite{Meiksin:1998mu,Scoccimarro:1999kp}
\beqa
\label{eq:CP_trisp_general}
&&C^P_{\rm NG}(k, k' ) \nn \\
&=&  k_{\rm F}^3  \int_{k} \frac{ d^3 p }{ V_{\rm s} (k) }  \int_{k'} \frac{ d^3 p' }{ V_{\rm s} (k') }  T( \mb{p} , - \mb{p} ,  \mb{p}' , - \mb{p}' ) .  
%% && V_{\rm F}^2 \int_{k_1} \frac{ d^3 k'  }{ V(k_1) }  \int_{k_2} \frac{ d^3 k'' }{ V(k_2) } T( \mb{k}' , - \mb{k}' ,  \mb{k}'' , - \mb{k}'' ) \Ddel( \mb{0} ) \nn \\
%% &=&  V_{\rm F}  \int_{k_1} \frac{ d^3 k' }{ V(k_1) }  \int_{k_2} \frac{ d^3 k'' }{ V(k_2) }  T( \mb{k}' , - \mb{k}' ,  \mb{k}'' , - \mb{k}'' ) .  
\eeqa
Only the parallelogram shape trispectrum contributes to the power spectrum covariance.  The non-Gaussian part does not depend on the bin width $\Delta k $.  Note that both the Gaussian and non-Gaussian covariance are inversely proportional to the volume of the box (through $k_{\rm F}^3$). We will see that the same is true for the bispectrum covariance.  We can trace back the factor $ k_{\rm F}^3 $ to the Dirac delta function in the definition of the polyspectrum, which arises from the statistical translational invariance of the field.  Because of this invariance, the amount of statistics is simply proportional to the volume. On the other hand, when a window function is imposed, the statistical translational invariance is broken. Indeed the supersample covariance term scales differently with volume \cite{Takada:2013bfn}.

Nonlinearity induces mode coupling and hence non-Gaussianity. The tree-level dark matter trispectrum $T$ has two distinct contributions, $T_1 $  and  $T_2$ \cite{1984ApJ...279..499F,Scoccimarro:1999kp}
\beqa
&&T_1(\mb{k}_1, \mb{k}_2, \mb{k}_3, \mb{k}_4 )\nn\\
 & = & 6 F_3 ( \mb{k}_2, \mb{k}_3, \mb{k}_4 )  P_{\rm L} (k_2)  P_{\rm L}(k_3)  P_{\rm L}(k_4)  + 3 \, \mathrm{cyc.},   \\
&&T_2(\mb{k}_1, \mb{k}_2, \mb{k}_3, \mb{k}_4 ) \nn \\
& = & [ 4 F_2 ( - \mb{k}_3, \mb{k}_{23} )  F_2( \mb{k}_4, \mb{k}_{23} )   P_{\rm L}(k_{23})  P_{\rm L}(k_3)  P_{\rm L} (k_4)  \nn \\
&+& ( \mb{k}_1 \leftrightarrow  \mb{k}_2 )    ]   +    5 \, \mathrm{cyc.} ,  
\eeqa
where cyc.~denotes cyclic permutations. $P_{\rm L }$ is the linear power spectrum, and $ F_2$ and $F_3 $  are the coupling kernels in standard perturbation theory; see \cite{Goroffetal1986,PTreview}.

The $T_1$ and $T_2$ contributions to the covariance matrix are then given by 
\beqa
\label{eq:T1_covmat}
&& T_1( \mb{k}_1, - \mb{k}_1, \mb{k}_2, - \mb{k}_2 ) \nn \\
&=& 12 F_3( \mb{k}_1 ,  \mb{k}_2,  - \mb{k}_2 ) P^2_{\rm L}( k_2 ) P_{\rm L}( k_1) + 
( \mb{k}_1 \leftrightarrow  \mb{k}_2 ) , \nn \\
\label{eq:T2_covmat}
&& T_2(\mb{k}_1, - \mb{k}_1, \mb{k}_2, - \mb{k}_2 ) \nn \\
&=& 4 P_{\rm L}(| \mb{k}_1 - \mb{k}_2 |  ) [ F_2( - \mb{k}_1, \mb{k}_1 - \mb{k}_2 ) P_{\rm L}(k_1)   \nn \\
&+&  F_2( - \mb{k}_1, \mb{k}_1 - \mb{k}_2 ) P_{\rm L}(k_2)    ]^2  + ( \mb{k}_2 \rightarrow - \mb{k}_2    ). 
\eeqa
%We note that the both Eq.~\ref{eq:T1_covmat} and \ref{eq:T2_covmat} respect the symmetry  that they are invariant under $ \mb{k}_1 \rightarrow - \mb{k}_1  $ or  $ \mb{k}_2 \rightarrow - \mb{k}_2  $.  This is a useful check for the correctness of the final results. 

The resultant covariance matrix contribution reads \cite{Scoccimarro:1999kp}
\beqa
\label{eq:bar_trispectrum}
&&\bar{T}( k,k')  =  \int_{k} \frac{ d^3 p  }{ V_{\rm s} (k) } \int_{k'} \frac{ d^3 p'  }{ V_{\rm s}(k') } \nn \\
&\times  & \Big\{   12 F_3( \mb{p} ,\mb{p}' , - \mb{p}' ) P_{\rm L} (p) P_{\rm L}^2(p')  + ( \mb{p} \leftrightarrow \mb{p}'   )  \nn \\
  &+&  8P_{\rm L}(|\mb{p} - \mb{p}' |) [ F_2(- \mb{p},\mb{p} -  \mb{p}') P_{\rm L}(p) +   ( \mb{p} \leftrightarrow  \mb{p}'   )  ]^2    \Big\}. \nn \\
\eeqa

%% \red{check these and add DM prediction to the plots }
%% The $T_1$  contribution can be written as
%% \beqa
%% & & \int_{k_1} \frac{d^3 k'}{ V(k_1)} P(k') \int_{k_2} \frac{d^3 k'}{ V(k_2)}  12 F_3 ( \mb{k}' , \mb{k}'' , - \mb{k}'' ) P^2(k'') \nn \\
%% &=& 12  \int_{k_1} \frac{d^3 k'}{ V(k_1)} P(k') \frac{1}{V(k_2) } I_{F_{3}} (k_2, k'), 
%%  \eeqa
%% where $ I_{F_{3}} $ is defined as 
%% \beqa
%%  I_{ F_3} ( k, k' ) &=& \int_{ ( k - \frac{ \Delta k }{ 2 } ) / k' }^{ ( k + \frac{ \Delta k }{ 2 } ) / k' } dr P^2( k' r )  \frac{ \pi k'^{3} }{ 756  }   \Big[\frac{12}{ r^2 } -158 + 100 r^2  \nn \\
%% &-&42 r^4 + \frac{ 3 }{r^3} (r^2 -1 )^3 ( 7 r^2 + 2 ) \ln \Big(\frac{1 + r  }{|1 -r |} \Big)  \Big] . 
%% \eeqa
%% The $T_2 $ contribution can be written as 
%% \beqa
%% & &\frac{ 8 }{ V(k_1)V(k_2) } \int_{ k_1 - \frac{ \Delta k}{ 2 }  }^{  k_1 + \frac{ \Delta k}{ 2 }  } 4 \pi k^{' 2 } d k' \int_{k_2}  d^3 k'' P(|\mb{k}' - \mb{k}''|)  \nn \\
%% &\times &  [ F_2(- \mb{k}',\mb{k}' -  \mb{k}'') P(k') +   ( \mb{k}' \leftrightarrow  \mb{k}''   )    ]^2 , 
%% \eeqa
%% where for the inner integral, $ \mb{k}_1 $ is in the $z$-direction, and for the outer integral, it should not depend on where $ \mb{k}_1 $  points. Thus we can integrate out the solid angle. 

\begin{figure}[!htb]
\centering
\includegraphics[width=0.9\linewidth]{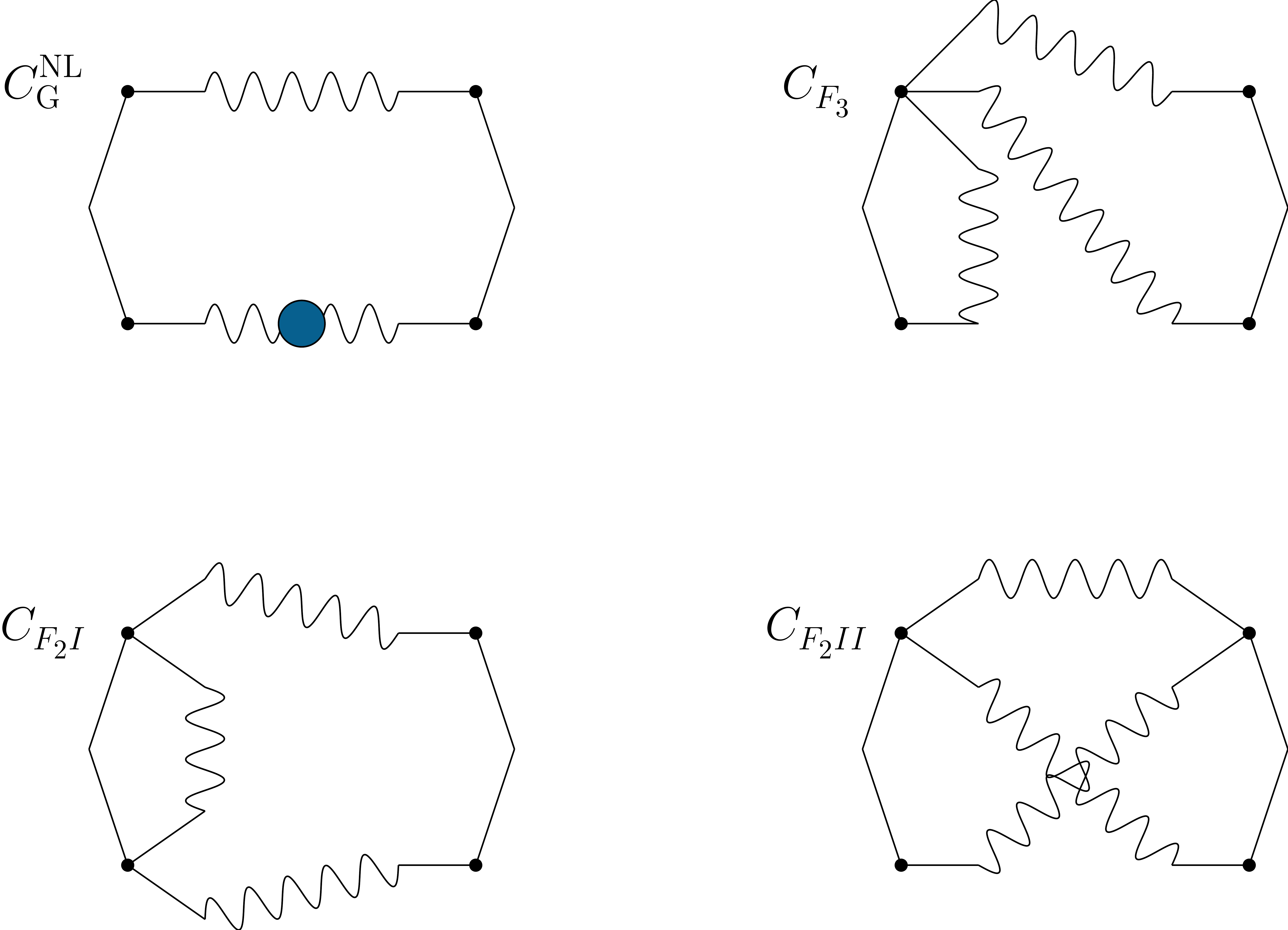}
\caption{   A  diagrammatic representation of the leading contributions to the covariance of the matter power spectrum. $ C_{\rm G}^{\rm NL} $ is the Gaussian term (top left), while the rest are non-Gaussian contributions.  The set of black dots on the left and right of each diagram represent the two $\delta $'s in each power spectrum estimator.    The legs branching from each dot represent the perturbation theory kernel, $F_1$, $F_2$, and $F_3$ respectively.  Each wavy line represents the linear power spectrum. In $C_{\rm G}^{\rm NL }$, the filled circle indicates that the linear power spectrum is replaced by the 1-loop power spectrum.     }
\label{fig:Covmat_Pk_diagrams}
\end{figure}

The contribution to the covariance of the power spectrum can be represented graphically, and the results are shown in Fig.~\ref{fig:Covmat_Pk_diagrams}.  We will apply similar techniques to the bispectrum shortly, and it will be more useful as there are more contributions to the bispectrum covariance. In each diagram two black dots on both sides represent the two $\delta $'s in the power spectrum estimator. The legs branching from each dot represent  the perturbation theory kernels. We use the wavy line to represent the linear power spectrum, while the additional dot on top means that the 1-loop power spectrum should be used instead. The term in  the top-left corner is the  Gaussian term, while the others are the non-Gaussian terms. The term $C_{F_3} $ gives the $T_1$ contribution, and  both  $C_{F_2 I}$ and  $C_{F_2 II}$ combine to give $T_2 $.    %We will see that similar diagrammatic representation works for the bipectrum covariance as well.  

%As stressed in \cite{Scoccimarro:1999kp}, the effects of the window function was ignored. Indeed, in \cite{Takada:2013bfn} the effects of the window function was taken into account and the super-sample variance, or beat coupling,  was  derived.  \red{Comment on super sample covarinace here.  }

It has also been shown that the large scale mode can modulate the small scale modes to cause the so-called beat coupling or supersample covariance \cite{Rimes:2005dz,Hamilton:2005dx,dePutter:2012,Takada:2013bfn,Li:2014sga}.     This can be a significant source of covariance at small scales in real surveys. However, it only arises when a window function is imposed such as in real surveys or when sub-parts of a huge simulation are considered. In the simulations with periodic boundary conditions, the wave vectors are sharp and the supersample covariance does not appear. When a window function is present, the wave vectors are broadened and the large scale long modes can contribute.  At tree level, there is an extra diagram contributing to the trispectrum in addition to the ones shown in Fig.~\ref{fig:Covmat_Pk_diagrams}, and this leads to beat coupling at large scales.

\subsubsection{Halos} 

We now turn to the halo power spectrum covariance. In this case, besides the complications due to dark matter nonlinearity and halo biasing, the discrete nature of halos contributes further to stochastic fluctuations.  In fact, shot noise is the major source of halo covariance as we will see below.   In Appendix \ref{sec:appendix_PoissonPk} we derive the Poisson shot noise contribution to the covariance matrix of the power spectrum using the Poisson model.  In the Poisson model, we assume the point particles are formed by Poisson sampling the underlying continuous field.  The Poisson fluctuations give rise to the whole hierarchy of the $n$-point correlations in general. In particular, the connected and disconnected 4-point function contribute to the power spectrum covariance.  We refer the readers to the Appendix \ref{sec:appendix_PoissonPk} for details on the derivation. Here we  summarize the key results.

There are both Gaussian and non-Gaussian contributions to the covariance due to Poisson shot noise. The Gaussian shot noise contribution, which can be represented diagrammatically by the two disconnected diagrams in Fig.~\ref{fig:Poisson_2_3_4_pt}, can be combined with the smooth Gaussian covariance Eq.~\ref{eq:CP_G_continuous} to be written in a compact form \cite{FeldmanKaiserPeacock1994,Smith:2008ut}
\beq
\label{eq:CG_Poisson}
C^P_{\rm G}(k,k') 
= \frac{2 k_{\rm F}^3 }{ V_{\rm s} (k) } \Big[  P_{\rm h} ( k) + \frac{1}{ \tilde{n} } \Big]^2 \delta_{  k, k' } ,
\eeq
where $ P_{\rm h} $ is the smooth halo power spectrum and  $\tilde{n}= (2 \pi)^3 \bar{n} $ with $\bar{n} $ being the mean number density of halos. \footnote{ The presence of the $(2 \pi)^3  $ factor is due to the Fourier convention used in this paper. We can convert the formula to  the perhaps more popular convention by replacing $\tilde{n} $ by  $\bar{n} $.   }  The non-Gaussian shot noise contribution is given by 
\beqa
\label{eq:CNG_Poisson_short}
C_{\rm NG}^P(k,k')  &=&  k_{\rm F}^3  \Big[ \frac{1}{ \tilde{n}^3 }  + \frac{2}{ \tilde{n}^2 } ( P_{\rm h} (k) + P_{\rm h} (k') ) \nn \\
& + &\frac{ 2}{\tilde{n}^2 }  \int_k \frac{ d^3 p }{ V_{\rm s}(k) }   \int_{k'} \frac{ d^3 p' }{ V_{\rm s}(k') } P_{\rm h} ( |\mb{p} +  \mb{p}' |)  \Big] \nn \\
&+& \dots  
 \eeqa
 where the dots denote the contribution associated with the halo bispectrum and also the smooth trispectrum contributions. See Eq.~\ref{eq:CNG_Poisson} for the full expression. These terms are due to the connected 4-point function in the Poisson model (Diagrammatically, the three types of terms in Eq.~\ref{eq:CNG_Poisson_short} can be represented by the first, second, and fourth diagrams in the third row of Fig.~\ref{fig:Poisson_2_3_4_pt}).  For the halo power spectrum we will not include the shot noise contribution associated with the bispectrum for simplicity because we find that the prediction without bispectrum works reasonably well.   A simple estimate suggests that this is small compared to the dominant Gaussian term but not negligible.

\begin{figure}[!htb]
\includegraphics[width=\linewidth]{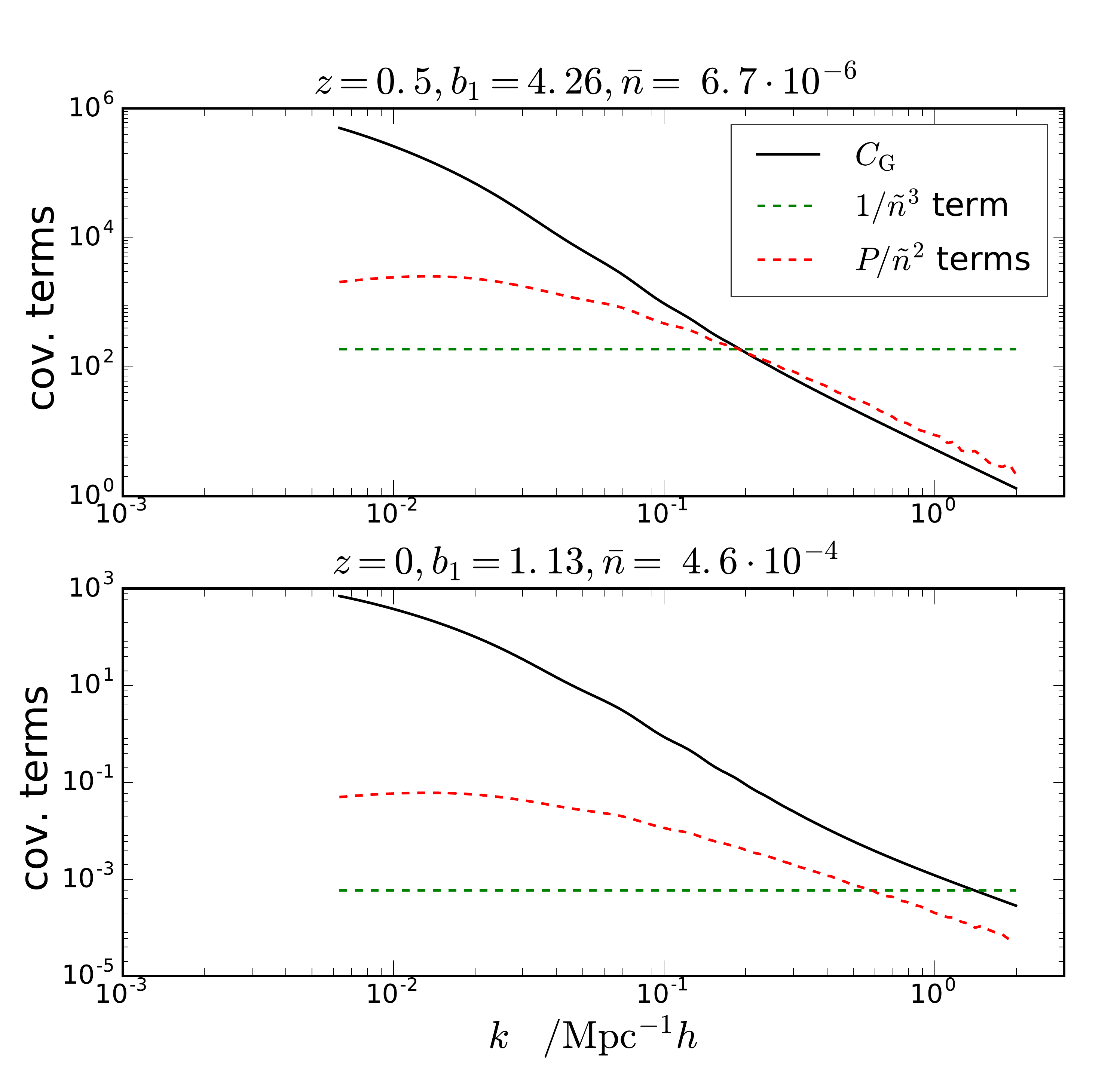}
\caption{   Various contributions to the diagonal elements of the covariance of the halo power spectrum for two selected halo groups. The Gaussian covariance (solid black) and the non-Gaussian terms proportional to $1/\tilde{n}^3$ (dashed green) and $P/\tilde{n}^2$ (dashed red).   }
\label{fig:Pkcov_Poisson_components}
\end{figure}

 In Fig.~\ref{fig:Pkcov_Poisson_components}, we plot the components of the Gaussian and non-Gaussian contributions  to the diagonal of the covariance. Note that the covariance depends on the volume of the simulation and the bin width $\Delta k$. For the theoretical computations, unless otherwise stated, we use box size $656.25 \MpcOh$, which corresponds to the Small or Hires set shown in Table \ref{tab:sim_detaials}. For the power spectrum, we use the binning width of $9.6 \times 10^{-3}  \hOMpc $, which is equal to the fundamental mode of the Small set.  We show the results for two representative halo groups, which correpsond to the Large group 4 at $z=0.5$ and the Hires group 2 at $z=0$ shown in Table \ref{tab:halo_group}.  We have combined  the terms proportional to $1/\tilde{n}^2$, which are similar in magnitude.  For the low number density group in the range $ k \lesssim  0.2 \hOMpc $, the Gaussian term, $C_{\rm G}^P  $ dominates, while for higher $k$, the non-Gaussian term $ 1/\tilde{n}^3 $ is the only significant term.  On the other hand, for the low bias and high number density sample, the Gaussian term is dominant up to $k\sim 1 \hOMpc$.

\subsection{Numerical results } 

In this paper we use the simulations from the DEUS-PUR project \cite{Rasera:2013xfa,Blot:2014pga}.  We consider three sets of simulations labelled as Large, Small, and Hires respectively. The detailed properties of these simulations are shown in Table \ref{tab:sim_detaials}.  A flat $\Lambda$CDM model with the WMAP7 cosmological parameters \cite{2007ApJS..170..377S} is adopted for these simulations. In particular, $h=0.72$, $\Omega_{\rm m} = 0.257$, $n_{\rm s } = 0.963$, and $\sigma_{8 } = 0.801 $. The Zel'dovich approximation is used to generate the Gaussian initial conditions  at $z_i = 105$ for the Large and Small sets, and  $z_i = 190$  for the Hires set.  The transfer function is computed with CAMB \cite{CAMB}.  The simulations are evolved using the adaptive mesh refinement solver RAMSES \cite{2002A&A...385..337T}. We consider simulation snapshots at $z=1$, 0.5, and 0 respectively. The Large and Small sets have the same mass resolution although the box size of the Large set is twice that of the Small set. The Hires has higher mass resolution than the Small set while their box sizes are the same.  For more details on the descriptions of the simulations, see Ref.~\cite{Blot:2014pga}. 

The halos used in this work are obtained using the friends-of-friends algorithm with linking length set to 0.2 times the mean inter-particle separation.  Only halos with at least 100 particles are used.   We divide the halos into four mass groups. The details of these mass groups are shown in Table \ref{tab:halo_group}.  The simulations of the Large and Small sets only have the highest mass group, Gr.~4, while the Hires set has groups 1, 2, 3, and 4. Moreover, the results at $z=1$, 0.5 and, 0 are available for the Large and Small sets, while  only $z=0 $  ones are available for the Hires set. We note that the number density of the halo groups considered here is low compared to the expected number density in future galaxy surveys, e.g.~the number density in Euclid is projected to be $\bar{n} \sim 10^{-3} (\MpcOh)^{-3} $.

We note that the output time of the simulations differs from the nominal output time slightly, e.g.~for the Small set at $z=1$, the fluctuation is about 0.3\% on the scale factor. For dark matter, we correct for this by multiplying by $D^2$ and $D^4$ to the power spectrum and bispectrum respectively ($D$ is the linear growth factor). As in the evolution model, the decay of the bias parameters roughly cancels the growth factor of the density \cite{ChanScoccimarroSheth2012}, the time dependence of the halo overdensity is expected to be weak and we do not apply any correction for the halos. The corrections reduce some noticeable differences between the Large and Small set results for the power spectrum covariance. We do not find any noticeable effects for the case of bispectrum.   However, the corrections are imperfect in the nonlinear regime, and this may explain that there are some differences between the Large and Small sets for their power spectrum covariance.

\begin{table*}[!htb]
\caption{ Details of the simulations.  }
\label{tab:sim_detaials}
\centering
\begin{ruledtabular}
\begin{tabular}{ c | c | c | c | l   }
Box label &  Box size ($\MpcOh$ ) & Number of particles & Redshift snapshots  & Number of realizations  \hspace{175pt}             \\
\hline 
Large & 1312.5    &  $ 512^3 $     & 1, 0.5, 0   &  512                  \\
Small & 656.25    &  $ 256^3 $     & 1, 0.5, 0   &  4096                 \\
Hires & 656.25    &  $ 1024^3 $     & 1, 0.5, 0   &  96                 \\
\end{tabular}
\end{ruledtabular}
\end{table*}

\begin{table*}[!htb]
\caption{ Properties of the halo groups.  }
\label{tab:halo_group}
\centering
\begin{ruledtabular}
\begin{tabular}{ c | c |c | c | c | l   }
Box label & Mass group    & Redshift snapshots &  Mass range ($10 ^{12} \Msun $)  &  Linear bias  &   Number density $ ( \MpcOh )^{-3}   $    \hspace{175pt}             \\
\hline 
Large, Small  & 4  &  1     & $ > 120 $         &  6.44     &  $ 1.94 \times 10^{-6 }  $             \\
Large, Small  & 4  &   0.5   & $ > 120 $         &  4.26    &   $ 6.74 \times 10^{-6 }  $                \\
Large, Small  & 4  &  0     & $ > 120 $         &  2.90     &   $ 1.57 \times 10^{-5 }  $                \\  \hline    
Hires         & 1  & 0    & $1.88  - 5.63 $   &  0.94     &   $ 9.77 \times 10^{-4 }  $            \\
Hires         & 2  & 0    & $5.63  - 18.8 $   &  1.13    &  $ 4.60 \times 10^{-4 }  $              \\
Hires         & 3  & 0    & $18.8  - 120 $   &  1.57      &  $ 1.77 \times 10^{-4 }  $          \\
Hires         & 4  & 0    & $ > 120 $        &  2.78     &   $ 1.80 \times 10^{-5 }  $           \\
\end{tabular}
\end{ruledtabular}
\end{table*}

\begin{figure}[!htb]
\includegraphics[width=\linewidth]{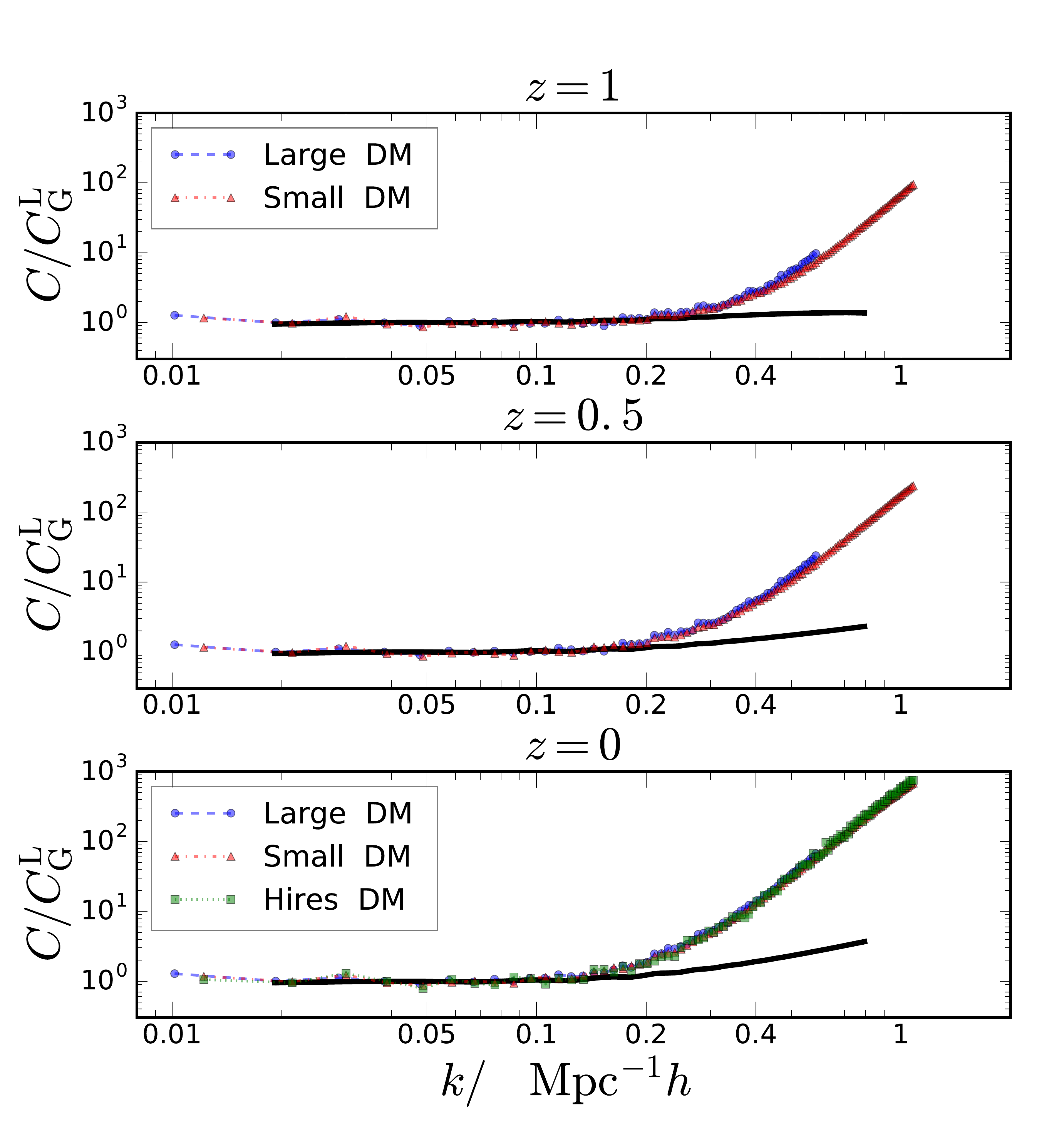}
\caption{   The diagonal elements of the dark matter power spectrum covariance at $z=1$, 0.5, and 0 (from top to bottom). The results are normalized with respect to the Gaussian covariance. The results from the Large (blue circles), Small (red triangles), and Hires (green squares) are shown.  The perturbation theory predictions are overplotted (black solid line).      }
\label{fig:Diag_PkCov_DM}
\end{figure}

\begin{figure*}[!htb]
\includegraphics[width=0.9\linewidth]{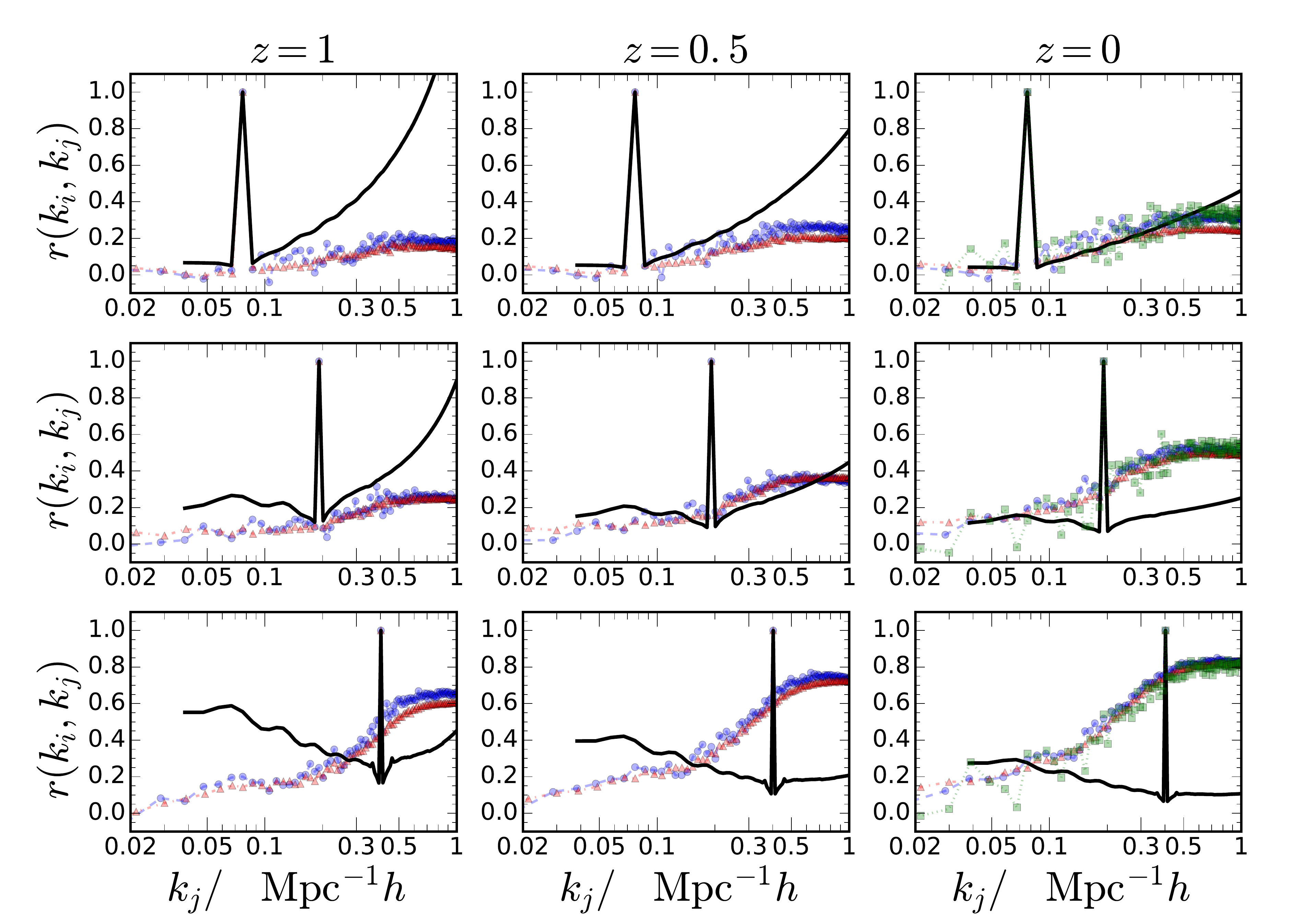}
\caption{     The correlation coefficients of the dark matter power spectrum covariance $r$  at $z=1$, 0.5, and 0 (from left to right). $r(k_i,k_j) $ as a function of $k_j$ for a list of fixed $k_i= 0.076 $, 0.19, and $0.40 \hOMpc $ (top to bottom).  The results from the Large (blue circles), Small (red triangles), and Hires (green squares) are shown.  The perturbation theory predictions are overplotted (black solid line). }
\label{fig:corr_coef_PkCov_DM}
\end{figure*}

\subsubsection{Dark matter}
 
The covariance of the dark matter power spectrum has been shown in Ref.~\cite{Blot:2014pga} using the same data set.  Here we only show the dark matter power spectrum results for completeness. We will also compare the results between the Small and Large set, while in Ref.~\cite{Blot:2014pga} only results from the Small set were shown.

The covariance depends on the volume and the binning.  For the power spectrum measurements, we choose the same band width $\Delta k = 9.6 \times 10^{-3} \hOMpc $ for all the simulations. We will normalize the covariance with respect to the Gaussian one. In this way, the volume dependence is expected to cancel out (see Eq.~\ref{eq:CP_G_continuous} and Eq.~\ref{eq:CP_trisp_general}).

%This means that even the Gaussian parts are different for the simulations of different box sizes.  Nonetheless we will often show the results as ratios with respect to the respective Gaussian covariance so that the results from different boxes agree with each other at least at large scales. 

We estimate the covariance of the power spectrum as
\beqa
&& C^P( k,  k' ) \nn \\
&=& \frac{1 }{ N-1 } \sum_{i=1}^N [ P_i(k) -\bar{P}(k) ][ P_i(k') -\bar{P}(k') ],
\eeqa
where $N$ is the number of realizations used and $\bar{P}$ is the mean of the power spectrum  measured from the simulations.

In Fig.~\ref{fig:Diag_PkCov_DM}, we plot the diagonal elements of $C^P$ for dark matter. We show the results using the three sets of simulations. The results are normalized with respect to the Gaussian covariance, in which the power spectrum is the linear one. The results from the Small and Large set agree with each other well.  At low $z$ the results from the Small and Hires are very similar, thus the mass resolution effects are small at low $z$, consistent with that reported  in Ref.~\cite{Blot:2014pga}.  We also plot the perturbation theory results, which include the nonlinear power spectrum correction to the Gaussian covariance and the trispectrum Eq.~\ref{eq:bar_trispectrum}.  The agreement between perturbation theory and simulation results seems to be worse than what shown in Ref.~\cite{Takahashi:2009ty}. A possible reason is that we have used the nonlinear power spectrum from simulations, which is lower than the 1-loop one in the nonlinear regime.

We now show the correlation coefficient $r_P$ defined as 
\beq
r_P (k_i, k_j)\equiv \frac{ C^P(k_i,k_j)  }{\sqrt{  C^P(k_i,k_i)  C^P(k_j,k_j) } }.  
\eeq
In Fig.~\ref{fig:corr_coef_PkCov_DM} we plot $r_P(k_i,k_j)$ as a function of $ k_j$, with $ k_i $ fixed at the values of $k_i = 0.076$, 0.19, and  $0.40 \hOMpc $ respectively. $r_p$ is expected to be independent of the simulation volume, and indeed we find that the results from a different simulation set agree with each other well.  When both $k_i $ and  $k_j$ are small, the agreement between the tree-level perturbation theory and the numerical results is reasonable, but it deteriorates for large $k_i$ or $k_j$ as the mode coupling becomes more significant in the nonlinear regime. See Refs.~\cite{Bertolini:2015fya,Mohammed:2016sre} for the improvement of the agreement by including the 1-loop corrections to the trispectrum.

For Gaussian distribution, the error of the mean goes like  $\propto 1/ \sqrt{ N_{\rm mode} N_{\rm real}}$, while the error of the covariance is $\propto 1/ \sqrt{ N_{\rm real}}$, where $N_{\rm mode} $ and $ N_{\rm real} $ are the number of modes in the bin and the number of independent realizations, respectively; see  e.g.~\cite{TaylorJoachimiKitching_2012}.   As the total volume of the Large set is the same as that of the Small set, we expect them to have a simiar error bar for the mean.  On the other hand, from Fig.~\ref{fig:corr_coef_PkCov_DM},  the correlation coefficients for the Large set is much noisier than that of the Small set because of small number of realizations.   We find that the small scale trispectrum contribution is largely insensitive to the simulation box and this is consistent with \cite{Mohammed:2016sre} (As we mentioned, the small differences between the Large and Small set could result from the output time fluctuations). Thus to get the small-scale covariance, we can run large number of small box size simulations to beat down the noise on the covariance without worrying about the volume effects.

\begin{figure}[!htb]
\includegraphics[width=\linewidth]{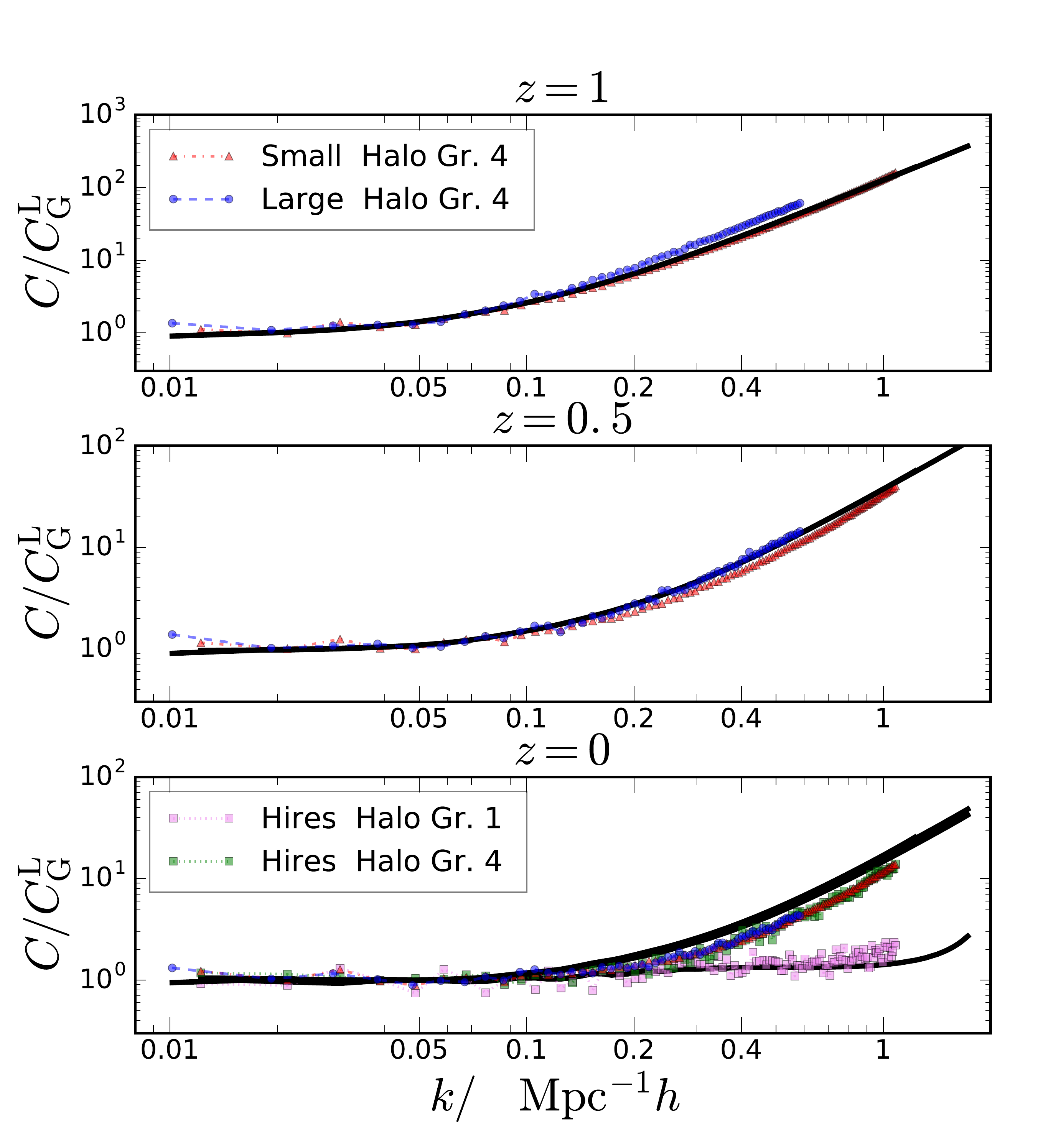}
\caption{  The diagonal elements of the halo power spectrum covariance. The mean Poisson shot noise is subtracted.   The results from Large Gr.~4 (blue circles), Small Gr.~4 (red triangles), and Hires Gr.~1 (violet squares) and  Gr.~4 (green squares) are shown. The predictions are computed using Eq.~\ref{eq:CNG_Poisson_short} (black lines).      }
\label{fig:Diag_PkCov_halo}
\end{figure}

\begin{figure*}[!htb]
\includegraphics[width=0.9\linewidth]{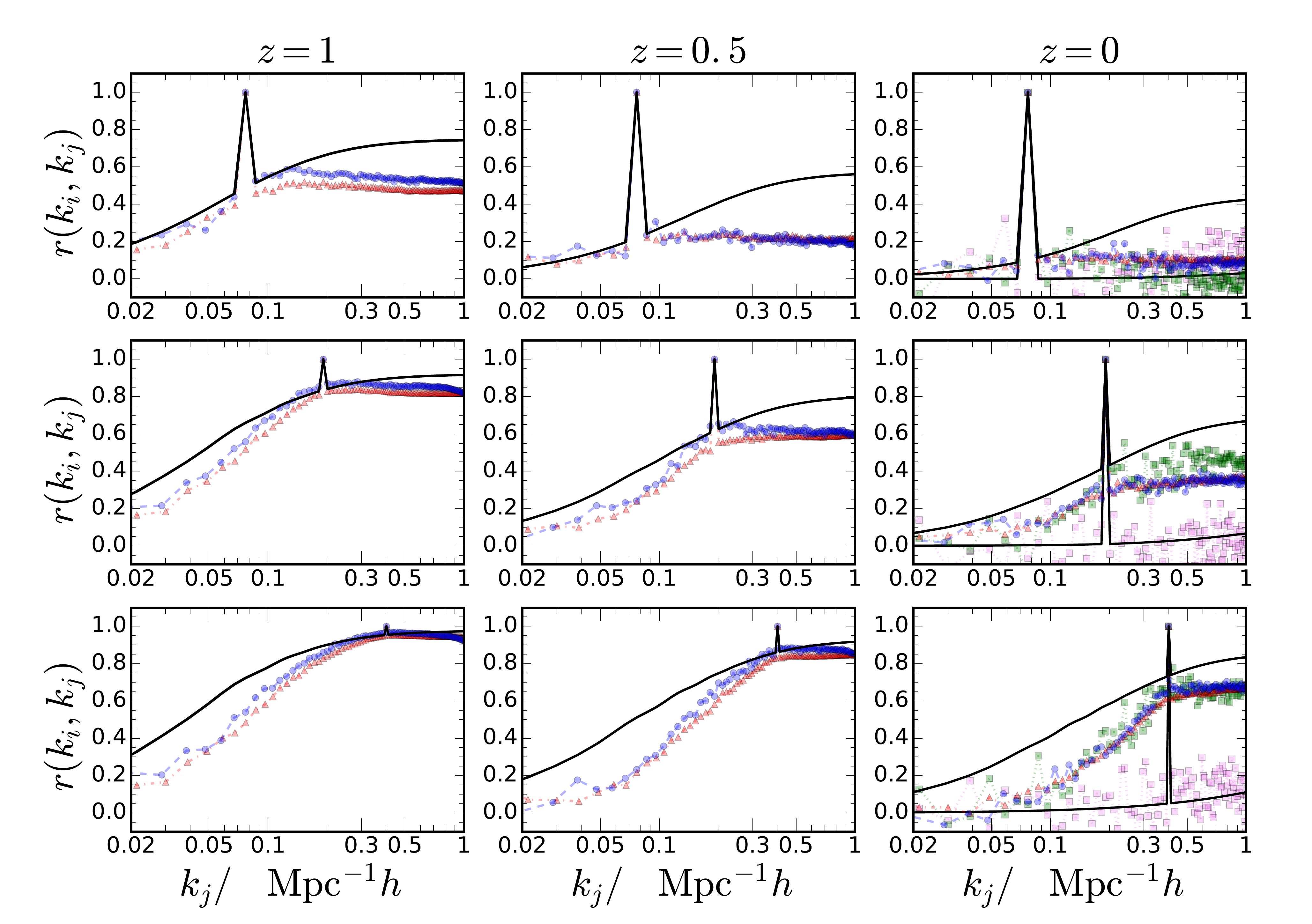}
\caption{   Similar to Fig.~\ref{fig:corr_coef_PkCov_DM}, except for the halo power spectrum covariance.  The mean Poisson shot noise is subtracted.   The results from Large Gr.~4 (blue circles), Small Gr.~4 (red triangles), and Hires Gr.~1 (violet squares) and  Gr.~4 (green squares) are shown.   The predictions are computed using Eq.~\ref{eq:CNG_Poisson_short} (black lines). At $z=0$, the lower curve is the theory prediction for Hires Gr.~1.       }
\label{fig:corr_coef_PkCov_halo}
\end{figure*}

%% \begin{figure}[!htb]
%% \includegraphics[width=\linewidth]{nva_epsilon2.pdf}
%% \caption{ $\epsilon^2 $, the fractional RMS variance of $n_{\rm va}$, for all the halo groups used in this work. The data from Large Gr.~4 (circles, blue), Small Gr.~4 (triangles, blue), and Hires Gr.~1 (squares, red), Gr.~2 (squares, green), Gr.~3 (squares, cyan), and  Gr.~4 (blue, cyan).  The measurements from the data is given at $z=0$, 0.5 and 1. The predictions with only Poisson fluctuation is given at $z+0.05$, while the one include both Poisson variance and sample variance is shown at  $z-0.05$.   }
%% \label{fig:nva_epsilon2}
%% \end{figure}

\begin{figure}[!htb]
  \includegraphics[width=\linewidth]{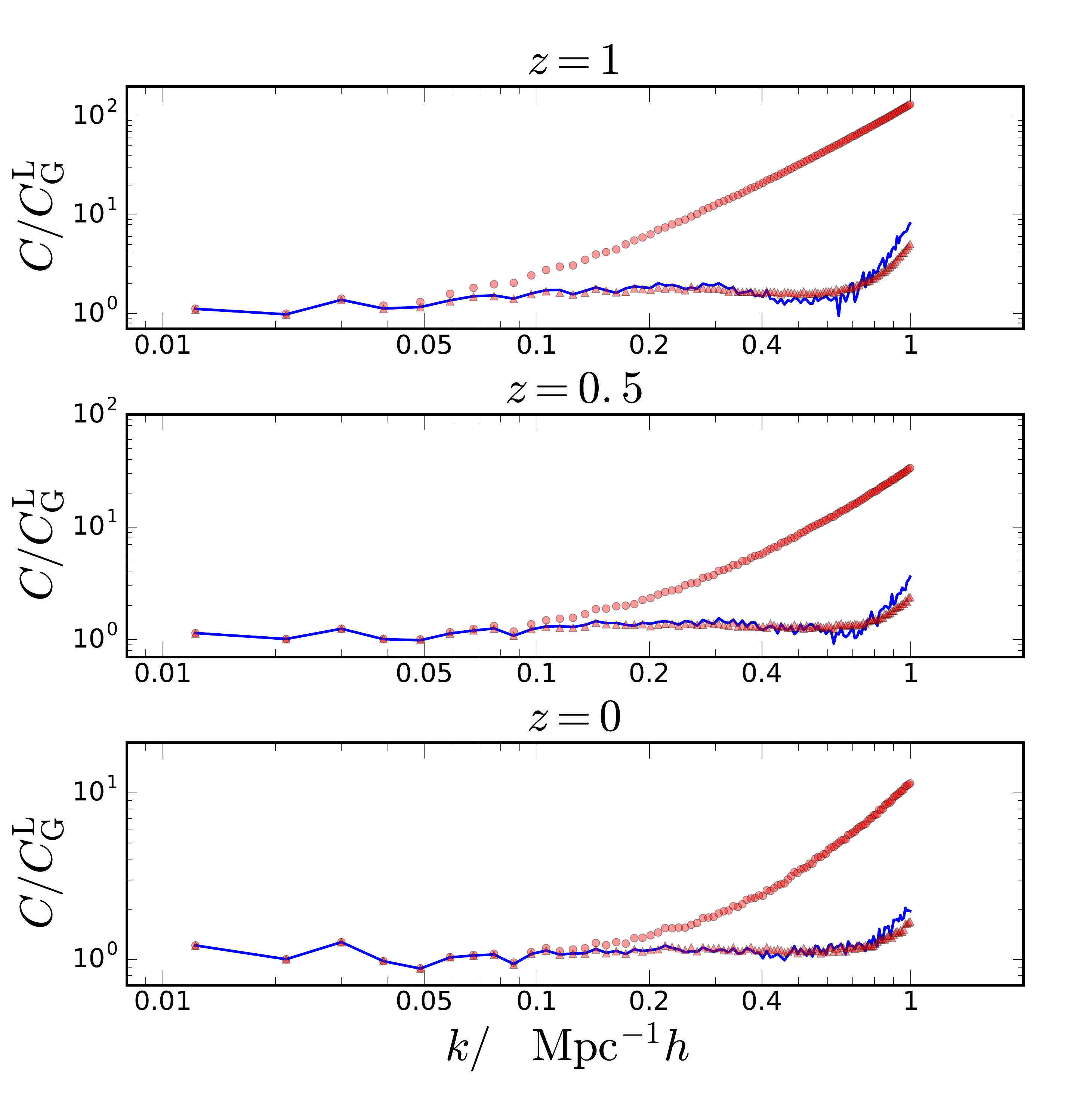}
  \caption{ The diagonal elements of the halo power spectrum with the mean shot noise subtracted (circles, upper data set) and the shot noise  estimated and subtracted from each realization (triangles, lower data set). The data from the Small set are used. The prediction using Eq.~\ref{eq:covP_Poisson_sub} is overplotted (solid blue).    }
\label{fig:Diag_PkCov_halo_shotsub}
\end{figure}

\begin{figure*}[!htb]
  \includegraphics[width=0.9\linewidth]{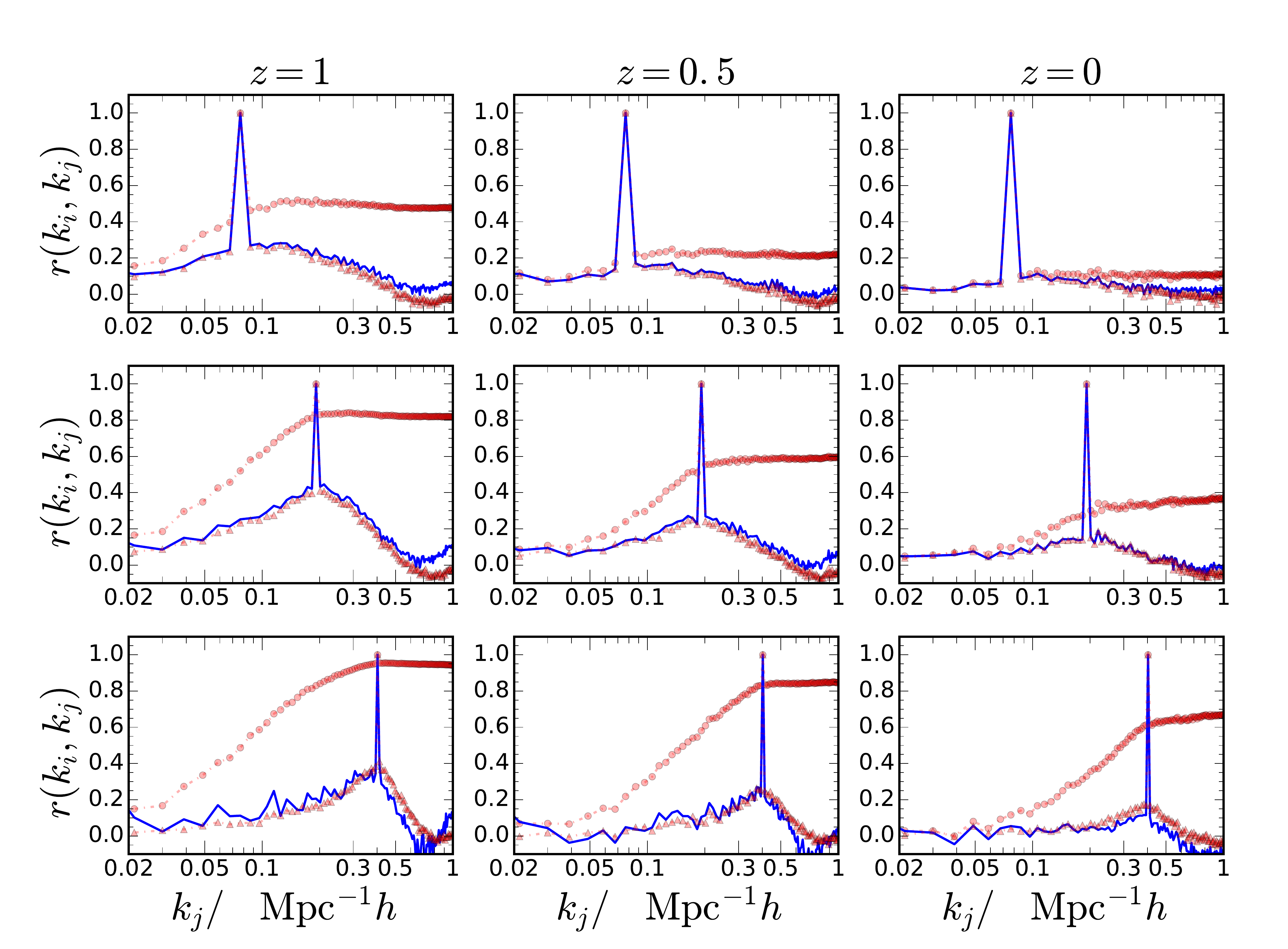}
\caption{ The correlation coeffcients for the halo power spectrum with the mean shot noise subtracted (circles, upper data set) and the shot noise  estimated and subtracted from each realization (triangles, lower data set). The data from the Small set is used.    The prediction using Eq.~\ref{eq:covP_Poisson_sub} is overplotted (solid blue).     }
\label{fig:corr_coef_PkCov_halo_shotsub}
\end{figure*}

\subsubsection{Halos} 
\label{sec:Pk_Halo_cov_numerical}

We now look at the the covariance of the halo power spectrum.    We are usually interested only in the continuous halo power spectrum signal, and the Poisson shot noise is subtracted using Eq.~\ref{eq:Pk_discrete}.  Caution must be taken for the $\bar{n}$ appearing in the shot noise formulas.  The mean density of halos $\bar{n}$ is obtained by ensemble average, but it is usually estimated using the volume-averaged density measured in a particular simulation/survey. We will see that using the volume average number density results  in a  substantially smaller covariance.

We first look at the case when the ensemble averaged number density is used. This corresponds  directly to the Poisson model prediction given in Appendix   \ref{sec:appendix_PoissonPk}.   The  ensemble averaged number density is obtained by further averaging the volume-averaged ones over the realizations of the simulations. It is clear that in this case the covariance of the power spectrum with the Poisson shot noise subtracted is the same as that of the raw power spectrum. Here we use raw power spectrum to refer to the one measured directly from simulation without any shot noise subtraction.  

In Fig.~\ref{fig:Diag_PkCov_halo}, we show the diagonal elements of the covariance of the halo power spectrum. We normalize the results using the Gaussian covariance in Eq.~\ref{eq:CG_Poisson}.  Halo groups from Large, Small, and Hires sets are used.  Similar to the dark matter case, we find that the results from different sets are similar. The differences between the Large and Small may be due to the output time fluctuations.  Because of the low number of realizations available, the trends for the low bias groups from the Hires set are noisy. For clarity, we only show the results from Hires group 1 and 4. Thanks to the smallness of the non-Gaussian corrections for the abundant halo group  as expected from Fig.~\ref{fig:Pkcov_Poisson_components}, the model agrees well with simulation for Group 1.  Overall, the Poisson model including the non-Gaussian corrections gives a reasonably good agreement with the numerical covariance up to $k \sim 1\hOMpc $.

%Nonetheless, the model appears to underpredict the covariance.  \red{ This is surprising given that the non-Gaussian correction correction is expected to be small? Do you use the halo power specturum measured from simulation?   

%For larger $k$, the model underpredicts the covariance. This is also expected because the non-Gaussian corrections are more significant in this regime  as can be seen from Fig.~\ref{fig:Pkcov_Poisson_components}. Moreover, the other terms left out in the prediction may play a non-negligible role in this regime. 

To compute the prediction we use a simple linear bias model 
\beq
\label{eq:linear_bias_Ph}
P_{\rm h}(k)  = b_1^2 P(k).  
\eeq
The linear bias $b_1$ is obtained by fitting the model to the mean of the auto power spectrum  up to $k=0.05 \hOMpc$. The best-fit values are shown in Table \ref{tab:halo_group}. To compute the prediction in the Poisson model we should use the fully nonlinear polyspectra, but for the high mass group 4, we find that the results are  similar even if we use the nonlinear power spectrum measured from simulations instead. However, for group 1, the nonlinear power spectrum must be used because the magnitude of the halo power spectrum is comparable to the Poisson shot noise for this abundant group.

In Fig.~\ref{fig:corr_coef_PkCov_halo}, we plot the correlation coefficient $r_P$ for the halo power spectrum. We choose the same sets of $k_i$ as in Fig.~\ref{fig:corr_coef_PkCov_DM}. We note that for dark matter, $r_P( k_i,k_j )$  generally increases as $k_j$ goes beyond the pivot scale $k_i $, while for halos, $r_P( k_i,k_j )$  tends to level off or increases very mildly beyond the pivot scale. When $k_i $ is small, the model can predict $r_P(k_i, k_j)$ near the pivot scale, but tends to overpredict it when the separation from the pivot is large. We also note that the model performs worse at low $z$ than at high $z$. This is  because at low $z$ higher order correlators are more important.

We now consider the case when the Poisson shot noise obtained with the volume average number density is subtracted from each realization. The covariance can then be expressed as 
\beqa
\label{eq:covP_Poisson_sub}
&& \mathrm{cov} \big[ P_{\rm raw} (k) - P_{\rm Pois}, P_{\rm raw} (k') - P_{\rm Pois} \big] \nn \\
%% & =& \mathrm{cov} \big(P_{\rm raw} (k), P_{\rm raw} (k')  \big) -  \mathrm{cov} \big(P_{\rm raw} (k), P_{\rm Pois} (k')  \big) \nn \\
%% & -&   \mathrm{cov} \big(P_{\rm Pois} (k), P_{\rm raw} (k')  \big)  +   \mathrm{cov} \big(P_{\rm Pois} (k), P_{\rm Pois} (k')  \big),   \nn \\
&\approx &  \mathrm{cov} \big[ P_{\rm raw} (k), P_{\rm raw} (k')  \big] - \mathrm{var} \big( P_{\rm Pois} \big),
\eeqa 
where $ P_{\rm Pois}   $ is defined as
\beq
\label{eq:P_Pois}
 P_{\rm Pois} = \frac{ 1 }{ (2 \pi)^3   n_{\rm va} }, 
\eeq
with $ n_{\rm va} $ being  the volume-averaged number density obtained in a particular realization. In Eq.~\ref{eq:covP_Poisson_sub}, we have assumed that
\beq
\label{eq:PoisRaw_correlation}
 \mathrm{cov} \big[ P_{\rm Pois} , P_{\rm raw} (k')  \big] \approx  \mathrm{var} \big( P_{\rm Pois}  \big). 
\eeq
We shall see that this is indeed a good approximation. We find that the fluctuations of the number density of the halo groups in the simulation volume can be modelled as a Poisson fluctuation. %% i.e.

In Fig.~\ref{fig:Diag_PkCov_halo_shotsub} we plot the diagonal elements of the halo power spectrum covariance when the Poisson shot noise from the individual realization is subtracted. For clarity we only show the results from the Small set.  Note that we still normalize the covariance using   Eq.~\ref{eq:CG_Poisson}. First, we find that the resultant covariance becomes much closer to the Gaussian one. This is good news  because for the more realistic scenario, where the mean density is computed as a volume average over the survey/simulation, the halo power spectrum covariance  is  reduced and is easier to predict. This makes sense intuitively since when the number density is estimated from the realization, part of the fluctuations is absorbed in the shot noise term.  The distinction between the local volume-averaged density and the global ensemble one  is analogous to the effects arising from defining the density contrast with the local or global average density found in Ref.~\cite{dePutter:2012}.  We show the results obtained using Eq.~\ref{eq:covP_Poisson_sub}, in which the variance of the number density is measured from simulations. We see that it agrees with simulation results very well. This validates Eq.~\ref{eq:PoisRaw_correlation} and shows that the Poisson shot noise does not correlate with the clustering.

Similarly, when the Poisson shot noise is subtracted from the individual realizations correlation coefficients are also significantly reduced as shown in Fig.~\ref{fig:corr_coef_PkCov_halo_shotsub}. The predictions using Eq.~\ref{eq:PoisRaw_correlation} also result in very good agreement with the simulation results. However, even though from Fig.~\ref{fig:Diag_PkCov_halo_shotsub}, it appears that after subtraction of the individual shot noise, the diagonal elements are very close to the Gaussian one,  it is clear from   Fig.~\ref{fig:corr_coef_PkCov_halo_shotsub} that after subtracting the variance of $P_{\rm Pois } $, there are still large cross correlation coefficients.

In summary, our results show that the difference between the two different ways to subtract the Poisson shot noise simply arises from the fluctuation in the number density of the sample, and its effect can be modelled by a fluctuating $ P_{\rm Pois}$.  The fluctuations in the Poisson shot noise account for large amount of non-Gaussianity in the covariance matrix.

In the plots of this section the covariances are always normalized with respect to the Gaussian covariance. Here we comment on the magnitudes of the matter and halo power spectrum covariances. For the massive halos in group 4, the power spectrum covariance is 3 to 5 order of magnitudes higher than the matter one, depending on the redshift in question. Thus for these kinds of halos, the shot noise contribution to the covariance completely dwarfs the matter power spectrum covariance.   However, as the number density of halos increases, the shot noise contribution to the power spectrum decreases. For the Hires group 2, we still find that the halo power spectrum covariance is one order of magnitude higher than that of the dark matter. When the number density is as high as that of the Hires group 1, we find that the halo covariance is comparable in magnitude to that of the dark matter. These explain why the simple linear bias model works well except for the most abundant groups.  In Euclid, the number density of galaxies is expected to reach $ \gtrsim 10^{-3} (\MpcOh)^{-3} $ \cite{EuclidRedBook}, hence the covariance is expected to get non-negligible contributions from dark matter nonlinearity and galaxy biasing.

Finally we point out that \cite{Smith:2008ut} also studied the covariance of halo power spectrum using $N$-body simulations although using only 30 simulations of box size $1500 \MpcOh$. Our results are similar to those in \cite{Smith:2008ut} regarding the effects of different shot noise subtraction procedures to the covariance.  Here we go on to show that the fluctuating Poisson shot noise term accounts for most of the non-Gaussianity.

% In particular,  our finding that the Gaussian shot noise model works reasonably well for the samples at $z=0$ with the shot noise subtracted from each realization agrees with Ref.~\cite{Smith:2008ut}.

\section{ Covariance of the bispectrum }
\label{sec:bispectrum_covariance}
In this section, we first discuss the theory of the bispectrum covariance for dark matter and halo.   Then we present the numerical covariance results and the comparison with theory. 
 
\subsection{ Theory of the bispectrum covariance }

The theoretical discussion on the bispectrum covariance in previous works has been mainly limited to the Gaussian contribution to the covariance \cite{SCFFHM98,Scoccimarro:2003wn}.  This is partly because the bispectrum covariance has a relatively large number of elements and the number of available realizations in most of the existing simulation sets is not large enough to get good signal to noise.  Here we will consider the  non-Gaussian  contribution as well. We will see shortly that this is crucial to get good agreement with the simulation results.

Given the definition of the bispectrum
\beq
 \langle   \delta(\mb{k}_1)  \delta(\mb{k}_2)  \delta(\mb{k}_3)  \rangle = B(k_1,k_2,k_3) \Ddel( \mb{k}_{123} )  ,
\eeq
we can construct an estimator as \cite{SCFFHM98,Scoccimarro:2003wn}
\beqa
\label{eq:Bisp_estimator}
  \hat{B}(k_1,k_2,k_3)  
& = & \frac{ k_{\rm F}^3 }{ V_{123}  }\int_{ k_1 }d^3 p \int_{ k_2 }d^3 q \int_{ k_3 }d^3 r  \,   \  \nn \\  
& \times &  \Ddel( \mb{p} + \mb{q} + \mb{r} )   \delta(\mb{p})  \delta(\mb{q} )  \delta(\mb{r} ) , 
\eeqa
where $k_i$ indicates the integration is over a spherical shell of width $[ k_i - \Delta k /2 , k_i + \Delta k /2 ) $, with $\Delta k $ being the width of the bin in Fourier space.  The term $V_{123}$ counts the number of modes satisfying the triangle constraint:
\beq
\label{eq:V123}
V_{123} = \int_{k_1} d^3 p \int_{k_2} d^3 q \int_{k_3} d^3 r \,   \Ddel( \mb{p} + \mb{q} + \mb{r} ).
\eeq 
%In Eq.~\ref{eq:Bisp_estimator}, we have used the relation $\Ddel( \mb{k}=\mb{0} ) = 1 /  k_{\rm F}^3 $, where $ k_{\rm F}$ is the fundamental model of the box. 
In fact, we can compute $ V_{123} $ analytically to get \cite{Scoccimarro:2003wn}
\beq
\label{eq:V123_final}
V_{123} =   8 \pi^2 k_1 k_2 k_3 ( \Delta k )^3 \beta(\Delta ) , 
\eeq
where $\Delta  $ is defined as 
\beq
\label{eq:Delta_triangle}
\Delta = \hat{\mb{k}}_1 \cdot \hat{\mb{k}}_2 =  \frac{ k_3^2 - k_1^2 -  k_2^2 }{ 2 k_1 k_2  },
\eeq
and $\beta (\Delta) $ is given by
\beq
\label{eq:beta_Delta}
\beta(\Delta )  = \begin{cases} \frac{1}{2}  \, & \mbox{if } \Delta  =  \pm 1    \\
                                1               &\mbox{if }  0  < \Delta < 1   \\
                                0               &\mbox{otherwise }  
              \end{cases} .
\eeq
 For more details on the derivation of Eq.~\ref{eq:V123_final}, see Appendix \ref{sec:derivation_zoo}.  In Appendix \ref{sec:bisp_estimaotr_distribution}, we check the probability distribution of $\hat{B} $. We find that it is close to Gaussianly distributed but with non-negligible skewness and kurtosis as well.

The covariance matrix  of $\hat{B}$, $C^B$  is defined as:
\beqa
&&  C^B( k_1, k_2, k_3, k_1', k_2', k_3'  ) \nn \\
& \equiv & \mathrm{cov}[ \hat{B}(k_1, k_2, k_3), \hat{B}(k_1', k_2', k_3')  ]   \nn \\
&=& \langle  \hat{B}(k_1, k_2, k_3) \hat{B}(k_1', k_2', k_3')  \rangle  \nn \\  
&& \quad     -  \langle  \hat{B}(k_1, k_2, k_3)  \rangle   \langle   \hat{B}(k_1', k_2', k_3')  \rangle.
\eeqa 
In the following we will skip the superscript $B$ in the notation for the sake of simplicity.   %The contribution to  $ C $ arises from both the disconnected and connected 6-point function in $ \langle \hat{B} \hat{B}' \rangle $.  

\subsubsection{Dark matter }
\label{sec:Bcov_DMtheory}
When $\delta $ is Gaussian, $\langle \hat{B}\rangle  $ is zero, however $ C $ does not vanish. 
The Gaussian covariance can be written as \cite{Scoccimarro:2003wn}
\beq
\label{eq:BCov_Gaussian} 
C_{\rm G}^{\rm L} =  \frac{ k_{\rm F}^3 }{V_{123} }  \delta_{k_1k_2k_3, k_1'k_2'k_3'  } s_{123}   P_{\rm L}(k_1) P_{\rm L}(k_2) P_{\rm L}(k_3) ,
\eeq
where  $\delta_{k_1k_2k_3, k_1'k_2'k_3'  }$  is non-vanishing only if the shape of the triangle  $ k_1k_2k_3 $ is the same as that of $ k_1'k_2'k_3' $. If none of the sides of the triangle are equal to each other,  $ s_{123} =1$. If the triangles are isosceles, $ s_{123} =2$. For equilateral triangles, we have  $ s_{123} = 6$. The derivation of Eq.~\ref{eq:BCov_Gaussian} is reviewed in Appendix \ref{sec:derivation_zoo}.   In Eq.~\ref{eq:BCov_Gaussian}, for Gaussian $\delta$,  $P_{\rm L}(k) $ is the  linear power spectrum. As we consider the non-Gaussian contribution below, we find that part of the contribution can be resummed if we use the 1-loop matter power spectrum instead of the linear one for one of the power spectra.  However, the 1-loop power spectrum overestimates the matter power spectrum from simulation in the weakly nonlinear regime already. Similar to the case of power spectrum, we shall use the nonlinear power spectrum measured from simulations in place of the 1-loop results.  We will use the notation $ C_{\rm G}^{\rm NL} $ to distinguish the case when the nonlinear power spectrum is used, that is
\beqa
\label{eq:CG_B_NL}
&&    C_{\rm G}^{\rm NL} = C_{\rm G}^{\rm L} +  \frac{ k_{\rm F}^3 }{V_{123} }  \delta_{k_1k_2k_3, k_1'k_2'k_3'  } s_{123}   \nn \\
& \times &[ P_{\rm L}(k_1) P_{\rm L}(k_2) (P_{\rm NL}(k_3) - P_{\rm L} (k_3) ) + 2 \cyc  ],
\eeqa
where cyc.~denotes cyclic permutations and $ P_{\rm  NL} $ denotes the nonlinear power spectrum.

\begin{figure}[!htb]
\centering
\includegraphics[width=0.9\linewidth]{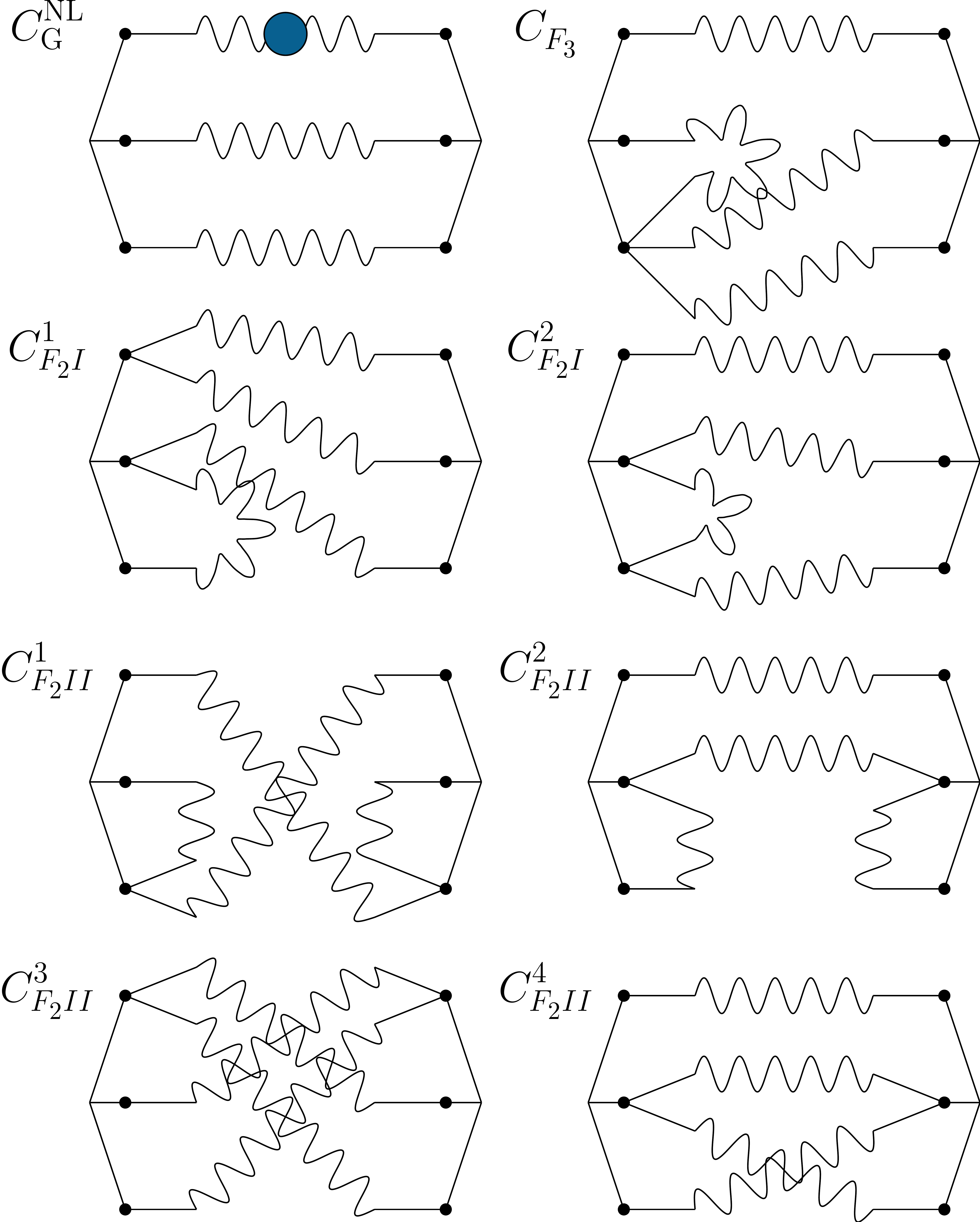}
\caption{  A  diagrammatic representation of the covariance of the matter bispectrum to the tree level order. $ C_{\rm G}^{\rm NL} $ is the Gaussian term, while the rest are non-Gaussian contributions.  The set of black dots on the left and right of each diagram represents the three $\delta $'s in each bispectrum estimator.      The legs branching from each dot represent the perturbation theory kernel, $F_1$, $F_2$, and $F_3$ respectively.  Each wavy line represents the linear power spectrum. In $C_{\rm G}^{\rm NL }$, the filled circle indicates that the linear power spectrum is replaced by the nonlinear power spectrum.  The Gaussian term couples only triangles of the same shape, while the non-Gaussian terms couple triangles with at least one side equal to each other. %The terms in the second column can be written as a product of power spectrum and trispectrum, and non-Gaussian terms in the first column can be expressed as bispectrum squared.
}
\label{fig:Covmat_Bk_diagrams}
\end{figure}

In the top-left  corner of Fig.~\ref{fig:Covmat_Bk_diagrams}, we show a diagrammatic representation of  $C_{\rm G}^{\rm NL}$.  The rules are the same as those in Fig.~\ref{fig:Covmat_Pk_diagrams}.   The black dots on the left and right hand side represent the three $\delta$'s in each of the bispectrum estimator $\hat{B} $. The curly line represents the linear power spectrum. We put a filled circle on one of the curly line to indicate that the nonlinear power spectrum is used instead of the linear one. We see that the Gaussian term  has the same structure as the Gaussian covariance for power spectrum  in Fig.~\ref{fig:Covmat_Pk_diagrams}. The Gaussian covariance for the bispectrum is also inversely proportional to the number of modes in the bins (and it scales with bin width as $1/(\Delta k )^3$).

Nonlinear evolution causes mode coupling and departure from Gaussianity.  At the tree-level order, there are both connected and disconnected contributions to the 6-point function. In this paper we shall evaluate the leading disconnected tree-level contributions only. We will estimate the leading connected contributions and show that they are subleading relative to the terms we consider.

%take the ansatz that the disconnected part is more important than the connected ones in the weakly nonlinear regime, and focus only on the disconnected non-Gaussian contributions to the 6-point function. This ansatz is partly motivated by the fact that the Gaussian contribution is also disconnected.

The disconnected non-Gaussian tree level contributions can arise either from the $F_3 $ or $F_2 $ kernel.   In Fig.~\ref{fig:Covmat_Bk_diagrams}, we show a graphical representation of these non-Gaussian contributions. There is one contribution due to $F_3$,  $C_{F_3}$, in which $F_3$  is represented as three legs branching from a black dot.  The non-Gaussian tree level can also arise from two $F_2$ kernels. We further classify these diagrams into type I if both $F_2$ kernels are on the same triangle, and type II if they are distributed on different triangles.  There are two type I diagrams and four type II.

%In Fig.~\ref{fig:F3_Bcovmat} we show the diagrams for the nonGaussian contribution arising from one $ F_3 $ kernel, while in Fig.~\ref{fig:F2I_Bcovmat} and \ref{fig:F2II_Bcovmat}, we show the type I and type II diagrams from two $F_2$ kernels. If two $F_2$ kernels are on the same side of the bispectrum, it is classified as type I, while if the two $F_2$ kernels are distributed on different sides of the bispectrum, it is type II.  Note that some of the diagrams are resummed when linear power spectrum in the Gaussian covariance are replaced by the nonlinear one. 

%% \begin{figure}[!htb]
%% \centering
%% \includegraphics[width=0.9\linewidth]{F3_Bcovmat.pdf}
%% \caption{  The $F_3$ contribution to the nonGaussian covariant matrix of the bispectrum.    }
%% \label{fig:F3_Bcovmat}
%% \end{figure}

%% \begin{figure}[!htb]
%% \centering
%% \includegraphics[width=0.9\linewidth]{F2I_Bcovmat.pdf}
%% \caption{  The type I of the $F_2$ contribution to the nonGaussian covariant matrix of the bispectrum.    }
%% \label{fig:F2I_Bcovmat}
%% \end{figure}

%% \begin{figure*}[!htb]
%% \begin{center}
%% \includegraphics[width=0.9\linewidth]{F2II_Bcovmat1.pdf}
%% \includegraphics[width=0.9\linewidth]{F2II_Bcovmat2.pdf}
%% \caption{ The type II of the $F_2$ contribution to the nonGaussian covariant matrix of the bispectrum. }
%% \label{fig:F2II_Bcovmat}
%% \end{center}
%% \end{figure*}

\begin{widetext}

First for convenience we define the notation
\beq
\label{eq:int_D_notation}
\int \mathcal{D} \equiv \frac{ k_{\rm F}^3 }{ V_{123}  V_{123}'  } \int_{k_1}d^3 p \int_{k_2}d^3 q \int_{k_3}d^3 r  \Ddel( \mb{p} + \mb{q} + \mb{r} )  \int_{k_1'}d^3 p' \int_{k_2'}d^3 q' \int_{k_3'}d^3 r'  \Ddel( \mb{p}' + \mb{q}' + \mb{r}' )  . 
\eeq
The contribution due to $F_3 $ reads
\beqa
\label{eq:NGB_CF3}
C_{F_3} = 6  \int \mathcal{D} \,  \Ddel( \mb{p} + \mb{p}' ) F_3(\mb{q} , \mb{q}' , \mb{r}' ) P(p) P(q) P(q') P(r')  +    [( 3!\times 3 + k_{123} \leftrightarrow k_{123}'   ) -1 ] \cyc 
\eeqa
By inspecting  the diagrammatic representation of  $ C_{F_3}$ in Fig.~\ref{fig:Covmat_Bk_diagrams}, it is easy to see that there are 3! permutations for $k_1k_2k_3 $ and 3 permutations for $k_1'k_2'k_3'$ as two of the legs are symmetric. There are  also additional contributions from interchanging  $k_1k_2k_3 $ and $k_1'k_2'k_3'$. Thus there are altogether 36 permutations. For other diagrams the number of permutations can be worked out in a similar manner.

The type I contributions due to $F_2 $ are 
\beqa
C_{F_2 I}^1 & =& 2\int \mathcal{D} \,  \Ddel(  \mb{p} - \mb{p}' ) F_2( \mb{q}, \mb{r}' )    F_2( \mb{p}', \mb{r} )  P_{\rm L}(q) P_{\rm L}(r') P_{\rm L}(p') P_{\rm L}(r) + [( 3!\times 3 + k_{123} \leftrightarrow k_{123}'    ) -1] \cyc,    \\
C_{F_2 I}^2 &=& 8 \int \mathcal{D} \, \Ddel(  \mb{p} + \mb{p}' ) F_2(- \mb{q}', \mb{q} +  \mb{q}' )    F_2( \mb{q} +  \mb{q}' , \mb{r}' )  P_{\rm L}(p) P_{\rm L}(q') P_{\rm L}(r') P_{\rm L}(| \mb{q} + \mb{q}'  |) \nn \\
& + & [( 3\times 3 + k_{123} \leftrightarrow k_{123}'    ) - 1 ] \cyc
\eeqa

The type II contribution reads 
\beqa
C_{F_2 II}^1 & =& 4 \int \mathcal{D} \, \Ddel(  \mb{p} -  \mb{p}' )  F_2( \mb{q}, \mb{p} )   F_2( \mb{q}', \mb{p} )  P_{\rm L}( q ) P_{\rm L}(p') P_{\rm L}( p ) P_{\rm L}(q'  )  +  [( 3! \times 3! ) -1] \cyc,   \\
C_{F_2 II}^2 & =& 4 \int \mathcal{D} \,  \Ddel(  \mb{p} +  \mb{p}' )  F_2( \mb{r}, \mb{p} )   F_2( -\mb{p}, \mb{r}' )  P_{\rm L}( p ) P_{\rm L}(r) P_{\rm L}( |\mb{q} + \mb{r} | ) P_{\rm L}(r')  +  [( 3! \times 3! ) -1] \cyc  \\
C_{F_2 II}^3 & =& 4 \int \mathcal{D} \,  \Ddel(  \mb{p}  -  \mb{p}' )  F_2( \mb{q}, \mb{r} )   F_2( \mb{q}', \mb{r}' )  P_{\rm L}( q ) P_{\rm L}(q') P_{\rm L}( r ) P_{\rm L}(r')  +  [( 3 \times 3 ) -1 ] \cyc  \\
\label{eq:NGB_CF2II_4}
C_{F_2 II}^4 & =& 4 \int \mathcal{D} \,  \Ddel(  \mb{p}  +  \mb{p}' )  F_2( -\mb{r}',  \mb{q} + \mb{r}' )   F_2( \mb{q} + \mb{r}', \mb{r} )  P_{\rm L}( p ) P_{\rm L}(r') P_{\rm L}( r ) P_{\rm L}(| \mb{q} + \mb{r}'   |)  +  [( 3! \times 3! ) -1] \cyc
\eeqa
\end{widetext}
Note that the expressions are tree-level only, thus there is no loop integration, and the high-dimensional integrals result from the bin width integration. The leading non-Gaussian terms we consider here are all of the order  $P_{\rm L}^4 $.  All the terms contain a factor of Dirac delta function  $\Ddel( \mb{p} \pm \mb{p}')  $, and this implies that these terms couple only triangles with at least one side equal to each other.

In each of the non-Gaussian terms  there are four Dirac delta functions: two due to the triangle constraints $\Ddel(\mb{p} + \mb{q} + \mb{r})  $ and  $\Ddel(\mb{p}' + \mb{q}' + \mb{r}')  $, and another two imposed on two disjoint sets of $\mb{p}$, $ \mb{q}$, $ \mb{r}$, $\mb{p}'$, $\mb{q}'$, and   $\mb{r}' $. These can also be seen by inspecting the non-Gaussian diagrams in Fig.~\ref{fig:Covmat_Bk_diagrams}.   Hence it is clear that one of these latter  two  Dirac delta functions is redundant, and it gives $\Ddel(\mb{0} ) $.  That is why in  $\int \mathcal{D} $ there is a factor of $k_{\rm F}^3 $ only. Hence the non-Gaussian terms have the same volume dependence as the Gaussian one, and we will use this observation to compare simulation results from different boxes.  The other Dirac delta function relates two vectors, each coming from one of the bispectrum estimator, which is $\Ddel( \mb{p} \pm \mb{p}')  $ in Eq.~\ref{eq:NGB_CF3}--\ref{eq:NGB_CF2II_4}.  This Dirac delta function is analytically integrable, and thus these integrals are non-vanishing only for triangles with at least  one side equal to each other.  We will make use of the pattern of the Dirac delta functions discussed here to construct an efficient Monte Carlo integration method.

Except for the Dirac delta $\Ddel(  \mb{p} \pm  \mb{p}' )  $, the other two remaining Dirac delta functions in general cannot be integrated analytically. Thus the resultant 15-dimensional integral is hard to compute. High dimensionality causes problems for numerical integrators in general. The presence of the remaining two Dirac delta functions makes the integrand non-vanishing only in narrow regions. Although high dimensional integrals can often be attacked by  Monte Carlo integration method, generic Monte Carlo integration would fail as they would miss the narrow peaks in the high dimensional space.

Here we present a Monte Carlo method that can efficiently sample the points that satisfy the Dirac delta function constraints. We first note that the vectors that fulfil the triangle constraint must be some small perturbations of the triangle  $k_1k_2k_3$  and  $k_1'k_2'k_3'$. Thus instead of sampling all the points in the full integration domain, we can proceed as follows. We first generate a vector $\mb{p} $ in the  $k_1$ shell randomly. For $\mb{q} $ we must have  $\hat{ \mb{p} }\cdot \hat{ \mb{q} } \equiv \mu  \approx \Delta $.   To determine the allowed variation of $  \mu $, we consider
\beq
d \Delta = \frac{ k_1^2 - k_3^2 - k_2^2  }{ 2 k_1 k_2^2 } dk_2 + \frac{k_3  }{k_1 k_2  } d k_3. 
\eeq
Hence we sample $\mu $ uniformly in the range  $\pm \Big[ \Big( \frac{ k_1^2 - k_3^2 - k_2^2  }{ 2 k_1 k_2^2 } \Delta k \Big)^2   + \Big(  \frac{k_3  }{k_1 k_2  } \Delta  k  \Big)^2 \Big]^{1/2} $. To further fix  $\mb{q}$, we sample the polar angle of  $\mb{q}$,  $\theta_2 $,  uniformly in the interval $[0,\pi]$. The azimuthal angle of  $\mb{q}$, $\phi_2 $ is fixed by the relation
\beq
\cos( \phi_1 - \phi_2 ) = \frac{ \mu - \cos \theta_1 \cos \theta_2 }{ \sin \theta_1 \sin \theta_2  },
\eeq 
where $\theta_1$ and $\phi_1 $ are the spherical coordinates of $\mb{p} $.    If the length $|\mb{p} + \mb{q} | $ falls within the interval $[ k_3 -\Delta / 2,  k_3 + \Delta / 2  )$, the vector $\mb{r} $ is assigned to be $- (\mb{p} + \mb{q} )$, and the three vectors $\mb{p}$, $\mb{q}$, and  $\mb{r} $ are accepted, otherwise the procedure is repeated until the proposed vectors are accepted.  We find that the acceptance rate can reach about 20\% and it does not vary much with the triangle configuration considered.  For the triads  $\mb{p}'$, $\mb{q}'$, and  $\mb{r}' $, we make use of the Dirac delta function $\Ddel( \mb{p} \mp  \mb{p}') $ and assign   $ \mb{p}' = \pm \mb{p} $ accordingly. The construction of  $\mb{q}'$ and  $\mb{r}' $ is then similar to those for   $\mb{q}$ and  $\mb{r} $.

After developing an efficient algorithm to sample the vectors satisfying the constraints imposed by the three Dirac delta functions, we can attack the non-Gaussian integrals using the Monte Carlo method (see e.g.~\cite{Press:2007:NRE:1403886} for a review).  The integrals can  be schematically written as 
\beqa
\label{eq:CNG_MCintegral}
I_{\rm NG } &=& \int \mathcal{D} \Ddel( \mb{p} \mp  \mb{p}'  )  f( \mb{p}, \mb{q}, \mb{r} , \mb{p}', \mb{q}', \mb{r}' ) \nn \\
&=&  \langle  f( \mb{p}, \mb{q}, \mb{r} , \mb{p}', \mb{q}', \mb{r}' ) \rangle  \int \mathcal{D} \Ddel( \mb{p} \mp  \mb{p}'  ) , 
\eeqa
where the integrand $f$ is averaged over the points in the integration domain defined by the Dirac delta functions. The success of this method relies on the fact that the integration volume can be computed analytically and it reads
\beq
\label{eq:intD_integral}
\int \mathcal{D} \Ddel( \mb{p} \mp  \mb{p}' )  = \frac{ k_{\rm F}^3  }{ V_{123} V_{123}'  }  U( k_1, k_1' ), 
\eeq
where $U$ is given by 
\beqa
\label{eq:Uintegral}
&&U(k_1 , k_1';  k_2, k_3, k_2', k_3' )  \nn \\
& =&  2^4 \pi^3 k_2 k_3 k_2' k_3' (\Delta k)^5  \beta( \Delta ) \beta( \Delta' )  \delta_{k_1 , k_1'} .      
\eeqa
In the Appendix \ref{sec:derivation_zoo}, we show the derivation of $U$.

From Eq.~\ref{eq:CNG_MCintegral} and \ref{eq:intD_integral}, we find that the non-Gaussian integral corrections scale with the bin width $\Delta k $ as $ 1 / \Delta k $. In contrast the Gaussian term  $C_{\rm G} $ exhibits a stronger  scaling  $1/ (\Delta k )^3 $.   From Eq.~\ref{eq:intD_integral} and \ref{eq:Uintegral}, we explicitly see that the non-Gaussian terms scale with the volume of the simulation as $k_{\rm F}^3 $, the same as the Gaussian term. The volume dependence is the same as that for the power spectrum covariance. As we mentioned, this is a consequence of the statistical translational invariance.

\begin{figure}[!tb]
\begin{center}
\includegraphics[width=\linewidth]{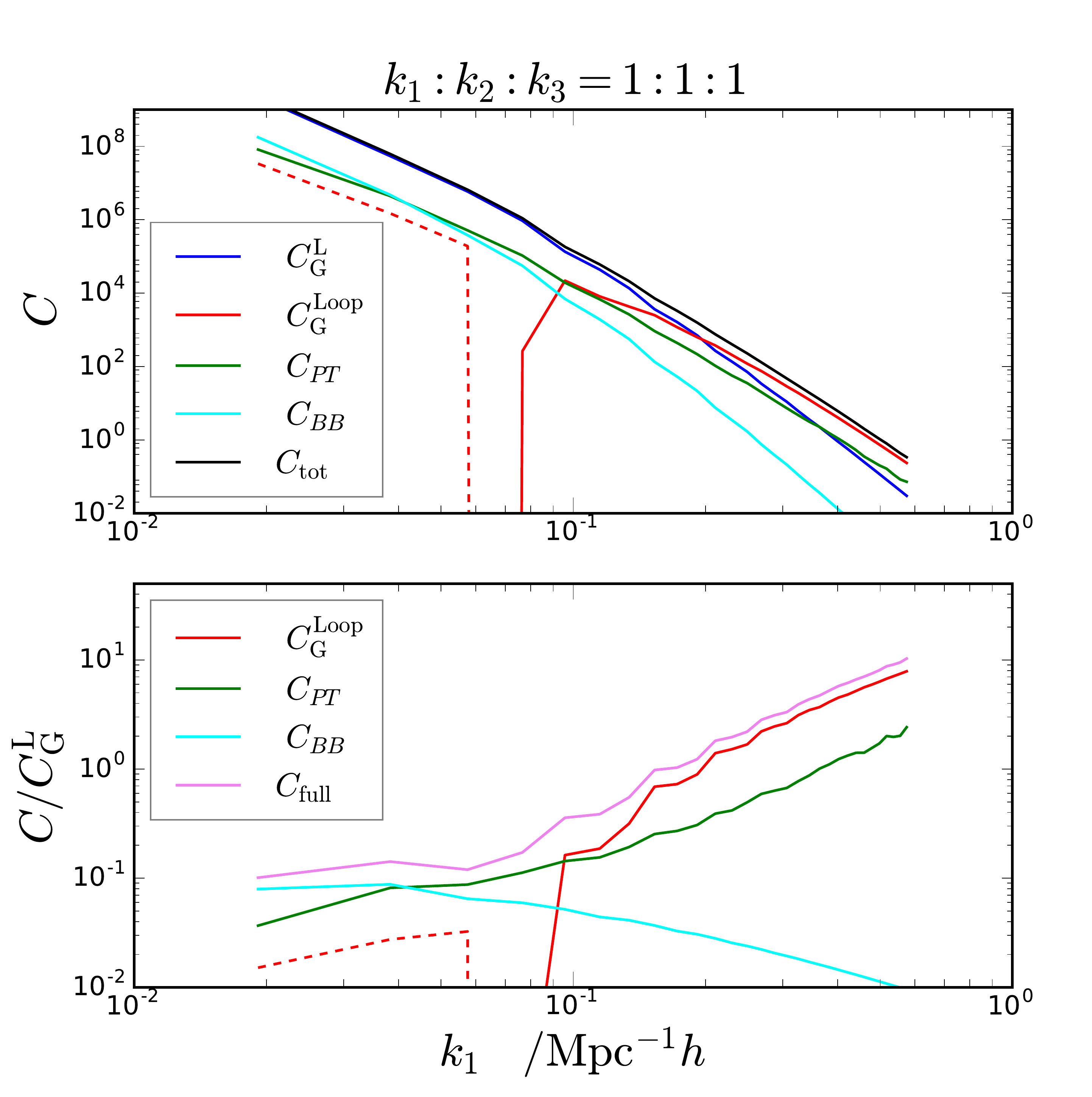}
\caption{  The leading tree-level contributions to the diagonal elements of the  dark matter bispectrum covariance matrix at $z=0$ for the equilateral triangle configuration.  The binning $\Delta k =  0.019  \hOMpc $ is used here and all the plots for bispectrum.   Upper panel: the Gaussian contributions $C_{\rm G}^{\rm L} $ (solid blue) and $C_{\rm G}^{\rm Loop}$ (red, solid for positive part, dashed for negative part), the non-Gaussian $BB$ contribution (solid, cyan) and $PT$ contributions (solid, green), and the sum of all the contributions (solid black).  Lower panel: the ratio of various terms normalized with respect to $C_{\rm G}^{\rm L} $. The violet line includes all the high order correction terms.    }
\label{fig:Covmat_G_NG_diag_111}
\end{center}
\end{figure}

Ref.~\cite{KayoTakadaJain_2013} computed the non-Gaussian contributions to the weak lensing bispectrum covariance. The structures of the terms are similar to the ones we consider here, with the main difference  that their results are for 2D field, while ours are 3D. The authors classified the non-Gaussian terms into groups of order bispectrum squared, $BB$,  a product of power spectrum and trispectrum, $PT$, and connected 6-point function.  Our terms $C_{F_3}$, $C_{F_2 I}^2 $, $ C_{F_2 II}^2  $, $C_{F_2 II}^4 $ would be classified as the order  $PT$. This can be easily seen from Fig.~\ref{fig:Covmat_Bk_diagrams} as there is a curvy line, which represents a power spectrum, directly connecting the two sides of the bispectrum  estimators on the top of each of these diagrams. The rest of the diagram has the same structure as the trispectrum terms shown in Fig.~\ref{fig:Covmat_Pk_diagrams} except $C_{F_2 II }^2 $.  The trispectrum part in $C_{F_2 II }^2 $ is in fact the contribution to the beat coupling for the power spectrum discovered in \cite{Hamilton:2005dx}. For the periodic boundary condition simulation we consider here, it does not contribute to the power spectrum covariance. However, it can exist in part of another diagram.  The remaining terms $ C_{F_2 I}^1$,  $ C_{F_2 II}^1$, and  $ C_{F_2 II}^3$  belong  to the group $BB$.  This is because each of the diagrams in this set can be decomposed into two bispectrum parts with each part consisting of two points from one  of the estimators and another one from the other estimator.

In addition, \cite{KayoTakadaJain_2013} also estimated the connected 6-point function using 1-halo term in the halo model.  The leading tree-level connected 6-point function is of the order of $P_{\rm L}^5$. Relative to those in  Eq.~\ref{eq:NGB_CF3}--\ref{eq:NGB_CF2II_4}, there is no explicit Dirac delta function such as $\Ddel( \mb{p} \pm \mb{p}' ) $ in the connected 6-point contributions, thus they couple triangles of different shapes. Furthermore they do not depend on the bin width. We can make a simple estimate by taking the equilateral triangle shape and assuming $F_n$ kernels are of order 1. Neglecting the symmetry factors, the connected 6-point function contribution from perturbation theory is $ \sim k_{\rm F}^3 P_{\rm L}^5 $. The magnitude of the terms in Eq.~\ref{eq:NGB_CF3}--\ref{eq:NGB_CF2II_4} is $\sim k_{\rm F}^3 P_{\rm L}^4 U /V_{123}^2 \sim  k_{\rm F}^3 P_{\rm L}^4 / (4 \pi k^2 \Delta k )$. Thus the 6-point contribution is subdominant. For example at $z=0$ and  $k = 0.2 \hOMpc $, it is 7\% of the non-Gaussian terms we considered here. Of course, for triangles of different shapes, the non-Gaussian terms we consider vanish, the connected 6-point contribution must be included.

Classifying the terms into groups $PPP$ (the Gaussian one), $BB$, $PT$, and the connected 6-point contributions provides a way to resum the perturbation series. As we mentioned, using the nonlinear power spectrum  in the Gaussian term we effectively resum part of the higher order contributions. Similarly, we can replace $P$, $B$ and $T$ with the nonlinear ones obtained either from simulation measurements or other analytic methods, such as halo model.  This approach was taken in \cite{KayoTakadaJain_2013}, but we will not pursue this further in this paper.

%In Fig.~\ref{fig:Covmat_G_NG_diag_111} and \ref{fig:Covmat_G_NG_diag_221}, we show the contribution to the covariance for the equilateral shape and isoceles triangle with side ratio 2:2:1 respectively.  We have plotted the high order correction to the Gaussian term and have grouped the non-Gaussian terms into $BB$ and $PT$.  We see that for $k \lesssim 0.06 \hOMpc $, the $BB$ term is the dominant one, but it is negligible for high $k$. The $PT$ contribution is the dominant non-Gaussian contribution for $ k\gtrsim  0.06 \hOMpc $.  Yet still the $PT$ term is small compared to hte loop contributiuon to the Gaussian for $k \gtrsim 0.1 \hOMpc $. We find the the behaviors for the equilateral triangle and isoceless triangle (2:2:1) are similar.  The non-Gaussian terms are relatively more important for the  isoceles case  mainly because the symmetry factor for the Gaussian terms and the non-Gaussian have different shape dependence. From equilateral to isoceles shape, the symmetry factor for the Gaussian term drops from 6 to 2, while for the non-Gaussian terms it only goes from 9 to 5.

In Fig.~\ref{fig:Covmat_G_NG_diag_111}, we show the contribution to the covariance for the equilateral shape.   We have compared the linear Gaussian term with  the high order correction to the Gaussian term and the non-Gaussian terms, which we have grouped into $BB$ and $PT$ respectively.  We see that for $k \lesssim 0.06 \hOMpc $, the $BB$ term is the dominant non-Gaussian term, but it is negligible for high $k$. The $PT$ contribution becomes the dominant non-Gaussian contribution for $ k\gtrsim  0.06 \hOMpc $.  Yet still the $PT$ term is small compared to the loop contribution to the Gaussian for $k \gtrsim 0.1 \hOMpc $.  For the isosceles shape, the results are qualitatively similar.

\begin{figure}[!tb]
\begin{center}
\includegraphics[width=\linewidth]{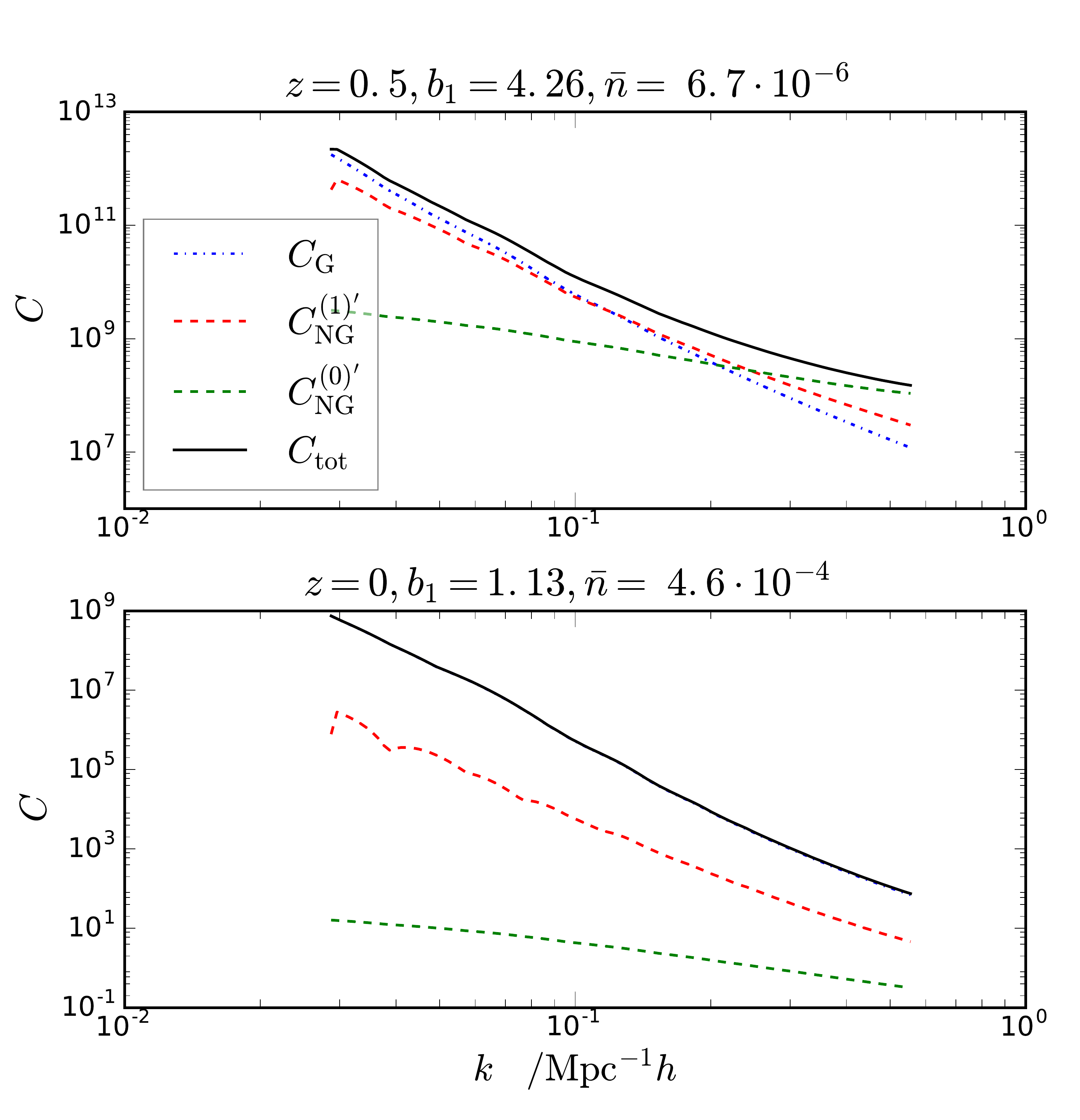}
\caption{ Various shot noise contributions to the covariance of the halo bispectrum. The diagonal elements for the equilateral triangle configuration are shown. We have plotted the results for two selected groups with the parameters used written on the plot. The Gaussian term (dotted-dashed, blue), the non-Gaussian contribution $C_{\rm NG} ^{(1)'} $ (dashed, red) and  $C_{\rm NG} ^{(0)'} $ (dashed, green), and the sum of the covariances (solid, black). In the low panel the Gaussian covariance overlaps with the total covariance.   }
\label{fig:Bk_eq_comparison}
\end{center}
\end{figure}

\begin{figure}[!tb]
\begin{center}
\includegraphics[width=\linewidth]{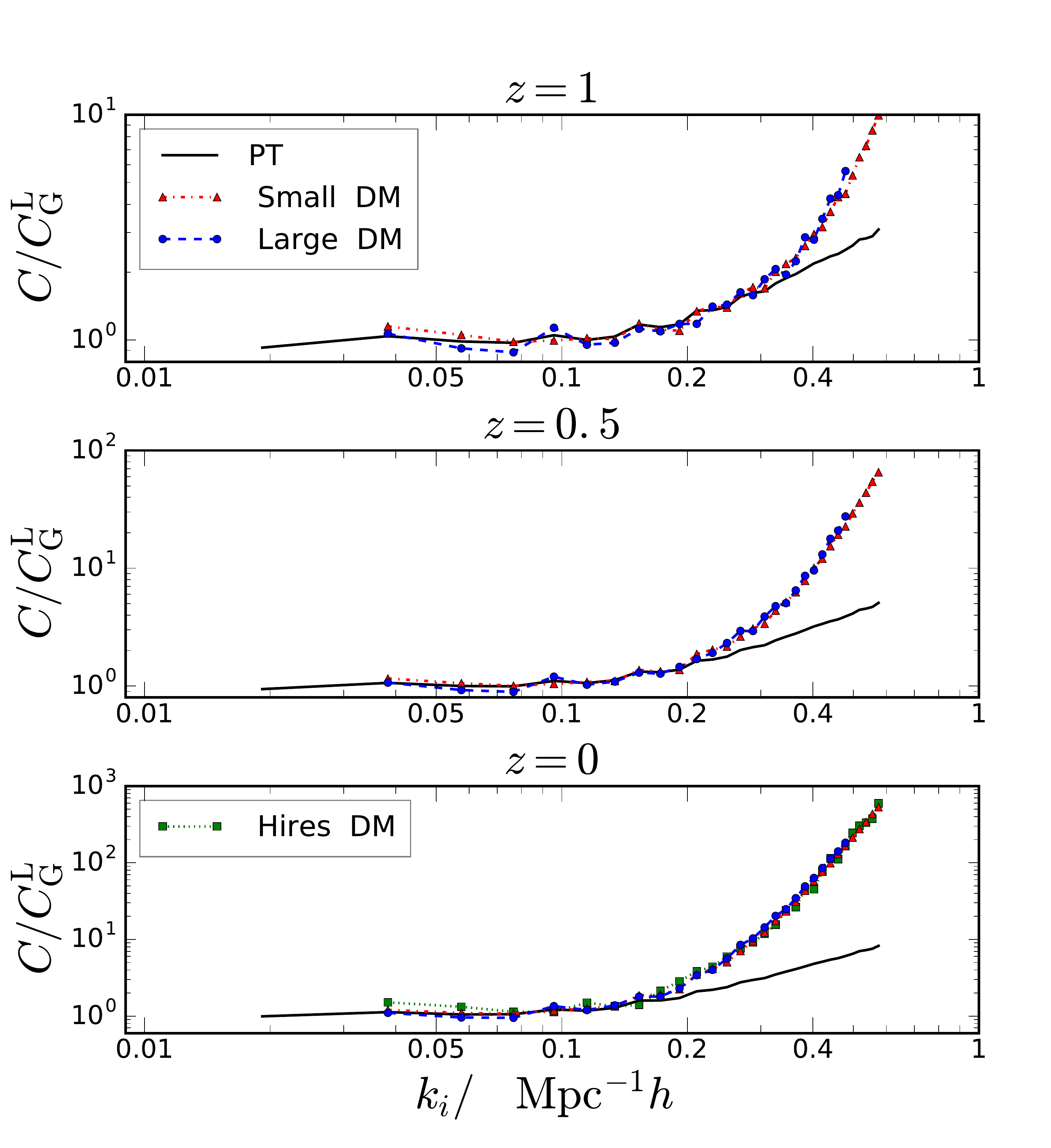}
\caption{  The diagonal elements of  the  dark matter bispectrum covariance matrix  for equilateral triangle configuration at $z=1$, 0.5, and 0 (from top to bottom). The results from the simulation set Small (red triangles), Large (blue circles), and Hires (green squares) are shown. The perturbation theory predictions (solid black lines) are also overplotted.
}
\label{fig:Bcov_111_diag_DM}
\end{center}
\end{figure}

\begin{figure*}[!tb]
\begin{center}
\includegraphics[width=0.9\linewidth]{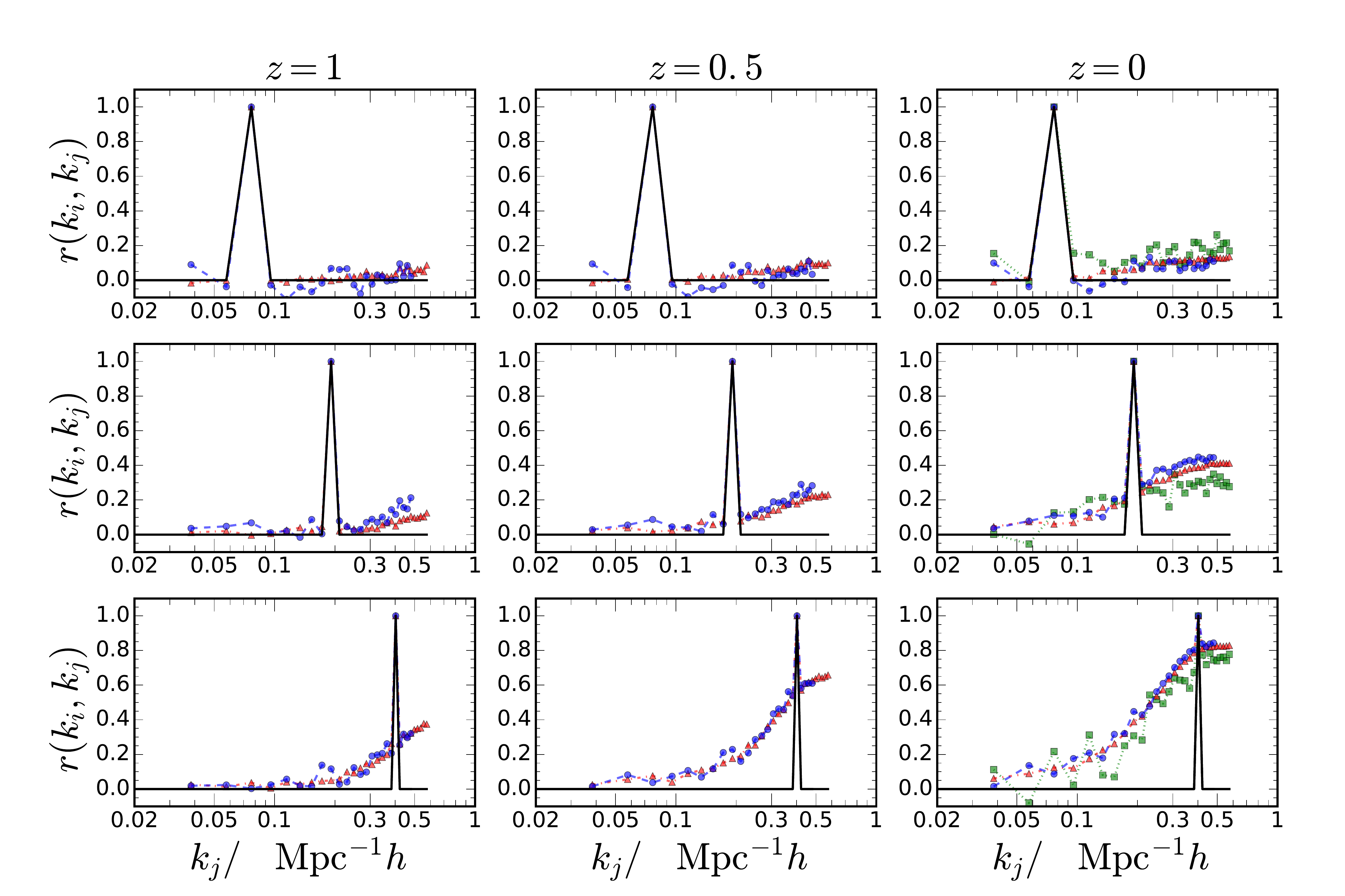}
\caption{ The correlation coefficient for the dark matter bispectrum. The equilateral triangle configurations are used. Results at $z=1$, 0.5, and 0 (left to right columns) are shown. The three rows correspond to the results obtained with $k_i$ fixed to be $0.076$,  $0.19$,  and  $0.40 \hOMpc $ respectively.  The results from the simulation set Small (red triangles), Large (blue circles), and Hires (green squares) and the prediction (solid black) are shown.  }
\label{fig:corr_coef_111_DM}
\end{center}
\end{figure*}

\subsubsection{Halos} 
\label{sec:BkCov_Poisson_summary}

The discrete nature of halos causes stochastic fluctuations.  Similar to the case of power spectrum,  the shot noise is the main source of halo bispectrum covariance as we will discuss in next section.  In Appendix \ref{sec:appendix_PoissonBk}, we use the Poisson model to derive the contribution to the bispectrum covariance due to  Poisson fluctuations. In the Poisson model we assume that the point particles Poisson sample the underlying continuous density field.  Both the contact correlation function due to Poisson sampling and the intrinsic correlation of the continuous field contribute to the 6-point function.  This results in both connected and disconnected contributions to the 6-point function (see Eq.~\ref{eq:Y_d}), which are represented diagrammatically in Fig.~\ref{fig:Poisson_shot_6_pt}.  We can classify the terms (or diagrams) using the correlator expansion  mentioned in Sec.~\ref{sec:Bcov_DMtheory}. The diagrams that are disconnected with three separate components are the Gaussian terms, and there are altogether three such diagrams. I.e.~they are in the $PPP$ group. The diagrams with two disconnected components are either in the non-Gaussian group $PT$ or $BB$. The connected diagrams represent the connected 6-point function contributions. Furthermore, within each group, there are terms with various power of $1/\tilde{n}$, we can regard it as an expansion in  $1/\tilde{n}$.  We refer the readers to the Appendix for the details. Here we summarize the main results.

There are three terms in the Poisson model that couple only triangles of the same shape, as the Gaussian term  in Eq.~\ref{eq:BCov_Gaussian}. These terms contain two Dirac delta functions in the 6-point function Eq.~\ref{eq:Y_d} (or three disconnected components in Fig.~\ref{fig:Poisson_shot_6_pt}).  They can be combined with Eq.~\ref{eq:BCov_Gaussian} as
\beqa
\label{eq:BCov_Gaussian_d} 
C_{\rm G} &=&  \frac{ k_{\rm F}^3 }{V_{123} }  \delta_{k_1k_2k_3, k_1'k_2'k_3'  } s_{123}   \Big[  P_{\rm h} (k_1)  + \frac{1}{ \tilde{n}} \Big]    \nn \\
 &\times&  \Big[ P_{\rm h} (k_2)   + \frac{1}{ \tilde{n}}    \Big] \Big[ P_{\rm h} (k_3)  + \frac{1}{ \tilde{n}} \Big].
\eeqa
Eq.~\ref{eq:BCov_Gaussian_d} agrees with \cite{SCFFHM98}.  Although the Poisson contributions do not arise from Gaussian fluctuations, they are on the same footing as the smooth Gaussian term,  we call them Gaussian terms as well.

There are also non-Gaussian contributions due to Poisson fluctuations. The non-Gaussian terms with one Dirac delta function in Eq.~\ref{eq:Y_d} (with two disconnected pieces in Fig.~\ref{fig:Poisson_shot_6_pt}) belong to the $PT$ or $BB$ group. As they contain the continuous correlator up to the halo tripsectrum, in this paper we shall only explicitly evaluate the terms with halo power spectrum or some terms with bispectrum. The terms with one Dirac delta function that we evaluate, denoted by $ C_{\rm NG}^{(1)'} $ here, are the ones up to the seventh line in Eq.~\ref{eq:C_NG_1_d}. The non-Gaussian terms due to the connected 6-point function are the ones without any Dirac delta  Eq.~\ref{eq:Y_d} (diagrammatically, they are the connected diagrams in Fig.~\ref{fig:Poisson_shot_6_pt}).  We  only explicitly evaluate the first line of Eq.~\ref{eq:C_NG_0_d}. We will denote this by $C_{\rm NG}^{(0)'} $.

We plot the shot noise contributions to the covariance of the halo bispectrum in Fig.~\ref{fig:Bk_eq_comparison} for  two selected halo groups, which correspond to Gr.~4 of the Large/Small simulation set at $z=0.5$ and Gr.~2 of the Hires set at $z=0$. We plot the diagonal elements for the equilateral triangle configuration.    The Gaussian covariance (Eq.~\ref{eq:BCov_Gaussian_d}) is compared with the non-Gaussian contributions $C_{\rm NG}^{(1)'} $ and   $C_{\rm NG}^{(0)'} $. For the low density sample,  $C_{\rm NG}^{(1)'} $ is comparable to the Gaussian one, while  $C_{\rm NG}^{(0)'} $ is sub-dominant until $k\sim 0.2\hOMpc $. For the more abundant sample, the Gaussian covariance is the dominant one comparable to the non-Gaussian contributions. Thus when the number density is high $\bar{n} \gtrsim 5 \times 10^{-4  }  (\MpcOh)^{-3} $, Gaussian covariance is a good approximation. Recall that for the same sample, Gaussian covariance is also a good approximation for the power spectrum.   %Although in the Poisson model the continuous correlators are the nonlinear ones, we have compared the results obtained using the measured  $P_{\rm h}$ and Eq.~\ref{eq:linear_bias_Ph} and found that the differences are negligible for our purpose here. 

%We have split the Gaussian covariance into the smooth and the Poisson parts. For the high bias group, the Gaussian Poisson component is dominant over the smooth Gaussian part, while for the low bias group they are comparable. We find that for the high mass group, the non-Gaussian components are  non-negligible, especially  $C_{\rm NG}^{(1)} $. The analytic part $C_{\rm NG}^{(3)'}$ also suggests that other terms in the bispectrum contribution $C_{\rm NG}^{(3)}$ are small but  not  negligible in comparison with the Gaussian contribution.  On the other hand, for the low bias group the non-Gaussian contributions can be neglected. 

\subsection{Numerical results }

%First in Fig.~\ref{fig:Bk_eq_comparison}, we show the mean bispectrum measurement for the equilateral triangle configuration for both box sizes. The results have been normalized with respect to the dark matter tree level bispectrum. We find that the results from both boxes agree with each other well. 

%As a comparison, we also show the 1-loop matter bispectrum prediction from standard perturbation theory \cite{1997ApJ...487....1S,SCFFHM98,PTreview}. Overall the 1-loop results agree with the measurment well. However, we note that the 1-loop results are slightly higher than the numerical measurements in the mildly nonlinear regime $0.1-0.3 \hOMpc$.  

%% \begin{figure*}[!htb]
%% \begin{center}
%% \includegraphics[width=\linewidth]{Bk_eq_comparison.pdf}
%% \caption{ The dark matter bispectrum for equilateral triangle configuration as a function of the wavenumber of the side of the triangle. Redshift from three redshifts 1, 0.5 and 0 are plotted.  The results from the large box (blue circles) and small one (red triangles) are shown. The 1-loop standard perturbation theory prediction is also plotted (solid green). The results are normalized with respect to the tree level matter bispectrum  $B_{\rm tree} $.    }
%% \label{fig:Bk_eq_comparison}
%% \end{center}
%% \end{figure*}

\subsubsection{Dark matter}

We first look into the covariance of the dark matter bispectrum. The covariance is estimated from the available realizations as 
\beqa
&& C( k_1,k_2,k_3,  k_1',k_2',k_3' ) \nn \\
&=& \frac{1 }{ N-1 } \sum_{i=1}^N [ B_i(k_1,k_2,k_3) -\bar{B}(k_1,k_2,k_3) ]  \nn \\
& & \quad \times  [ B_i(k_1',k_2',k_3') -\bar{B}(k_1',k_2',k_3') ],
\eeqa
where $N$ is the number of realizations used and $\bar{B}$ is the mean of the bispectrum measured from the simulations.

We show the diagonal elements of the bispectrum covariance matrix for the equilateral triangle configuration in Fig.~\ref{fig:Bcov_111_diag_DM}.  The results are normalized with respect to the Gaussian covariance $C_{\rm G}$, Eq.~\ref{eq:BCov_Gaussian}. As we noted before, for both the Gaussian and non-Gaussian covariance, the volume of the simulation factors out. Thus by dividing by the Gaussian covariance, we are able to compare the simulation results obtained from different box sizes.  The results from the Large, Small, and Hires simulation sets agree with each other well.  We find that the non-Gaussian contribution to the covariance increases rapidly as $z$ decreases from 1 to 0 in the mildly nonlinear regime $k \sim 0.4 \hOMpc $. The covariance is  within 20\% from $ C_{\rm NG}^{\rm L}$  up to $k \sim 0.2 \hOMpc$ at $ z=1 $ and $k \sim  0.1  \hOMpc$  at $ z=0 $.  We also plotted the non-Gaussian prediction described in Sec.~\ref{sec:Bcov_DMtheory}. The non-Gaussian correction gives the predictions agreeing with the simulation results up to $k\sim 0.3 \hOMpc $ at $z=1 $ and  $0.16 \hOMpc  $ at $z=0$.

Similar to the case of the power spectrum, the cross-correlation coefficient $ r $ is defined as
\beqa
 && r ( k_1,k_2,k_3,  k_1',k_2',k_3' )  \nn \\
&=& \frac{ C ( k_1,k_2,k_3,  k_1',k_2',k_3') }{ \sqrt{  C ( k_1,k_2,k_3,  k_1,k_2,k_3)      C ( k_1',k_2',k_3',  k_1',k_2',k_3') } }. \nn \\
\eeqa
We plot $ r $ for the  equilateral triangle configuration in Fig.~\ref{fig:corr_coef_111_DM}.  In each subplot, we fix one of the lengths of the triangle $k_i$ and vary the length of the  other one, $k_j $.  In these plots, we have fixed $k_i$ to be $0.076$,  $0.19$,  and  $0.40 \hOMpc $ respectively. 

As we noted previously, the Gaussian covariance couples only the triangle of the same shape, while the leading non-Gaussian  corrections couple triangles with at least one side equal to each other.  Thus both the Gaussian and disconnected non-Gaussian tree-level corrections cannot give rise to cross correlations between equilateral triangles of different sizes. We see that indeed when  both $k_i$ and $k_j$  are small, the correlation coefficients are consistent with being zero.  As one of the wavenumbers increases $r$ also  starts to increase. The larger the value of the wavenumbers the larger the covariance. This is expected from  the fact that nonlinearity increases the coupling of different wave modes.  The accuracy of the perturbation theory prediction is qualitatively similar to that for the power spectrum covariance shown in Fig.~\ref{fig:corr_coef_PkCov_DM}.

\begin{figure}[!tb]
\begin{center}
\includegraphics[width=\linewidth]{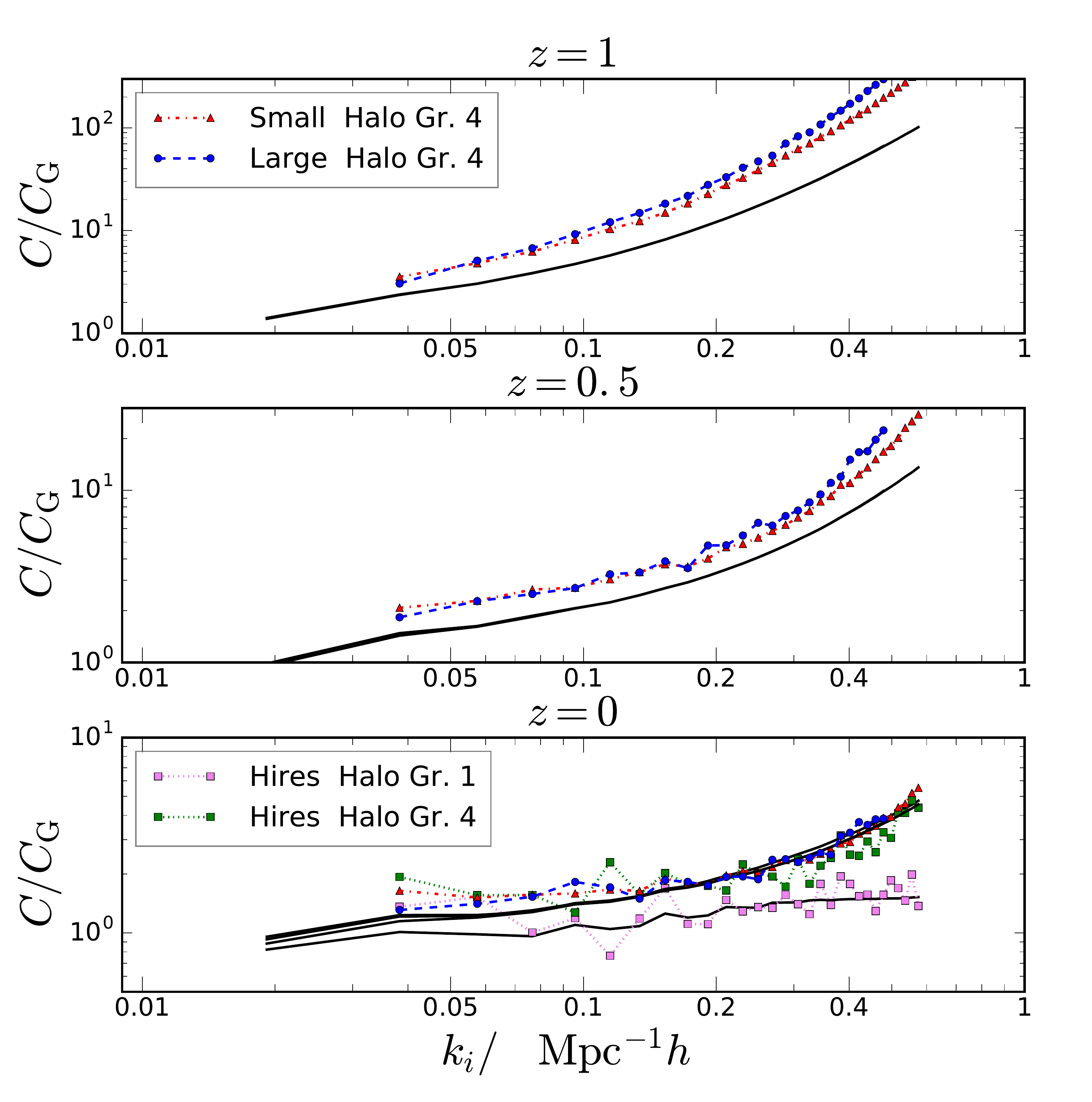}
\caption{  The diagonal elements of the halo bispectrum for the equilateral triangle configuration.   The mean Poisson shot noise is subtracted. The halo groups from the  Large (blue circles), Small (red triangles), and  Hires [violet squares (group 1) and green squares (group 4)]  simulation sets are shown.  The theory prediction (solid black lines) includes the Gaussian covariance and the non-Gaussian terms $ C_{\rm NG}^{(0)'} $ and  $ C_{\rm NG}^{(1)'} $.   }
\label{fig:Bcov_111_diag_halo_ShotSubFlag1}
\end{center}
\end{figure}

\begin{figure*}[!htb]
\begin{center}
  \includegraphics[width=0.9\linewidth]{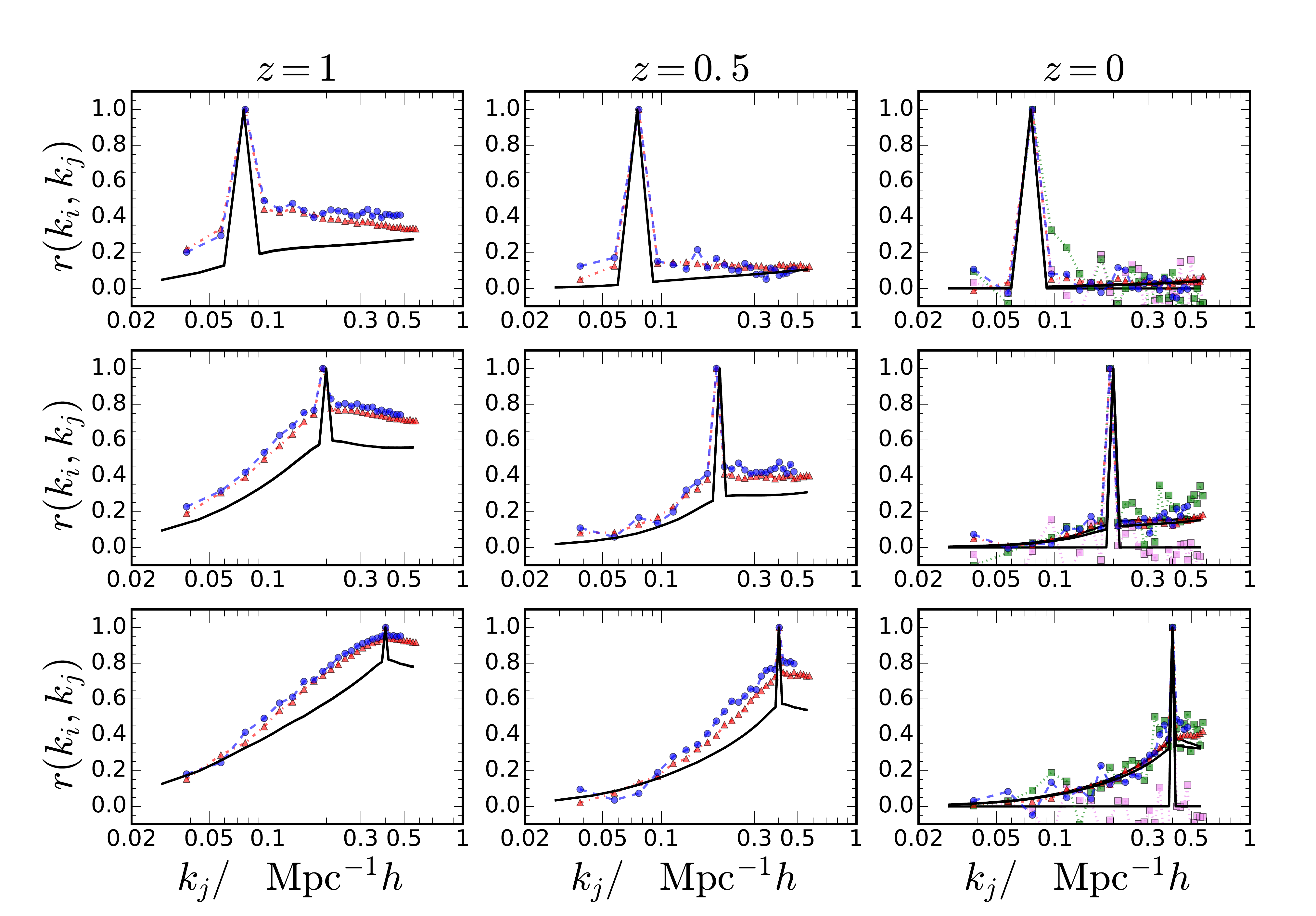}
\caption{   The correlation coefficient for the halo bispectrum of the equilateral triangle shape.  The mean Poisson shot noise is subtracted.  The halo groups from the  Large (blue circles), Small (red triangles), and Hires [violet squares (group 1) and green squares (group 4)]  simulation sets are shown.  The predictions (solid black lines) include   Gaussian covariance and non-Gaussian corrections $ C_{\rm NG}^{(0)'} $ and  $ C_{\rm NG}^{(1)'} $.     }
\label{fig:corr_coef_111_halo_ShotSubFlag1}
\end{center}
\end{figure*}

\begin{figure}[!htb]
\begin{center}
  \includegraphics[width=\linewidth]{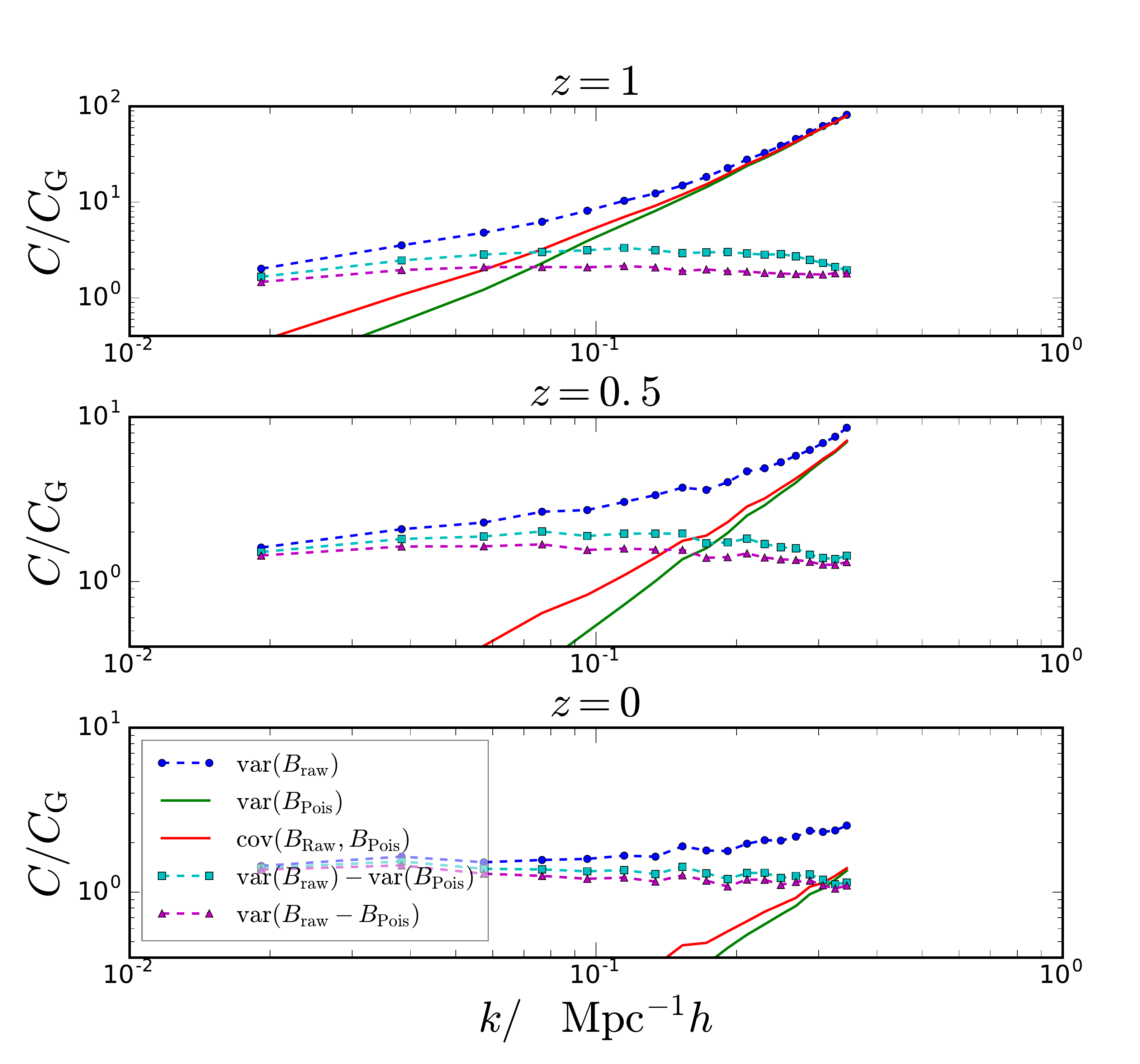}
  \caption{  The diagonal elements of equilateral triangle halo bispectrum obtained with the mean Poisson shot noise subtraction (blue circles) and individual shot noise subtraction (violet triangles) are compared. The variance of the Poisson shot noise, $\mathrm{var}(B_{\rm Pois} ) $ (solid green line), the covariance between the raw halo bispectrum and the Poisson shot noise, $\mathrm{cov}(B_{\rm raw}, B_{\rm Pois} ) $ (solid red line), and the prediction using the ansatz in Eq.~\ref{eq:cov_Braw_BPois} (cyan squares) are also plotted.      }
\label{fig:Bcov_111_diag_halo_ShotSubFlag2}
\end{center}
\end{figure}

\begin{figure*}[!htb]
\begin{center}
  \includegraphics[width=0.9\linewidth]{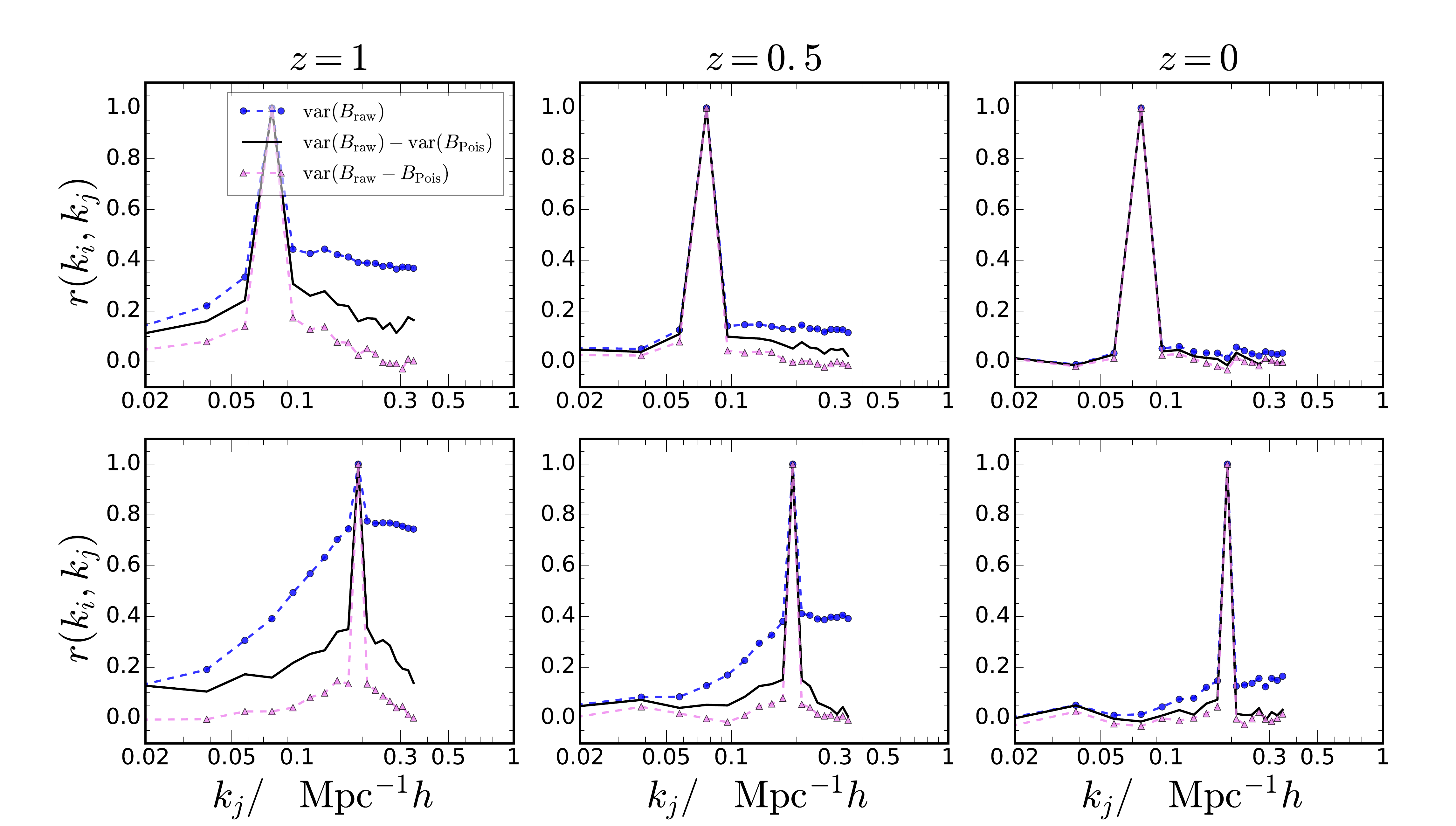}
  \caption{ The correlation coefficients for the  equilateral triangle halo bispectrum with the mean shot noise subtraction (blue circles) and individual shot noise subtraction (violet triangles). The black solid line shows the prediction using the ansatz in Eq.~\ref{eq:cov_Braw_BPois}.   Only the results for $ k_i = 0.076$ and $0.19 \hOMpc $  are shown.   }
\label{fig:corr_coef_111_halo_ShotSubFlag2}
\end{center}
\end{figure*}

\subsubsection{Halos}

We now move to the halo bispectrum covariance. The Poisson shot noise contribution is given by Eq.~\ref{eq:Bk_discrete}, i.e.
\beq
B_{\rm Pois}( k_1, k_2, k_3 )= \frac{1}{\tilde{n}}\Big[ P_{\rm h}(k_1) + P_{\rm h}(k_2) + P_{\rm h}(k_3)  \Big] 
+ \frac{1}{ \tilde{n}^2} , 
\eeq
where $P_{\rm h}$ is the smooth halo power spectrum.   As we are interested only in the smooth correlation function signal, the contribution to the halo bispectrum due to Poisson fluctuations is usually subtracted.  Similar to the case of the power spectrum, we distinguish between the mean Poisson shot noise subtraction  and the Poisson shot noise  estimated  and subtracted from each realization.

We first show the results when the mean Poisson shot noise is subtracted.   This case corresponds  precisely to the derivation done in Appendix \ref{sec:appendix_PoissonBk} and briefly summarized in Sec.~\ref{sec:BkCov_Poisson_summary}.   In Fig.~\ref{fig:Bcov_111_diag_halo_ShotSubFlag1}, we plot the diagonal elements  of the covariance  for the equilateral triangle configuration. The results are normalized with respect to the Gaussian covariance Eq.~\ref{eq:BCov_Gaussian_d}. We have also plotted the theory prediction, which includes the Gaussian term and the non-Gaussian corrections $ C_{\rm NG }^{(0)'} $ and $ C_{\rm NG }^{(1)'} $.   We find that the deviation of the numerical results from the Gaussian covariance is significant even at low $k$ for the rare mass groups Large/Small Gr.~4 at $z=0.5$ and 1. The model tends to underestimate  the covariance in comparison with the numerical results. The reason is that we have only evaluated parts of the 6-point function contributions in the Poisson model.  On the other hand, for the relatively more abundant groups, the deviation from the Gaussian result is mild and the model agrees with the data well. This is mainly because for these low bias groups the non-Gaussian corrections are small as we  have seen in Sec.~\ref{sec:BkCov_Poisson_summary}. Note that for group 1, we have to use the measured nonlinear power spectrum to compute the Gaussian covariance as Eq.~\ref{eq:linear_bias_Ph} is not adequate.

 In Fig.~\ref{fig:corr_coef_111_halo_ShotSubFlag1}, we plot the correlation coefficient for the halo bispectrum with the mean Poisson shot noise subtracted. The triangle configurations are equilateral.  One of the wavenumbers, $k_i$ is   fixed  to be $0.076$,  $0.19$,  and  $0.40 \hOMpc $ respectively. The theory predictions are also overplotted for comparison. As  $C_{\rm NG}^{(1)'} $ couples triangles with at least one side equal to each other, it vanishes for equilateral triangles of different shapes.  Hence only  $C_{\rm NG}^{(0)'} $ contributes to the correlation coefficient.   Similar to the diagonal case, for the rare halo groups at $ z=1$ and  $0.5$ the model underestimates the covariance compared to the simulation results, while the model agrees reasonably well with the data for the more abundant groups at $z=0$.

 We now consider the scenario when the Poisson shot noise is subtracted from individual realizations. The covariance of these two scenarios is related by
 \beqa
 \label{eq:cov_Braw_BPois}
&& \mathrm{cov} ( B_{\rm raw} - B_{\rm Pois} ,  B_{\rm raw}' - B_{\rm Pois}' )  \nn \\
 &=&  \mathrm{cov} ( B_{\rm raw} ,  B_{\rm raw}' ) - \mathrm{cov} ( B_{\rm raw} ,  B_{\rm Pois}' )  \nn \\
 && \quad   - \mathrm{cov} ( B_{\rm Pois} ,  B_{\rm raw}' ) +  \mathrm{cov}( B_{\rm Pois}, B_{\rm Pois}') \nn \\
 & \stackrel{?}{\approx} & \mathrm{cov} ( B_{\rm raw} ,  B_{\rm raw}' ) -  \mathrm{cov}( B_{\rm Pois}, B_{\rm Pois}') . 
 \eeqa
In the last line we assume that 
\beq
\label{eq:covB_raw_Pois_ansatz}
 \mathrm{cov} ( B_{\rm raw} ,  B_{\rm Pois}' )  \stackrel{?}{\approx}   \mathrm{cov}( B_{\rm Pois}, B_{\rm Pois}') . 
\eeq
We will shortly check how good this ansatz is.

Unlike the former case, which is equivalent to no shot noise subtraction at all, we have to rely on the accuracy of the Poisson model here. We find that the shot noise subtracted equilateral bispectrum goes negative for $k \gtrsim 0.35 \hOMpc $ at $z=1$. At $z=0$, for Gr.~4, this occurs at $ k\sim 0.21 \hOMpc$.  Although there seems to be no fundamental reason that the smooth halo bispectrum must be positive, this may well indicate that the Poisson model is not accurate enough. Therefore we shall not show the results for $k$ beyond  $0.35 \hOMpc$.   In Fig.~\ref{fig:Bcov_111_diag_halo_ShotSubFlag2}, we compare  the diagonal elements of the covariance obtained with the mean Poisson shot noise subtraction and the individual shot noise subtraction. We normalize  the covariance  by the Gaussian covariance Eq.~\ref{eq:BCov_Gaussian_d}.  Similar to the power spectrum case, the subtraction of shot noise from the individual realizations significantly reduces the covariance, and it gets much closer to the Gaussian one. From Fig.~\ref{fig:corr_coef_111_halo_ShotSubFlag2}, we find that the off-diagonal elements also exhibit substantially lower level of correlation, especially for the groups at $z=1$ and 0.5. The correlations are roughly consistent with zero for the groups at $z=0.5$ and 0 within the scatter.

In Fig.~\ref{fig:Bcov_111_diag_halo_ShotSubFlag2}  we also show the covariance $  \mathrm{var}( B_{\rm Pois} ) $ and  $  \mathrm{cov}( B_{\rm raw}, B_{\rm Pois}) $. At low $k$, they are quite different, but they approach each other at high $k$.  Because we found that the number density fluctuation does not correlate with the continuous clustering, i.e.~halo power spectrum in Sec.~\ref{sec:Pk_Halo_cov_numerical}, the difference between them must come from the continuous power spectrum in  $  B_{\rm Pois}$. On the other hand, at high $k$ the number density fluctuation dominates, so these two covariances agree. To test the ansatz in Eq.~\ref{eq:cov_Braw_BPois}, we plot the prediction using $ \mathrm{var} ( B_{\rm raw} ) -  \mathrm{var}( B_{\rm Pois}) $ in Fig.~\ref{fig:Bcov_111_diag_halo_ShotSubFlag2} and \ref{fig:corr_coef_111_halo_ShotSubFlag2},  which yields a covariance close to the one obtained by subtracting shot noise from individual realizations. This demonstrates that  Eq.~\ref{eq:covB_raw_Pois_ansatz} is a reasonable approximation. Our results show that most of the non-Gaussianity in the halo bispectrum covariance arises from  the fluctuations in  $B_{\rm Pois}$.

We now comment on the magnitudes of the dark matter and halo bispectrum covariances. For the rare group Gr.~4, depending on the redshift, the covariance of the halo bispectrum is nine to five orders of magnitude larger than that of the dark matter bispectrum covariance. The differences decrease when the number density of the sample increases. Even for the abundant group, Hires Gr.~2, the covariance of the halo bispectrum is still an order of magnitude higher than that of the dark matter. When the number density reaches as high as that of the Hires Gr.~1 [$\bar{n} \sim 10^{-3} (\MpcOh)^{-3} $], the magnitude of the halo covariance is comparable to that of the dark matter. Hence the relative differences are quite similar to the power spectrum covariance. This suggests that shot noise is the dominant contribution to the halo bispectrum covariance.

Gaussian covariance for the bispectrum is often used in forecast e.g.~\cite{Xu:2014bya,Welling:2016dng,XuHamannChen2016}. When the mean number density is used, our results show that the Poisson fluctuation gives rise to large non-Gaussianity in the covariance unless the number density of the sample is sufficiently high $ n \gtrsim 5 \times 10^{-4} (\MpcOh)^{-3}$.  Fortunately, in reality, the number density is estimated from the local average and by subtracting the  shot noise from individual realizations, the covariance gets much closer to the Gaussian covariance.  This is because most of the non-Gaussianity is due to the fluctuations in the Poisson shot noise $B_{\rm Pois} $. However, there is still significant amount of non-Gaussianity left. We will see in in Sec.~\ref{sec:signal_to_noise} that use   of the Gaussian covariance severely overestimates the signal-to-noise ratio.

\begin{figure}[!htb]
\begin{center}
\includegraphics[width=\linewidth]{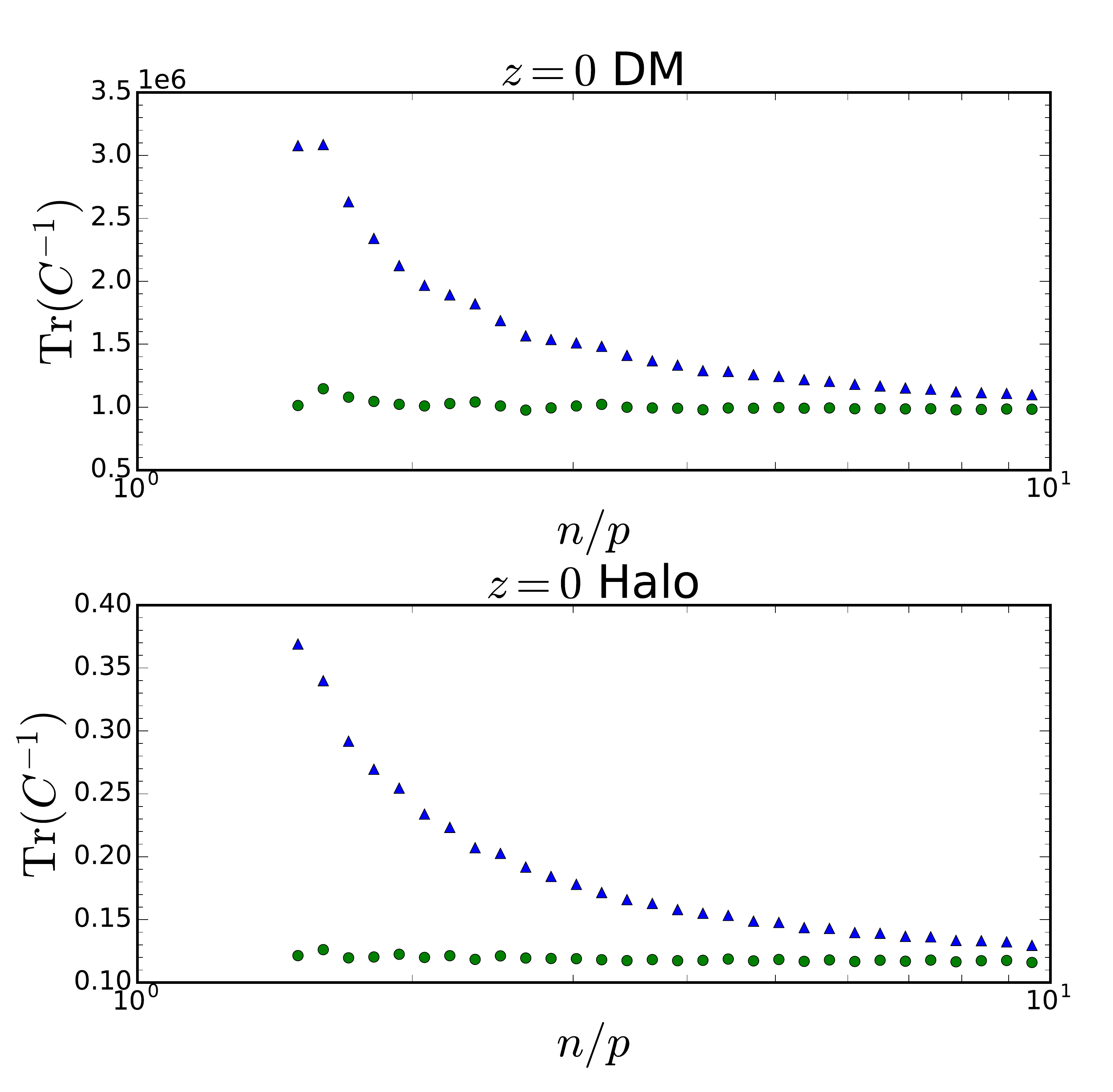}
\caption{   The trace of the precision matrix as a function of $n/p$, where $n$ is the number realizations used to estimate the covariance matrix and $p$ is the number of the bispectrum bins, which is equal to 429 here.  Both the results from dark matter and halo (Gr.~4) at $z=0$  are shown. The results from the naive estimate $ C^{-1}_{\rm sample}$ (blue triangles) and  bias-corrected estimate  $ C^{-1}_{\rm unbiased}$  (green circles) are compared.     }
\label{fig:invC_converge_check}
\end{center}
\end{figure}

\begin{figure*}[!htb]
\begin{center}
\includegraphics[width=0.9\linewidth]{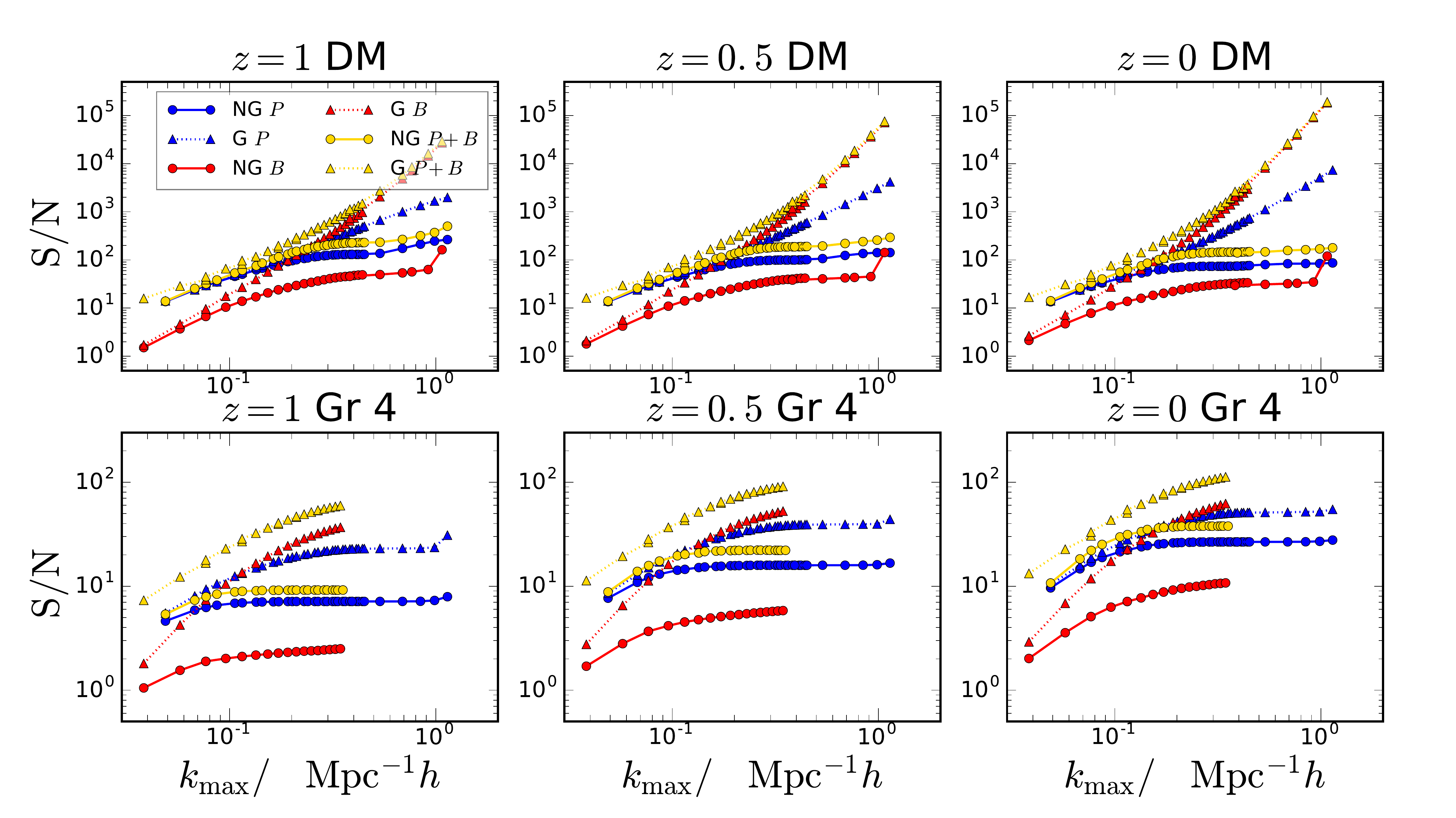}
\caption{ The signal-to-noise ratio for dark matter and halo power spectrum and bispectrum.  Dark matter and halo group 4 data from Small simulation set are used. The signal-to-noise ratio of the power spectrum (blue), and bispectrum (red), and the sum of them (yellow) are shown.  The results obtained with the full non-Gaussian covariance results (circles, solid line) are compared with the Gaussian covariance (triangles, dotted line) ones.   }
\label{fig:SignalNoise_ratio}
\end{center}
\end{figure*}

%\section{ The signal-to-noise ratio of the power spectrum and bispectrum  }
\section{ The information content of the power spectrum and bispectrum  }
\label{sec:signal_to_noise}

    As the power spectrum has been well explored, and the bispectrum becomes the next frontier in large scale structure, it is crucial to address how much information one can gain by going to higher order correlators. The signal-to-noise ratio, S/N, is often used to quantify the information content in the power spectrum and bispectrum, e.g.~\cite{Takahashi:2009ty,Scoccimarro:2003wn,Sefusatti:2004xz,Blot:2014pga}.  The Fisher information \cite{Tegmark:1996bz} is an alternative way to characterize the information content. However,  we shall not consider it here as it involves the derivatives of the polyspectra with respect to the cosmological parameters, which would require a dedicated set of simulations with varying cosmological parameters.

The Gaussian covariance is usually used to make forecast and quantify the information content.  In particular, the forecast based on the Gaussian covariance suggests that there is a lot of information in the bispectrum (\cite{Sefusatti:2004xz}, see also below).  However, since we have seen in the previous sections that the non-Gaussian contributions significantly increase the covariance already in the  weakly nonlinear regime, it is important to ask how the results are modified when the non-Gaussian covariance is taken into account.  

The signal-to-noise ratio is defined as 
\beq
( \mathrm{S/N} )^2 = S_i C^{-1}_{\quad ij} S_j, 
\eeq
where $C^{-1}$ is the inverse of the covariance matrix, also called the precision matrix, and $S$ is the signal.  Here we use the non-Gaussian covariance measured from simulations to quantify the information content in the power spectrum and bispectrum. For dark matter, the signal $S$ is simply the mean of the measured power spectrum and bispectrum respectively.  For halos, we use the power spectrum and bispectrum with the Poisson shot noise subtracted, using Eq.~\ref{eq:Pk_discrete}  and \ref{eq:Bk_discrete} respectively, with the Poisson shot noise estimated and subtracted from each realization.  At high $k$ the Poisson shot noise subtracted power spectrum and bispectrum can go negative, which is deemed unphysical, hence we shall only show the results that are reliable.  We also show the results obtained using the Gaussian covariance for comparison.

There is one more complication because we require the precision matrix rather than the covariance matrix. Ref.~\cite{Hartlap:2006kj} pointed out that for a  $p \times p$  covariance matrix, $p$  must be smaller than the number of realizations, $n$, for the covariance matrix estimated from the realizations to be invertible. The basic reason is that we can regard each realization as an independent random vector in $p$-dimensional space.  When $n$ is larger than $p$, these vectors are sufficient to span the space of dimension $p$, otherwise the rank of the covariance matrix is less than $p$. See \cite{Hartlap:2006kj,Anderson_Statisitcs} for a formal proof. Moreover, even when the covariance matrix is invertible, because inversion is a nonlinear operation, the inverse of an unbiased estimator is in general biased. If the distribution of the estimator is Gaussian, the bias-corrected estimator for the precision matrix reads \cite{Hartlap:2006kj,Anderson_Statisitcs} 
\beq
\label{eq:Cinv_correct}
C^{-1}_{\rm unbiased} = \frac{n - p - 2}{ n-1 } C^{-1}_{\rm sample},   
\eeq
where $ C_{\rm sample}$ is the unbiased sample covariance matrix. 

For the power spectrum the number of bins, $p$, is typically smaller than the number of realizations available in our case, e.g.~the maximum number of bins for the power spectrum is 90. Ref.~\cite{Blot:2015cvj} checked that Eq.~\ref{eq:Cinv_correct} works very well for the power spectrum precision matrix when $n/p \gtrsim 2$.  On the other hand, because of the large number of configurations, the number of bins of the measured bispectrum can be comparable to or even larger than the number of realizations that we use.  For example, for the binning of the Small simulation set that we used in the previous sections, there are 2825 bins.\footnote{ We can compute the total number of distinct triangle configurations (including the folded ones) with the formula $\sum_{j=1}^n \Big\lceil \frac{j+1}{ 2} \Big\rceil \Big\lfloor \frac{j+1}{ 2} \Big\rfloor   + \Big\lfloor  \frac{ j  }{ 2 }  \Big\rfloor $, where $\lceil \, \rceil $ and  $\lfloor \,  \rfloor $ are the ceiling and floor functions.  This formula is derived from the integer sequence A002620 in https://oeis.org/A002620 (see also \cite{JenkynsMuller2000}). In this case we use  $n=30$ bins, and we get 2825 configurations. } As Eq.~\ref{eq:Cinv_correct} assumes that the distribution of the bispectrum estimator is Gaussian, a priori it is not clear how well it works for the bispectrum covariance.  In Appendix \ref{sec:bisp_estimaotr_distribution}, we check that the distribution of the bispectrum estimator follows the Gaussian distribution reasonably well, although there are also some non-negligible deviations.  To test  Eq.~\ref{eq:Cinv_correct}, we shall use the Small simulation set as it has the largest number of realizations.  Because the key parameter in  Eq.~\ref{eq:Cinv_correct} is the ratio $n/p$, we rebin the bispectrum into wider bins so that the sides of the bispectrum are in units of  $8 k_{\rm F} $ instead of $2 k_{\rm F} $ that we have been using so far. After rebinning, there are altogether $p=429$ bins of bispectrum configurations.   As the correction is only an overall factor, following \cite{Hartlap:2006kj}, we plot the trace of the precision matrix against $n/p$ in Fig.~\ref{fig:invC_converge_check}.  We show the dark matter and halo data from the Small set at $z=0$.  Each data point in this figure corresponds to the results obtained with a precision matrix estimated by randomly choosing $n$ realizations from the total 4096 ones. We compare the results from the naive estimate $ C^{-1}_{\rm sample}$ and the bias-corrected estimate  $C^{-1}_{\rm unbiased} $.   As $n$ increases, the naive estimate decreases and approaches the bias-corrected estimate. For $n/p \gtrsim 2 $, the correction works very well already. Note that for $ n/p \lesssim 1 $ we encounter difficulties in inverting the covariance matrix due to the reason mentioned above. As Eq.~\ref{eq:Cinv_correct} works very well, we shall use the bias-corrected precision matrix for both power spectrum and bispectrum.

Finally, we are ready to present the S/N of the power spectrum and bispectrum. We will only show the results from the Small set due to its large number of realizations available. However, we will also comment on the results from Hires.  In Fig.~\ref{fig:SignalNoise_ratio}, we plot the $\mathrm{S/N}$ for both dark matter and halo power spectrum against the maximum $k$ used to compute the S/N, $k_{\rm max}$. For the power spectrum, the S/N does not depend on the binning width $\Delta k $ used to the lowest order. Of course there is some binning dependence if the field varies appreciably across the bin, but this dependence is of higher order in the binning width. A similar statement was also made in \cite{Takahashi:2009ty}. We verify it in Appendix \ref{sec:BinningDependence}.

  We find that for the power spectrum of dark matter, the information content increases as  $k_{\rm max}$  increases  in the linear regime.  The S/N then starts to level off at the weakly nonlinear regime $k_{\rm max} \sim 0.2 \hOMpc $. The flattening of the S/N becomes more and more serious as the redshift decreases. In particular at $z=0$, there is almost no increase in the S/N beyond $k_{\rm max}\sim 0.2 \hOMpc$.  In contrast, the $\mathrm{S/N}$ obtained with Gaussian covariance keeps on increasing with the phase volume.   The saturation of the information content of the dark matter $P(k)$ in the weakly nonlinear regime has been observed by many authors \cite{Rimes:2005xs,Neyrinck:2006xd,Lee:2008qy,Takahashi:2009ty,Blot:2014pga}.  This casts doubts on the efforts to model the nonlinear power spectrum accurately beyond the BAO scales and has motivated using alternative statistics to extract information from the large scale structure, notably the log transform \cite{ColesJones1991,NeyrinckSzapudiSzalay2009,CarronNeyrinck2012}.  Nevertheless, Fisher analysis seems to suggest that the information content on cosmological parameters is not completely erased in the nonlinear regime \cite{Blot:2015cvj}.

  We find that the Poisson shot noise subtracted halo power spectrum goes negative for $k$ in between 0.5 and $0.8 \hOMpc $ for Gr.~4. Because the power spectrum must be non-negative, this is mathematically inconsistent. This happens when the signal is so low that the theoretical uncertainty of the Poisson model is larger than or comparable to the signal. As we believe the contribution to the cumulative signal-to-noise from this range of data is negligible, we will show it as well.  For the halo power spectrum, the trends are qualitatively similar to that of the dark matter and they are roughly constant  at  $k_{\rm max} \sim 0.1 - 0.2 \hOMpc $ depending on the number density of the sample. The saturation of the information content of the halo power spectrum is also hinted in \cite{Angulo:2007fw,Smith:2008ut}.   We find that the signal-to-noise is in between a factor of a few to one order of magnitude lower than that of the dark matter  for the rare halo group Gr.~4.  Reassuringly for the more abundant groups from Hires, the S/N is comparable to that of the dark matter.  We  find that the Gaussian approximation overestimates the S/N by  a factor of two to a few at  $k_{\rm max} \sim 0.4\hOMpc $.

We now turn to the  $\mathrm{S/N}$  for the bispectrum.   We consider all the triangle configurations with the sides of the triangle less than certain $k_{\rm max} $, against which is plotted  in Fig.~\ref{fig:SignalNoise_ratio}.  In order to sample the low $k$ modes well, and also to be able to probe the S/N to high $k$, we combine the results from two different binnings for the bispectrum measurements. For the low $k$ results we use $\Delta k = 2 k_{\rm F} = 0.019 \hOMpc  $, while for high $k$ we bin the bispectrum with $\Delta k = 8 k_{\rm F} = 0.077 \hOMpc  $. In Appendix  \ref{sec:BinningDependence}, we verify that the S/N is invariant to rescalings of the bin width to the lowest order.

For the dark matter bispectrum, already at $ k_{\rm max} \sim 0.2 \hOMpc $ the S/N increases significantly slower than what the Gaussian predictions suggest.  The departure from Gaussian prediction sets in at lower and lower  $k$  as the redshift decreases.  We also find that the deviation generally occurs at lower $k$ than the case of the power spectrum, thus suggesting that the Gaussian approximation is worse for the bispectrum.  The non-Gaussian contribution significantly degrades the S/N for the bispectrum. For example, at $ z=0$, according to the Gaussian covariance predictions, the S/N of the dark matter bispectrum should surpass that of the power spectrum at $k_{\rm max} = 0.14 \hOMpc $, while the full non-Gaussian case shows that the S/N of the bispectrum is only 30\% of the power spectrum value.  However, it is encouraging to find that relative to the dark matter power spectrum, whose S/N already saturates at  $k_{\rm max} \sim 0.2 \hOMpc $, the S/N of the bispectrum still keeps on increasing mildly up to $k_{\rm max} \sim 0.5 \hOMpc $.  Interestingly, at $ k_{\rm max} \sim 1 \hOMpc $, the S/N increases sharply and overshoots the S/N of the power spectrum.  Thus by delving into the nonlinear regime, the information gain of the bispectrum is higher than that from the power spectrum.  We note that the information content of the dark matter power spectrum was found to increase sharply at $k_{\rm max} \sim 1.5 \hOMpc $ beyond the plateau \cite{Rimes:2005xs,Neyrinck:2006xd}. We suspect that this sharp rise in bispectrum S/N also due to  the same reason. However, we shall not investigate this further here.

After the Poisson shot noise subtraction, the halo bispectrum of Gr.~4 at $z=1$ starts to go negative for $k_{\rm max } \gtrsim 0.35 \hOMpc $.  For Gr.~4 at $z=0$, it happens at even lower $k$,  $k \gtrsim 0.22 \hOMpc $.      This occurs at lower $k$ than that for the halo power spectrum. This first happens for the triangles of shape close to the equilateral. As $k_{\rm max} $ increases, we find that more squeezed shapes also become negative.  Although there appears to be no fundamental reason that the bispectrum must be positive, we believe these negative values could well indicate that the Poisson model is not reliable as the bispectrum signal gets too small. To be conservative, we shall not show the halo bispectrum results beyond    $k_{\rm max } =  0.35 \hOMpc $.  For the halo bispectrum case, the overall trends are also similar to that of the power spectrum. Again for these rare groups shown, the Gaussian approximation significantly overestimates the S/N, even more seriously than for the dark matter bispectrum.  At $k_{\rm max} \sim 0.35 \hOMpc$, the S/N can be overestimated by as much as an order of magnitude. However, the Gaussian approximation  gets better for the more abundant groups, and the overestimation is narrowed to a factor of a few at  $k_{\rm max} \sim 0.35 \hOMpc$.

We also show the S/N obtained by combining the power spectrum and bispectrum measurements. We have properly included the cross covariance between the power spectrum and bispectrum in the joint covariance matrix. The covariance between the power spectrum and bispectrum are in general non-negligible. We will leave it for future work to model the cross covariance in details.  The joint S/N coincides with the power spectrum one in the low $k$ regime. In the mildly nonlinear regime, the total S/N is slightly higher than that of the power spectrum, this suggests that the power spectrum is the main source for the total  S/N. The S/N  also saturates for $k \gtrsim 0.2 \hOMpc$, similar to the trend of the power spectrum.

\begin{figure}[!htb]
\begin{center}
\includegraphics[width=\linewidth]{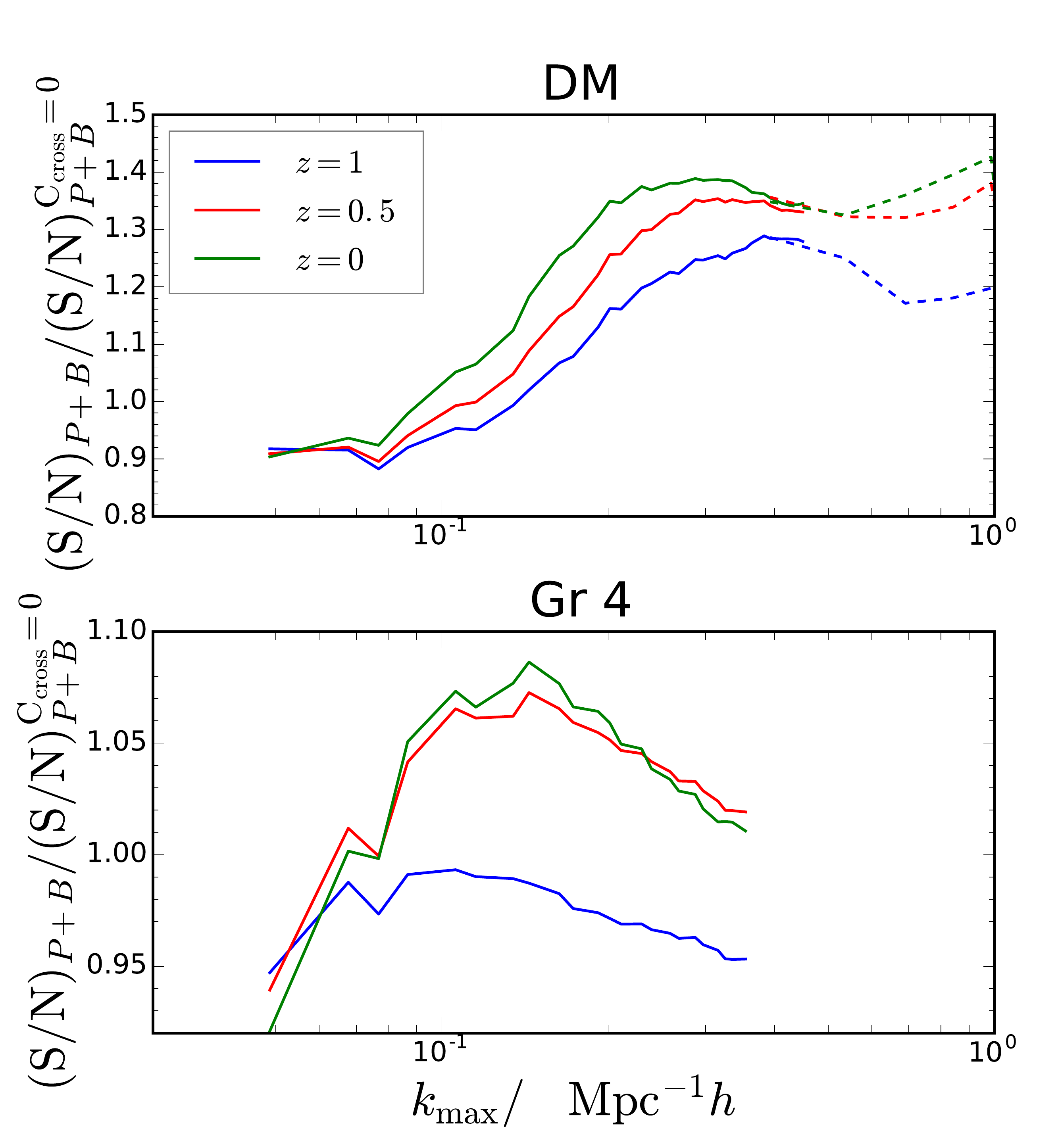}
\caption{  The combined S/N ratio obtained with the cross covariance between $P$ and $B$ properly taken into account and the cross covariance ignored. The upper panel is for dark matter, while the lower one is for halo group 4. The results at $z=1$ (blue), 0.5 (red), and 0 (green) are compared. Both the narrow bispectrum binning results ($2k_{\rm F}$, solid line) and the wide binning one ($8k_{\rm F}$, dashed line) are shown. }
\label{fig:SNratio_PBcross}
\end{center}
\end{figure}

As a contrast, we compare with the case that the cross covariance between $P$ and $B$ ignored in Fig.~\ref{fig:SNratio_PBcross}. For dark matter, the cross covariance lowers the total S/N at low $k_{\rm max} $ by 10\%, while in the high $k$ regime, it is enhanced by 20 to 40\%. We find that the effect of the cross covariance is smaller  for the case of halos. When the shot noise is large ($z=1$), the cross covariance lowers the total S/N by a few per cent up to  $k_{\rm max} \sim 0.3 \hOMpc $. For higher abundance, the trend is qualitatively similar to the dark matter case, and the cross covariance enhances the S/N  by a few per cent in the mildly nonlinear regime.   On the other hand, the amount of S/N in the nonlinear regime is much less than that suggested by the Gaussian covariance approximation.  Our results shows that adding the bispectrum information to the power spectrum only improves the total information content mildly.

In the context of weak lensing, the authors of \cite{KayoTakadaJain_2013} compared the S/N ratio for power spectrum and bispectrum, and the total, using the non-Gaussian covariance. Their results are consistent with ours; in particular, they also found that non-Gaussian covariance significantly degrades the S/N in the nonlinear regime.    

In cosmology, we are ultimately interested in how well the measurements of the polyspectra can put constraints on the cosmological parameters, and this can be estimated using the Fisher analysis. Furthermore, it has previously been shown that the Fisher analysis results may not be easily interpreted from a simpler signal-to-noise analysis. For example, ref.~\cite{Repp:2015} found that the power spectrum can still place strong constraint on the parameters which are not sensitive to the amplitude of the power spectrum in the nonlinear regime (see also \cite{Blot:2015cvj}).  We will leave a Fisher analysis for the combined power spectrum and bispectrum for future work (for weak lensing, see \cite{Sato:2013mq,Kayo:2013aha}).

%% Before closing this section, because Gaussian covariance is often used, it would be useful to have a rule-of-thumb estimate for the maximum wavenumber that one can safely use the Gaussian covariance. 
%% \red{ For DM, provide a rule-of-thumb for validity of Gaussian approximation for teh bispectrum covariance. Give 20\%  and a factor 50\% deviation from  the Gaussian approximation as a function of redshift. For halos, it significant depends on the number density, not provided here.  For the rule-of-thumb use the linear theory instead of the 1-loop as it is not very stable.  }

%% \begin{figure}[!htb]
%% \begin{center}
%% \includegraphics[width=0.9\linewidth]{k_{\rm max}max_z_BcovNGestimate.pdf}
%% \caption{ The wavenumber  $k_{\rm max}$ that the leading non-Gaussian correction reach 10\% and 20\% of the Gaussian covariance $C_{\rm G}^{\rm L} $.          }
%% \label{fig:kmax_z_BcovNGestimate}
%% \end{center}
%% \end{figure}
%% The halo bisepctrum covariance depends a lot on the shot noise, and the the model does not work...

\section{Conclusions }
\label{sec:conclusions}
Gaussian covariance is often assumed in making a forecast. In this paper, we have used a large suite of simulations from the DEUS-PUR project to study the covariance of the power spectrum and bispectrum, paying special attention to quantifying the effects of the non-Gaussian contributions to the covariance.    This work is the first to use such a large number of $N$-body simulations (altogether 4704) to estimate the covariance of the bispectrum.

We find that the non-Gaussianity is significant in the dark matter bispectrum covariance.  The diagonal elements of the covariance of the dark matter bispectrum already deviate from the Gaussian  covariance at $k \sim 0.1 \hOMpc $ by 10 \% at $ z=0 $. The correlation increases as the redshift decreases and we find that at $z=0$  the correlation coefficient $r(k_i,k_j) $ is within 20\% if $k_i$ and $k_j$ are less than $ 0.2 \hOMpc$  for the equilateral triangle configurations.  The covariance of the dark matter bispectrum significantly increases in the mildly nonlinear regime. To compare with the simulation results, we have  computed the leading disconnected non-Gaussian corrections in the 6-point function. %Because the integrals involved are high dimensional yet narrowly peaked, conventional methods would not be able to integrate them. To this goal, we have developed an algorithm to efficiently sample the points that satisfy the Dirac delta function constraints.
Including these non-Gaussian corrections we find that the predictions give good agreement with the simulation results in the weakly nonlinear regime. For the equilateral triangle configurations, the diagonal term agrees  with the simulation results up to $ k \sim 0.3 \hOMpc $ at $z=1$ and $0.16 \hOMpc $ at $z=0 $.

We have also studied the covariance matrix of the halo power spectrum and bispectrum. We distinguished between the case when the mean Poisson shot noise is subtracted and the Poisson shot noise is estimated and subtracted from each realization.   On the theory side we used the Poisson model to derive  the Poisson shot noise contribution to the covariance of the power spectrum and bispectrum. The model corresponds to the scenario in which the mean Poisson shot noise is subtracted.  For the power spectrum, the Poisson model describes the diagonal elements of the covariance matrix reasonably well, but it tends to overpredict the correlation coefficients. For the bispectrum, the model underpredicts the covariance, especially when the number density of the sample is low.  The covariance of the power spectrum and bispectrum depend on the 4-point and 6-point functions respectively. The expansion contains large number of terms and also various high order correlators. In the prediction, we only computed the terms with the power spectrum terms and some of the bispectrum terms. This can cause part of the disagreement between the model and simulation results.

On the other hand, in simulations or observations, we often have to estimate the shot noise using the volume-averaged density. When the individual Poisson shot noise is subtracted, we find the halo covariance is significantly reduced and gets close to the Gaussian covariance.  These hold for both the halo power spectrum and bispectrum.  This is because most of the non-Gaussian covariance arises from the fluctuations in the Poisson shot noise term.   Therefore, although the shot noise covariance is large, in reality because we use the number density estimated from the simulation/survey directly to subtract the shot noise, the halo covariance is close to the Gaussian covariance.

%We find that for both the halo power spectrum and bispectrum when the Poisson shot noise is subtracted in each realization, the resultant covariance is substantially reduced and gets closer to the Gaussian limit. It makes sense because part of the fluctuation is absorbed in the shot noise in this case.  For example, at $z=0$, the diagonal elements of the bispectrum covariance for the equilateral triangle configurations is roughly within 50\% from the Gaussian covariance up to $k \sim 0.35 \hOMpc$, and the cross correlation coeffcients are within 20\% when $ k_i $ and $k_j $ are less than $ 0.2 \hOMpc $. However we note that there is still significant amount of non-Gaussianity left.  We will leave for future work to model the covariance in this case. 

We note that the magnitudes of the halo power spectrum and bispectrum covariances are generally higher than those of the dark matter ones. The magnitudes of the halo power spectrum and bispectrum relative to those of the dark matter are quite similar.  The shot noise contribution decreases as the number density of halos increases. Even when the number density is $5 \times 10^{-4}  (\MpcOh)^{-3} $, the halo covariance is still higher than that of the dark matter by one order of magnitude. Thus matter nonlinearity and halo biasing are not expected to play an important role in this case.  Only when the number density is as high as $ 10^{-3}  (\MpcOh)^{-3} $, the covariance of the matter polyspectrum is comparable to that of the halo.

In this work we consider simulations with periodic boundary conditions; thus the supersample covariance  does not contribute. The study of the impact of the bispectrum supersample covariance will be presented elsewhere. We have compared the simulation results obtained from the Large and Small box sizes, and found that the power spectrum and bispectrum covariances are not sensitive to the volume of the box as expected from theory. Hence, to efficiently beat down the noise on the covariance, we can simulate the small scale non-Gaussianity using small box size.

As the power spectrum has been well explored, it is important to ask how much information one can gain by studying the bispectrum in large scale structure. In the second part of the paper, we have quantified the information content of the power spectrum and the bispectrum with the S/N ratio using the non-Gaussian covariance matrix measured from simulations.

The S/N of the dark matter power spectrum reaches a plateau in the mildly nonlinear regime.  This is because in this regime the covariance increases faster than the signal, causing the S/N to reach a plateau.  At $z=0 $, it flattens at $k_{\rm max} \sim 0.15\hOMpc $. This is in line with the findings of previous works. Similarly, we find that the S/N of the halo power spectrum flattens in the regime $k_{\rm max} \sim 0.1-0.2 \hOMpc$ depending on the number density of the samples.  The S/N of the halo power spectrum increases as the number density of the sample increases. We find that  at $z=0$  only the samples with number density $\gtrsim  5 \times 10^{-4} (\MpcOh )^{-3} $ yield the S/N comparable to that of dark matter. In contrast,  the Gaussian covariance approximation overestimates the S/N by a factor of two to a few at $ k_{\rm max}\sim 0.4 \hOMpc $, depending on the number density of the sample.

 For the bispectrum, we have computed the S/N using all the triangle configurations with lengths less than certain  $k_{\rm max}$.  For the case of the dark matter bispectrum, the S/N increases much slower than the Gaussian approximation suggests. For example, at $z=0$, the S/N at  $k_{\rm max} = 0.2 \hOMpc $ is an order of magnitude lower than the Gaussian result.   Although the Gaussian covariance suggests that the S/N of the matter bispectrum surpasses that of the matter power spectrum at $k_{\rm max} \sim 0.14 \hOMpc $ at $z=0$, using the non-Gaussian covariance we find that the S/N  is only 30\% of the Gaussian one at this scale.  In the nonlinear regime, the S/N of the dark matter bispectrum is still mildly increasing but it  stalls at $k_{\rm max} \sim 0.4 \hOMpc $.  The S/N of the halo bispectrum shares similar trends as that of the matter bispectrum. The Gaussian covariance approximation significantly overestimates the S/N. We find that the overestimation varies from an order of magnitude for the rare sample [$n \sim 10^{-5} (\MpcOh)^{-3} $] to a factor of a few for the abundant sample [$n  \sim  5 \times  10^{-4} (\MpcOh)^{-3} $] at  $k_{\rm max} \sim 0.3 \hOMpc $ for the redshift range considered.

We conclude that the bispectrum S/N is degraded more seriously by nonlinearities and shot noise relative to the power spectrum S/N.  Thus the bispectrum only adds a small amount of increment to the total S/N when the bispectrum is combined with the power spectrum.

Despite more than a decade of efforts to measure the 3-point statistics in Fourier space \cite{Feldman:2000vk,Scoccimarro:2000sp,Verde:2001sf,Gil-Marin:2014sta}  and configuration space \cite{Jing:2003nb,GaztanagaScoccimarro2005,Wang:2004km,Marin_2011,Marin_etal2013,Slepian:2015hca,Guo:2016ohx}, the information gain that we get is still modest compared to that from the 2-point statistics. It is well-known that the 3-point statistics are more sensitive to nonlinearities and halo biasing, both the local \cite{FryGaztanaga} and nonlocal \cite{ChanScoccimarroSheth2012, Baldaufetal2012} ones.  This is both a blessing and curse.  On one hand, it is easier to estimate the nonlinear coupling and higher order bias parameters using  bispectrum.  On the other hand, it suffers from stronger nonlinear effects and is harder to model.  In this sense our analysis simply reveals the cursing part that strong nonlinearities cause large information loss.

Given the low S/N of the bispectrum, it is not very useful to constrain the cosmological parameters alone.  However, there are subtle effects for which the bispectrum analysis is particularly useful.  When the power spectrum is combined with the bispectrum, some degeneracies can be broken.   For example the degeneracy between the linear bias  $b_1 $ and the growth rate can be broken when the halo power spectrum is combined with the halo bispectrum.  The bispectrum is also an important tool to constrain primordial non-Gaussianities.   These subtle effects are not reflected in the general signal-to-noise analysis.

As we are ultimately interested in how well the polyspectra can constrain the cosmological parameters, the Fisher matrix analysis is preferable. Previous works found that the power spectrum in the nonlinear regime can still constrain some of the cosmological parameters which are not sensitive to the amplitude of the power spectrum \cite{Repp:2015} and that the weak lensing bispectrum can yield strong constraints on the cosmological parameters even though its S/N is relatively low \cite{Sato:2013mq,Kayo:2013aha}. We leave the Fisher analysis for future work.

Even though we have only analysed the cases of the power spectrum and bispectrum, we speculate that higher point correlators, such as the trispectrum, may suffer information loss due to nonlinearities and shot noise even more seriously than the Gaussian approximation suggests. If this is true, then it is not a fruitful program to keep on measuring the correlation hierarchy. We can instead consider alternative ways  to extract information. Some of the interesting methods include log transformation \cite{ColesJones1991,NeyrinckSzapudiSzalay2009,CarronNeyrinck2012}, the clustering of voids \cite{2014PhRvL.112d1304H,2014PhRvD..90j3521C,2016MNRAS.456.4425C,2016PhRvL.116q1301K,2016PhRvL.117i1302H} and trying to recover information from the phases of the density field \cite{Chiang:2002,Wolstenhulme:2015,Eggemeier:2015}.

%\red{ Here we have used a simple estimator for the power spectrum and bispectrum. The signal-to-noise may be enhanced by using an optimal estimator, such as FKP, mass weighting,  can optimal weighting reduces the covariance?  } \blue{FKP assumes gaussian covariance}

 \begin{figure*}[!tb]
\centering
\includegraphics[width=0.8\linewidth]{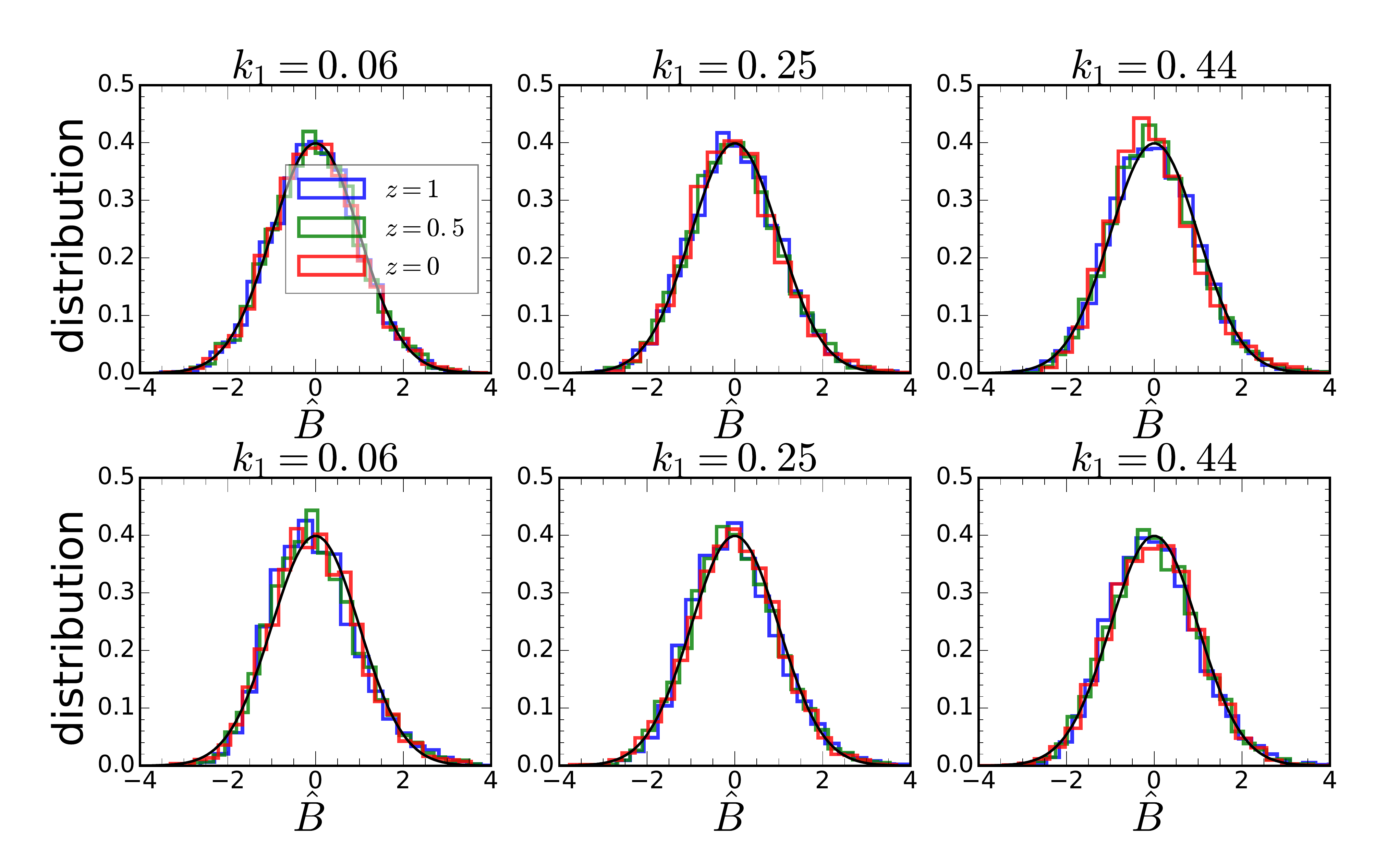}
\caption{ The distribution of the bispectrum estimator. The results from the Small simulation set at $z=1$ (blue), 0.5 (green) and 0 (red) are shown. The upper panels correspond to the results from dark matter while the lower panels are for halo Gr.~4.  The equilateral triangle configuration is used. Results at three different wave  numbers $k = 0.06$, 0.25 and $0.44 \hOMpc  $ (from left to right) are displayed.  The data have been transformed to the standard variable by Eq.~\ref{eq:std_Gaussian_variable}.  The Gaussian distribution with zero mean and unity variance is overplotted (solid black line).  }
\label{fig:BkEstimator_dist_111}
\end{figure*} 

\begin{figure}[!htb]
\centering
\includegraphics[width=0.9\linewidth]{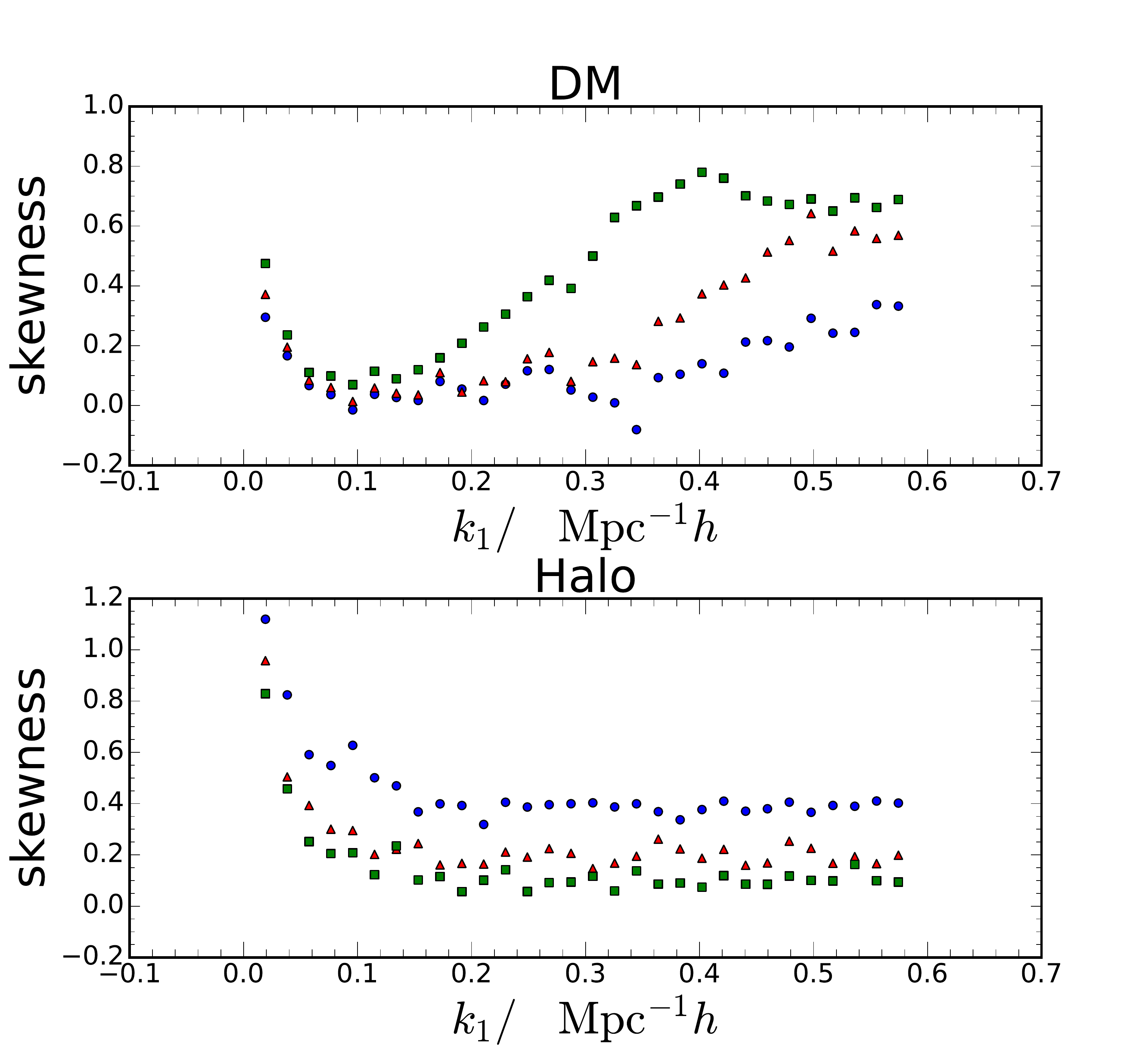}
\caption{  The skewness of the bispectrum estimator. The results from the dark matter (upper panel) and halo Gr.~4 (lower panel) of the Small simulation set are shown. Only equilateral triangle configuration is used. The results from $z=1$ (blue circles), 0.5 (red triangles) and 0 (green squares) are compared.  }
\label{fig:skewness_Bk_111}
\end{figure}

\begin{figure}[!htb]
\centering
\includegraphics[width=0.9\linewidth]{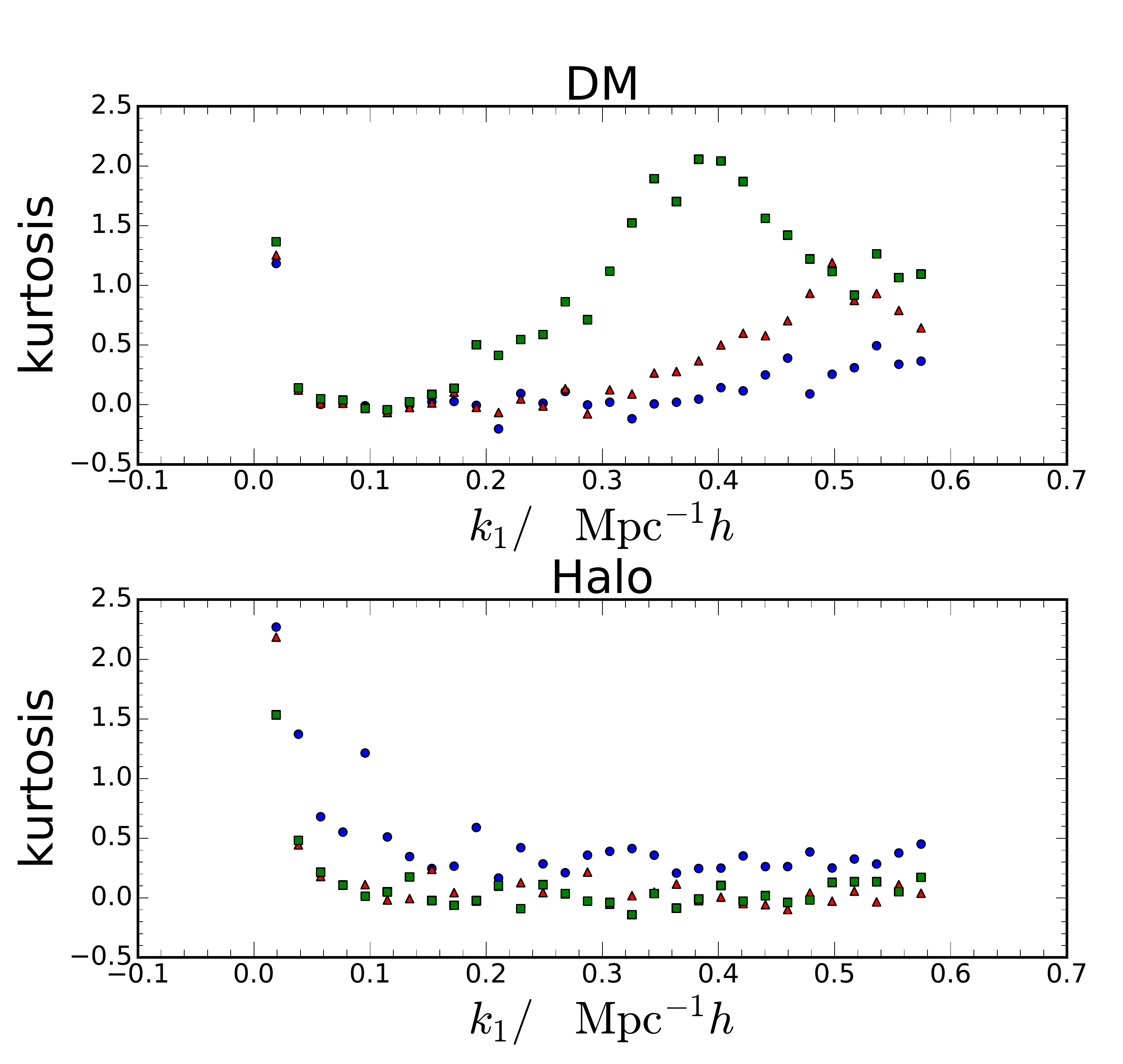}
\caption{ Similar to Fig.~\ref{fig:skewness_Bk_111} except for kurtosis.   }
\label{fig:kurtosis_Bk_111}
\end{figure}

\section*{Acknowledgment} 
We are grateful to Mart\'in Crocce, Azadeh Moradinezhad, Jorge Nore\~na, and Ravi Sheth for useful discussions, and Yann Rasera for help with the simulations. We thank Masahiro Takada and the referee for comments on the draft. We thank the members of the DEUS Consortium\footnote{\url{http://www.deus-consortium.org}} for sharing the data with us. This work was granted access to HPC resources of IDRIS through allocations made by GENCI (Grand \'Equipement National de Calcul Intensif).  K.C.C. and L.B. acknowledge the support from the Spanish Ministerio de Economia y Competitividad grant ESP2013-48274-C3-1-P.

\appendix

\section{ The distribution of the bispectrum estimator } 
\label{sec:bisp_estimaotr_distribution}

The distribution of the estimator being Gaussian is crucial for many analytic results, e.g.~the bias correction formula Eq.~\ref{eq:Cinv_correct}. Refs.~\cite{Takahashi:2009ty,Blot:2014pga} checked that the power spectrum estimator for dark matter follows the Gaussian distribution well. The skewness and kurtosis of the estimator agree with the chi-square distribution, which is a consequence of the underlying density field being Gaussian. However, \cite{Blot:2014pga} also found that the skewness deviates from the chi-square distribution result at low $z$ for  $k \gtrsim 0.2 \hOMpc $.  In this section, we shall check the distribution of the bispectrum estimator $\hat{B}$ (Eq.~\ref{eq:Bisp_estimator}).

In Fig.~\ref{fig:BkEstimator_dist_111}, we plot the distribution of the bispectrum estimator. We have transformed the measured data $\hat{B}_{\rm data} $  to the standard normal variable as
\beq
\label{eq:std_Gaussian_variable}
\hat{B}_{\rm snv} = \frac{ \hat{B}_{\rm data}  -  \mu  }{ \sigma },  
\eeq
where $\mu$ and $ \sigma $ are the mean and the standard deviation of the data. Here we use the data from the Small simulation set at $z=1$, 0.5 and 0, and only the results from the equilateral triangle configuration are shown. Upon comparison with the standard normal distribution, we find that the data across the three redshifts follow  the Gaussian distribution reasonably well. Furthermore, the results for both dark matter and halos are similar. The tendency for the distribution to be Gaussian results from the central limit theorem because in the estimator Eq.~\ref{eq:Bisp_estimator}, a large number of modes are averaged over. Nonetheless we note that there are some visible deviations from Gaussianity.

There are two counterinteracting effects at work. First, the central limit theorem works asymptotically for a large number of samples. When $k$ increases, there are more modes available to be averaged over in the estimator, see Eq.~\ref{eq:V123_final}. Thus we expect the central limit theorem to perform better for high $k$.  On the other hand, the underlying density field becomes more non-Gaussian and the modes couple with each other at high $k$. This violates the key assumption that the samples are independent in the central limit theorem.  Therefore we anticipate that the central limit theorem will fail at both the low $k$ and high $k$ regimes.

In order to quantify the deviation from Gaussianity, we compute the sample skewness and kurtosis as 
\beqa
S_3 &=& \frac{ \frac{1}{n}  \sum_{i=1}^n ( \hat{B}_i  - \bar{B})^3 }{ \Big[ \frac{1}{n}  \sum_{i=1}^n ( \hat{B}_i  - \bar{B})^2\Big]^{ \frac{ 3}{2 } }  } ,   \\
S_4 &=& \frac{ \frac{1}{n}  \sum_{i=1}^n ( \hat{B}_i  - \bar{B})^4 }{ \Big[ \frac{1}{n}  \sum_{i=1}^n ( \hat{B}_i  - \bar{B})^2\Big]^2  }  - 3 .  
\eeqa
We plot the skewness in Fig.~\ref{fig:skewness_Bk_111}. If the underlying density field is exactly Gaussian, as the skewness is  essentially the 9-point function of the underlying density field, it will vanish. Thus the finite value of the skewness is an indication of the deviation of the underlying density field from Gaussianity.  For dark matter, at low $k$ we find that the skewness is large precisely because the number of modes available to estimate the results are small. The skewness first decreases but eventually increases again  as $k$ increases. This is because the mode coupling in the underlying density field increases in the  nonlinear regime. The fact that the lower the redshift the larger the value of the skewness supports this interpretation. For the halo, we again find that the skewness is large at low $k$. The skewness increases with redshift and this can be attributed to the fact that there are larger Poisson fluctuations at high redshift due to lower number density.  On the other hand, we also find that as $k$ increases the skewness saturates to some constant value, which is harder to understand. 

We present the kurtosis in Fig.~\ref{fig:kurtosis_Bk_111}. For the dark matter bispectrum,  similar to the skewness, we find that kurtosis is large at low $k$, and then decreases to zero as $k$ increases, but eventually increases again at high $k$. Curiously, there is a bump in the kurtosis at  $k \sim 0.4 \hOMpc$ at $z=0$. We checked and confirmed that for other shapes, their kurtosis also exhibits a bump. This bump is hard to interpret without a concrete model.  Note that there is a small bump in skewness as well, however, this is not present in other shapes. The halo bispectrum kurtosis behaves in a similar way to the skewness. 

We have checked that the results are qualitatively similar for other shapes.

\begin{widetext}

\section{ A collection of derivations  }
\label{sec:derivation_zoo}
In this appendix, we present the derivations of some of the formulas used in the main text.

\subsection{$ V_{123}  $ }
We can analytically integrate  $V_{123}$ (Eq.~\ref{eq:V123_final}) as
\beqa
V_{123} &= & \int_{k_1} d^3 p \int_{k_2} d^3 q \int_{k_3} d^3 r \,   \Ddel( \mb{p} + \mb{q} + \mb{r} ) 
= \int_{k_1} d^3 p \int_{k_2} d^3 q \int_{k_3} d^3 r \, \int \frac{ d^3 x }{ (2 \pi)^3} e^{ i \mb{ x } \cdot ( \mb{p} + \mb{q} + \mb{r} ) }  \nn \\
&=&   \int \frac{ d^3 x }{ (2 \pi)^3}  \int_{k_1} dp 4 \pi p^2 j_0 (px)    \int_{k_2} dq 4 \pi q^2 j_0 (qx)      \int_{k_3} dr 4 \pi r^2 j_0 (rx) \nn \\ 
&=&   \int_{k_1} dp 4 \pi p^2  \int_{k_2} dq 4 \pi q^2    \int_{k_3} dr 4 \pi r^2  \frac{1}{ 2 \pi^2}    \int dx \, x^2 j_0 (px)  j_0 (qx)  j_0 (rx) . 
\eeqa
We now make use of an identity for  the integral of a product of three  spherical Bessel functions
\cite{Mehrem2011}
\beq
\label{eq:intg_produ_3j0}
\int_0^\infty  dr  \, r^2  j_{0 } (k_1 r) j_{0} (k_2 r) j_0 (k_3 r)  = \frac{\pi \beta(\Delta) }{4 k_1 k_2 k_3 }, 
\eeq 
where $\Delta  $ and $\beta (\Delta) $ are defined in Eq.~\ref{eq:Delta_triangle} and \ref{eq:beta_Delta} respectively.    Therefore we get 
\beqa
V_{123} =  8 \pi^2 k_1 k_2 k_3 ( \Delta k )^3 \beta(\Delta ) . 
\eeqa

\subsection{ Gaussian covariance of the bispectrum estimator $\hat{B}$ } 
For Gaussian $\delta $, the only non-vanishing contribution to the Gaussian covariance $C_{\rm G}^{\rm L} $ (Eq.~\ref{eq:BCov_Gaussian}) reads
\beq
C_{\rm G}^{\rm L} =  \int \mathcal{D}  \delta_{k_1k_2k_3, k_1'k_2'k_3'  } s_{123}  \Ddel( \mb{p} + \mb{p}' ) P_{\rm L}(p)    \Ddel( \mb{q} + \mb{q}' ) P_{\rm L}(q)  \Ddel( \mb{r} + \mb{r}' ) P_{\rm L}(r) . 
\eeq
 $\delta_{k_1k_2k_3, k_1'k_2'k_3'  }$  is non-vanishing only if the shape of the triangle  $ k_1k_2k_3 $ is the same as that of $ k_1'k_2'k_3' $.  If none of the sides are equal  $ s_{123} =1$. If the triangles are isosceles, $ s_{123} =2$. For the equilateral triangle, we have  $ s_{123} = 6$. Both  $\delta_{k_1k_2k_3, k_1'k_2'k_3'  }$ and  $s_{123} $ arise from the fact that the three Dirac delta functions must be satisfied and the number of contractions that they can be satisfied.

Then we can simplify it further as
\beqa
 C_{\rm G}
&=&  \frac{ k_{\rm F}^6 }{V_{123}^2 }  \delta_{k_1k_2k_3, k_1'k_2'k_3'  } s_{123} \int_{k_1}d^3 p \int_{k_2}d^3 q \int_{k_3}d^3 r  \Ddel( \mb{p} + \mb{q} + \mb{r} )    \Ddel( \mb{0} ) P_{\rm L}(p)  P_{\rm L}(q)  P_{\rm L}(r)  \nn \\
 &=&  \frac{ k_{\rm F}^3 }{V_{123}^2 }  \delta_{k_1k_2k_3, k_1'k_2'k_3'  } s_{123}  \int_{k_1}d^3 p \int_{k_2}d^3 q \int_{k_3}d^3 r   P_{\rm L}(p)  P_{\rm L}(q)  P_{\rm L}(r)     \int \frac{d^3x  }{(2 \pi  )^3 } e^{ i \mb{x }\cdot ( \mb{p} +  \mb{q} +  \mb{r} )}  \nn \\
&=&  \frac{ k_{\rm F}^3 }{V_{123}^2 }  \delta_{k_1k_2k_3, k_1'k_2'k_3'  } s_{123}   \int \frac{d^3x  }{(2 \pi  )^3 } \int_{k_1} dp 4 \pi p^2 j_0 (px)   \int_{k_2} dq 4 \pi q^2 j_0 (qx) \int_{k_3} dr 4 \pi r^2 j_0 (rx) P_{\rm L}(p) P_{\rm L}(q) P_{\rm L}(r)  \nn \\
&\approx&  \frac{ k_{\rm F}^3 }{V_{123} }  \delta_{k_1k_2k_3, k_1'k_2'k_3'  } s_{123}   P_{\rm L}(k_1) P_{\rm L}(k_2) P_{\rm L}(k_3) ,
\eeqa
where in the last line we have assumed that the bin is narrow and have taken the power spectra out of the integral. We note that in the first line, there is factor of $\Ddel (\mb{0}) $. This arises from the structure that there are two sets of Dirac delta functions, $  \Ddel( \mb{p} + \mb{q} + \mb{r} )  $ and   $  \Ddel( \mb{p}' + \mb{q}' + \mb{r}' ) $,  and  $ \Ddel( \mb{p} + \mb{p}' )$,  $\Ddel( \mb{q} + \mb{q}' )$, and $ \Ddel( \mb{r} + \mb{r}' )$. Hence one must be redundant,  it results in   $\Ddel (\mb{0}) $. We find a similar pattern for the dark matter bispectrum non-Gaussian covariance terms as well.

%In this derivation, we have assumed the power spectrum to be the linear one.    However,  when we consider the non-Gaussian contribution, we find that part of the contribution can be resummed if we use the 1-loop matter power spectrum instead of the linear one. \red{ Hence we will use nonlinear power measured from simulation for $ C_{\rm G}  $.  }

\subsection{ Integration domain volume  $ U $ } 
The volume of the integration domain defined by the Dirac delta functions in the  dark matter bispectrum non-Gaussian covariance terms can be computed analytically as
\beqa
U( k_1,k_1'; k_2,k_3,k_2',k_3')  &\equiv&  \int_{k_1}d^3 p \int_{k_2}d^3 q \int_{k_3}d^3 r  \Ddel( \mb{p} + \mb{q} + \mb{r} )  \int_{k_1'}d^3 p' \int_{k_2'}d^3 q' \int_{k_3'}d^3 r'  \Ddel( \mb{p}' + \mb{q}' + \mb{r}' )   \Ddel( \mb{p} + \mb{p}' ) \nn \\
& = &   \delta_{k_1, k_1'}   \int_{k_1} d^3 p  \int_{k_2} d^3 q   \int_{k_3} d^3 r   \int_{k_2'} d^3 q'    \int_{k_3'} d^3 r'  \Ddel( \mb{p} +   \mb{q} +   \mb{r} ) \Ddel( - \mb{p}  + \mb{q}' + \mb{r}' ) \nn \\
&=&  \delta_{k_1, k_1'}   \int  \frac{ d^3 x }{ (2 \pi)^3  }    \int  \frac{ d^3 y }{ (2 \pi)^3  }   \int _{k_1} dp 4 \pi p^2   j_0( p |\mb{x} - \mb{y} |)   \int _{k_2} dq 4 \pi q^2  j_0( qx)   \int _{k_3} dr 4 \pi r^2  j_0(  rx )   \nn \\
&\times & \int _{k_2'} dq' 4 \pi q'^2   j_0(q'y)   \int _{k_3'} dr' 4 \pi r'^2  j_0(r' y ). \nn  \\   
\eeqa
We can expand  $j_0(| \mb{x}-\mb{y} | r ) $ using the addition theorem for the spherical Bessel function
(Eq.~10.1.45 in \cite{AbramowitzStegun64})
\beq
j_0( | \mb{x}-\mb{y} | r ) = 4 \pi \sum_{l=0}^{\infty} \sum_{m = -l }^{ l }   j_l( xr) j_l(yr) Y_{lm}( \hat{\mb{x}}) Y_{lm}^*( \hat{\mb{y}}).  
\eeq
After taking the angular integrals of $\mb{x}$ and $\mb{y} $, we get 
\beq
U =   2^8 \pi \delta_{k_1, k_1'}  \int_{k_1} d p p^2  \int_{k_2} d q q^2  \int_{k_3} d r r^2  \int_{k_2'} d q'  q'^2    \int_{k_3'} d r'  r'^2   \int dx x^2 j_0( px) j_0(qx) j_0(r x)     \int dy y^2 j_0( py) j_0(qy) j_0(r y).    
\eeq
Using Eq.~\ref{eq:intg_produ_3j0}, we arrive at
\beq
U( k_1,k_1'; k_2,k_3,k_2',k_3') =  2^4 \pi^3 k_2 k_3 k_2' k_3' (\Delta k)^5  \beta( \Delta ) \beta( \Delta' )  \delta_{k_1, k_1'} .   
\eeq
The result is the same for $ \Ddel( \mb{p} - \mb{p}' ) $.

\section{Poisson shot noise contribution to the covariance of the halo power spectrum and bispectrum   }
\label{sec:Poisson_model_PkBk}
In this section, we derive the Poisson shot noise contribution to the covariance of the halo power spectrum and bispectrum. We consider the Poisson model in which the number density of the tracers is given by 
\beq
n(\mb{x} ) = \sum_i \Ddel( \mb{x} - \mb{x}_i ). 
\eeq
The discrete density contrast $\delta_{\rm d}$ is defined as 
\beq
\delta_{\rm d} (\mb{x}) =  \frac{ n( \mb{x} )  }{ \bar{n} }  - 1,
\eeq
where $ \bar{n} $ is the mean number density $\bar{n}  \equiv  \langle n(\mb{x} )   \rangle $.  In this section, all the smooth correlators such as $\xi$, $P$, $\zeta$, $B$ etc, refer to the  nonlinear correlators of the tracers.  The tracers can be unbiased such as the dark matter particles in $N$-body simulations. Halos are the prototypical example of biased tracers. The Poisson model can be applied to both kinds of tracers.

\subsection{Poisson shot noise contribution to the power spectrum covariance}
\label{sec:appendix_PoissonPk}
To get the Poisson shot noise contribution to the covariance of the power spectrum we will need the Poisson shot noise contribution to the 2-point and 3-point functions as well. As the computations are similar but less cumbersome, it is instructive to first review the derivations for the 2-point and 3-point functions. One can also include weighting, see Ref.\cite{ChanScoccimarro2012}, however we will not consider this here. In Ref.\cite{Matarrese:1997sk} the correlators including the Poisson shot noise are derived using an elegant functional method. Ref.~\cite{Smith:2008ut} applied the Poisson model to compute the shot noise contribution to the covariance of the cross power spectrum between matter and halo.  We will compare our results with theirs whenever possible.

The 2-point correlation of the discrete field is given by 
\beq
\xi_{\rm d}( \mb{x}_1, \mb{x}_2 )   =  \langle \delta_{\rm d}(\mb{x}_1 ) \delta_{\rm d}(\mb{x}_2 )    \rangle  
= \frac{1  }{ \bar{n}^2 } \langle  n( \mb{x}_1 )  n( \mb{x}_2 ) \rangle  - 1.  
\eeq
The 2-point correlator of $n$ can be written as  
\beqa
\label{eq:nn_correlator}
  \langle  n( \mb{x}_1 )  n( \mb{x}_2 ) \rangle   
&=& \langle \sum_i  \Ddel( \mb{x}_1 - \mb{x}_i )  \Ddel( \mb{x}_2 - \mb{x}_i )    \rangle 
+    \langle \sum_{i,j}  \Ddel( \mb{x}_1 - \mb{x}_i )  \Ddel( \mb{x}_2 - \mb{x}_j )    \rangle \nn \\
&=&\Ddel( \mb{x}_1 - \mb{x}_2 ) \bar{n} 
 +  \bar{n}^2 [ 1 + \xi( |\mb{x}_1  - \mb{x}_2 |) ]. 
\eeqa
In this section, all the dummy indices in the summation are unequal. For discrete points, we need to separate the part when two points are the same from the case when the points are different.  The latter case can be modelled by the smooth correlation function $ \xi $. Therefore the discrete correlation function can be written as
\beq
\label{eq:nn_correlation_funct}
\xi_{\rm d }( |\mb{x}_1  - \mb{x}_2 |)   = \xi( |\mb{x}_1  - \mb{x}_2 |)  + \frac{1}{ \bar{n} } \Ddel( \mb{x}_1 -  \mb{x}_2 ) . 
\eeq
Upon Fourier transforming, the discrete power spectrum reads \cite{Peebles}:
\beq
\label{eq:Pk_discrete}
P_{\rm d } (k ) = P(k) + \frac{ 1 }{ \tilde{n}  },  
\eeq
where $P(k)$ is the continuous power spectrum and $\tilde{n} \equiv  ( 2 \pi )^3 \bar{n} $. This is the well-known shot noise correction for the power spectrum (the presence of  $(2 \pi )^3$ is due to the Fourier convention used in this paper).    

Similarly the discrete 3-point function reads
\beqa 
\zeta_{\rm d}(\mb{x}_1,\mb{x}_2, \mb{x}_3) =  \langle \delta_{\rm d }(\mb{x}_1) \delta_{\rm d }(\mb{x}_2) \delta_{\rm d }(\mb{x}_3)   \rangle = \frac{ \langle n( \mb{x}_1 ) n( \mb{x}_2 ) n( \mb{x}_3 )  \rangle  }{ \bar{n}^3 }
 - \Big[ \frac{  \langle n( \mb{x}_1 ) n( \mb{x}_2 )   \rangle  }{ \bar{n}^2 }  + 2 \, \mathrm{cyc.}   \Big] + 2 ,
\eeqa
where cyc.~denotes cyclic permutations. As in Eq.~\ref{eq:nn_correlator}, we can express the 3-point correlator of $ n $ as
\beqa
\label{eq:nnn_correlator}
&& \langle n (\mb{x}_1) n(\mb{x}_2) n(\mb{x}_3)   \rangle  \nn \\
 & =  &
\langle \sum_i \Ddel( \mb{x}_1 - \mb{x}_i ) \Ddel( \mb{x}_2 - \mb{x}_i ) \Ddel( \mb{x}_3 - \mb{x}_i ) \rangle   +    
\Big[ \langle  \sum_{i,j} \Ddel( \mb{x}_1 - \mb{x}_i  ) \Ddel( \mb{x}_2 - \mb{x}_j  ) \Ddel( \mb{x}_3 - \mb{x}_j  )    \rangle  +2 \, \mathrm{cyc.} \Big]  \nn \\
&+&  \langle  \sum_{i,j, k}  \Ddel( \mb{x}_1 - \mb{x}_i  ) \Ddel( \mb{x}_2 - \mb{x}_j  ) \Ddel( \mb{x}_3 - \mb{x}_k  )   \rangle     \nn\\
&=& \Ddel(\mb{x}_1 - \mb{x}_2 ) \Ddel(\mb{x}_1 - \mb{x}_3 ) \bar{n}  + \big[ \Ddel( \mb{x}_2 - \mb{x}_3 ) \bar{n}^2 ( 1 + \xi_{12} ) + 2 \, \mathrm{cyc.}\big]  + \bar{n}^3 ( 1 + \xi_{12} + \xi_{23} + \xi_{31} + \zeta ) ,  
\eeqa
where $\zeta $ is the continuous 3-point function. For convenience, we have used $\xi_{12} $ to denote $\xi( |  \mb{x}_1 -   \mb{x}_2  |)  $, etc.

Using Eq.~\ref{eq:nn_correlator}  and \ref{eq:nnn_correlator}, we get 
\beq
\label{eq:nnn_correlation_funct}
\zeta_{\rm d}(\mb{x}_1,\mb{x}_2, \mb{x}_3) =  \frac{ 1 }{ \bar{n}^2 } \Ddel( \mb{x}_1 - \mb{x}_2  ) \Ddel( \mb{x}_1 - \mb{x}_3 ) + \Big[   \frac{\Ddel( \mb{x}_2 - \mb{x}_3 )  }{ \bar{n}  } \xi_{12} + 2 \, \mathrm{cyc.}     \Big] + \zeta_{123}.  
\eeq
In Fourier space we get the discrete bispectrum \cite{Matarrese:1997sk}
\beq
\label{eq:Bk_discrete}
B_{\rm d}( k_1, k_2, k_3) = \frac{1}{  \tilde{n}^2 }  + \frac{1}{ \tilde{n} } [ P(k_1) + 2 \, \mathrm{cyc.}  ]  + B( k_1, k_2, k_3),    
\eeq
with  $B$ being the continuous bispectrum.

The discrete 4-point correlation function is given by
\beqa
\eta_{\rm d} (\mb{x}_1,\mb{x}_2,\mb{x}_3,\mb{x}_4 )& =&  \frac{1}{\bar{n}^4 } \langle n(\mb{x}_1)n(\mb{x}_2) n(\mb{x}_3) n(\mb{x}_4 )  \rangle  - \Big[ \frac{1}{\bar{n}^3 } \langle n(\mb{x}_1)n(\mb{x}_2) n(\mb{x}_3)  \rangle  + 3 \, \mathrm{cyc.}  \Big]  \nn \\
&+& \Big[   \frac{1}{\bar{n}^2} \langle n(\mb{x}_1)n(\mb{x}_2)   \rangle  + 5 \, \mathrm{cyc.}  \Big]  - 3. 
\eeqa
The 4-point function of $n$ reads
\beqa
\label{eq:nnnn_correlator}
&& \langle  n(\mb{x}_1)n(\mb{x}_2) n(\mb{x}_3) n(\mb{x}_4 )  \rangle  \nn \\
& = & \bar{n} \Ddel( \mb{x}_1 - \mb{x}_2 )  \Ddel( \mb{x}_1 - \mb{x}_3 ) \Ddel( \mb{x}_1 - \mb{x}_4 )  +   [ \Ddel( \mb{x}_2 - \mb{x}_3 ) \Ddel( \mb{x}_2 - \mb{x}_4 ) \bar{n}^2 ( 1 + \xi_{ 12} ) +  3 \, \mathrm{cyc.} ]  \nn \\
&+ &   [ \Ddel( \mb{x}_1 - \mb{x}_2 ) \Ddel( \mb{x}_3 - \mb{x}_4 ) \bar{n}^2 ( 1 + \xi_{ 13} ) +  2  \, \mathrm{cyc.}  ]    + [ \Ddel( \mb{x}_1 - \mb{x}_2 ) \bar{n}^3 ( 1 + \xi_{23} + \xi_{24} + \xi_{34} + \zeta_{234} ) + 5 \, \mathrm{cyc.}] \nn \\
& + & \bar{n}^4 ( 1 + \xi_{12}  + \xi_{13} + \xi_{14} + \xi_{23}  + \xi_{24} + \xi_{34} + \zeta_{123} +  + \zeta_{124}  + \zeta_{134} +  + \zeta_{234} +  + \eta_{1234}  ),
\eeqa
where $ \eta $ is the continuous 4-point function.  

Using Eq.~\ref{eq:nn_correlator}, \ref{eq:nnn_correlator}, and \ref{eq:nnnn_correlator},   we get 
\beqa
\label{eq:nnnn_correlation_funct}
\eta_{\rm d} (\mb{x}_1, \mb{x}_2, \mb{x}_3, \mb{x}_4 )& =& \frac{1 }{\bar{n}^3 }  \Ddel( \mb{x}_1 - \mb{x}_2 ) \Ddel( \mb{x}_1 - \mb{x}_3 ) \Ddel( \mb{x}_1 - \mb{x}_4 ) +  [ \frac{1}{ \bar{n}^2 }   \Ddel( \mb{x}_2 - \mb{x}_3 ) \Ddel( \mb{x}_2 - \mb{x}_4 ) \xi_{12}  + 3 \, \mathrm{cyc.}] \nn \\
&+& \Big[  \frac{ 1 + \xi_{13} }{ \bar{n}^2 } \Ddel( \mb{x}_1 -\mb{x}_2  ) \Ddel( \mb{x}_3 -\mb{x}_4  )  + 2 \, \mathrm{cyc.}  \Big]
+  \Big[ \frac{ \Ddel( \mb{x}_1 - \mb{x}_2  )  }{ \bar{n} } ( \xi_{34} + \zeta_{234}  ) + 5\, \mathrm{cyc.} \Big]  + \eta_{1234} . 
\eeqa

In Fourier space, the 4-point correlator 
\beqa
\label{eq:T_connected_disconnectd}
\mathcal{T}_{\rm d} (k_1,k_2,k_3,k_4)  &=& \frac{ 1 }{ \tilde{n}^3 } + \frac{ 1 }{ \tilde{n}^2 } ( P(k_1) + 3 \, \mathrm{cyc.}  ) + \frac{1}{ \tilde{n}^2  } [  \Ddel( \mb{k}_{12}  ) + P( k_{12} ) +  \Ddel( \mb{k}_{13}  ) + P( k_{13} )  +   \Ddel( \mb{k}_{14}  ) + P( k_{14} )  ]  \nn \\
&+& \Big\{ \frac{1 }{ \tilde{n} } [ \Ddel( \mb{k}_{12} ) P(k_3)  + B(k_{12}, k_3, k_4 ) ] + 5 \, \mathrm{cyc.} \Big\}  +  \mathcal{T}( k_1,k_2, k_3,k_4 ),  
\eeqa
where $ \mathcal{T} $ is the continuous 4-point correlator in Fourier space. Note that $\mathcal{T}$ is not the trispectrum as it usually refers to the connected part of the 4-point function only, while  $\mathcal{T}$  contains the disconnected part as well.  Ref.~\cite{Matarrese:1997sk} wrote down the connected trispectrum, which corresponds to the terms without $\Ddel $ in Eq.~\ref{eq:T_connected_disconnectd}.

Recall that the covariance of the power spectrum is given by 
\beq
\label{eq:CP_covmat}
C^P(k,k') = k_{\rm F}^6   \int_{k} \frac{ d^3 p }{ V_{\rm s} (k) }  \int_{k'} \frac{ d^3 p' }{ V_{\rm s} (k') } \Big[   \langle \delta_{\rm d} ( \mb{p} ) \delta_{\rm d} ( - \mb{p} )    \delta_{\rm d} ( \mb{p}' ) \delta_{\rm d} ( - \mb{p}' )  \rangle  -   \langle \delta_{\rm d} ( \mb{p} ) \delta_{\rm d} ( - \mb{p} ) \rangle \langle   \delta_{\rm d} ( \mb{p}' ) \delta_{\rm d} ( - \mb{p}' )  \rangle       \Big].
\eeq

The covariance operator in Eq.~\ref{eq:CP_covmat} is now given by 
\beqa
\label{eq:cov_mat_operator}
&&   \langle \delta_{\rm d} ( \mb{p} ) \delta_{\rm d} ( - \mb{p} )    \delta_{\rm d} ( \mb{p}' ) \delta_{\rm d} ( - \mb{p}' )  \rangle  -   \langle \delta_{\rm d} ( \mb{p} ) \delta_{\rm d}( - \mb{p} ) \rangle \langle   \delta_{\rm d}( \mb{p}' ) \delta_{\rm d}( - \mb{p}' )  \rangle  \nn \\
&= & \frac{1}{k_{\rm F}^3 }  \bigg\{  \frac{1  }{ \tilde{n}^2 }[ \Ddel( \mb{p} + \mb{p}' ) +  \Ddel( \mb{p} - \mb{p}' )] +  \frac{2  }{ \tilde{n} }  [ \Ddel( \mb{p} + \mb{p}' ) + \Ddel( \mb{p} - \mb{p}' )    ] P(p)   \nn \\
&+&  \frac{1}{\tilde{n}^3 } +   \frac{1}{\tilde{n}^2 } (2 P(p) +2P(p') + P(|\mb{p} + \mb{p}'|) +  P(|\mb{p} - \mb{p}'|)  )   \nn \\
 & + &       \frac{1}{ \tilde{n} } [ 2B(| \mb{p} + \mb{p}'  |,p,p' ) +  2 B(| \mb{p} - \mb{p}'|,p,p' ) + B(0,p,p ) + B(0,p',p' )  ] \nn\\
&+&  \mathcal{ T} ( \mb{p} , - \mb{p}, \mb{p}' , - \mb{p}' ) - P(p) P(p') \bigg\} .         
\eeqa
The first line of  the RHS of Eq.~\ref{eq:cov_mat_operator} are the Gaussian terms.  Although we use the terminology ``Gaussian'' here, these terms are not related to the Gaussian distribution. In fact in the Poisson model, the discrete particles are Poisson distributed. They are called Gaussian because they  contribute only to the diagonal covariance as the smooth Gaussian terms. The second and third lines are the non-Gaussian terms and they can couple different bins. The last line is the continuous part of the 4-point function.

We can easily integrate over the Gaussian terms in the first line to get 
\beq
 \frac{ 2 k_{\rm F}^3  }{ V_{\rm s} (k) } \delta_{k,k'} \Big[  \frac{2 P(k) }{\tilde{n} } + \frac{  1} {  \tilde{n}^2  }  \Big] . 
\eeq
This term can be combined with the Gaussian contribution from the continuous part as
\beq
\label{eq:CG_P_noise}
C_{\rm G }^P (k,k')= \frac{ 2 k_{\rm F}^3  }{ V_{\rm s} (k) } \delta_{k,k'} \Big[ P(k) + \frac{  1} {  \tilde{n} }  \Big]^2 . 
\eeq
Eq.~\ref{eq:CG_P_noise} agrees with Ref.\cite{FeldmanKaiserPeacock1994}. In other words, the Gaussian covariance of the power spectrum for the halo is the same as the case for dark matter except with the dark matter power spectrum replaced by the halo one plus the shot noise term.

The non-Gaussian contribution due to the Poisson shot noise is given by 
\beqa
\label{eq:CNG_Poisson}
C_{\rm NG}^P (k,k')  &=&  k_{\rm F}^3  \Big[ \frac{1}{ \tilde{n}^3 }  + \frac{2}{ \tilde{n}^2 } ( P(k) + P(k') )  +\frac{ 2}{\tilde{n}^2 }  \int_k \frac{ d^3 p }{ V_{\rm s}(k) }   \int_{k'} \frac{ d^3 p' }{ V_{\rm s}(k') } P( |\mb{p} +  \mb{p}' |)  \Big]  \nn \\
&+&   \int_k \frac{ d^3 p }{ V_{\rm s}(k) }   \int_{k'} \frac{ d^3 p' }  { V_{\rm s}(k') } \Big\{ \frac{ 1}{\tilde{n} }  [ 4B(| \mb{p} + \mb{p}'  |,p,p' )  +  B(0,p,p ) + B(0,p',p' ) ] \nn \\
&+&  \mathcal{T} ( \mb{p} , - \mb{p}, \mb{p}' , - \mb{p}' ) - P(p) P(p') \Big\}.
 \eeqa
Ref.~\cite{Smith:2008ut} also derived the shot noise contribution to the power spectrum covariance.  Comparing Eq.~\ref{eq:CNG_Poisson} to the result in Ref.~\cite{Smith:2008ut}, besides the minor difference that we have assumed small binning width to simplify the expressions, we note that the terms  $ B(0,p,p ) + B(0,p',p' )$ are missing in \cite{Smith:2008ut}. These terms do not vanish in general.  Suppose the tree level halo bispectrum is used for $ B(0,p,p )$, although the local $b_1$-term and the nonlocal bias term \cite{ChanScoccimarroSheth2012} both vanish because they are generated by large-scale gravitational evolution, the local nonlinear bias term gives finite contribution $b_1^2 b_2 P^2(p) $.

\subsection{Poisson shot noise contribution to the bispectrum covariance }
\label{sec:appendix_PoissonBk}

The complexity of the perturbation series increases rapidly when the number of points in the $n$-point function increases. We will first summarize a set of diagrammatic rules to represent the $n$-point correlation  function for $n=2$, 3, and 4. Apart from the Dirac delta function, the rules are similar to the continuous case.   In real space, we can represent the Dirac delta function $\Ddel( \mb{x}_i -  \mb{x}_j )  $ by a new link between the  points  $  \mb{x}_i $ and $  \mb{x}_j $. This link is analogous to the continuous correlation function. We can further simplify the diagram by shrinking all the points linked by the Dirac delta functions to a dot. Graphical representations of the 2-point, 3-point, and 4-point correlation functions for  Eq.~\ref{eq:nn_correlation_funct}, \ref{eq:nnn_correlation_funct}, and \ref{eq:nnnn_correlation_funct} are shown  in Fig.~\ref{fig:Poisson_2_3_4_pt}.  For example the first two diagrams in Fig.~\ref{fig:Poisson_2_3_4_pt} denote the two terms in Eq.~\ref{eq:nn_correlation_funct}. The first circle-dot represents the two points connected by a $\Ddel$;  the wavy line in the second diagram denotes the continuous correlation function $\xi$.  In  the second line, the diagrams represent the three terms in Eq.~\ref{eq:nnn_correlation_funct}. The first diagram denotes the three points connected by  Dirac delta functions. In the second diagram, the circle-dot represents the two points connected by the Dirac delta function and  they are connected to the third point by a correlation function. The last one represents the three points connected by the continuous 3-point function. By comparing Eq.~\ref{eq:nnnn_correlation_funct} with the diagrams for the 4-point function  in Fig.~\ref{fig:Poisson_2_3_4_pt}, it is clear that similar rules apply. The terms that contribute to the Gaussian covariance of the power spectrum are the two disconnected diagrams in the 4-point function.  Among the non-Gaussian terms, the one as a circle-dot corresponds to $1/\tilde{n}^3 $, and the second and fourth diagrams in the third row of Fig.~\ref{fig:Poisson_2_3_4_pt} represent the terms with power spectrum in the first line of Eq.~\ref{eq:CNG_Poisson}.

We now apply the rules to the 6-point function and the results are shown in Fig.~\ref{fig:Poisson_shot_6_pt}.  They are arranged based on the number of Dirac delta functions, ranging from 5 to 0.  From these diagrammatic representations, it is straightforward to write down the discrete 6-point function $\sigma_{\rm d}  $
\begin{align}
\label{eq:nnnnnn_correlation_funct}
\sigma_{\rm d }( \mb{x}_1,\mb{x}_2,\mb{x}_3,\mb{x}_4,\mb{x}_5,\mb{x}_6 ) & = \frac{1  }{ \bar{n}^5} \Ddel( \mb{x}_1 - \mb{x}_2 )   \Ddel( \mb{x}_1 - \mb{x}_3 ) \Ddel( \mb{x}_1 - \mb{x}_4 ) \Ddel( \mb{x}_1 - \mb{x}_5 )  \Ddel( \mb{x}_1 - \mb{x}_6 ) \nn  \\ 
&+  \frac{ 1 }{ \bar{n}^4  }     \Ddel( \mb{x}_1 - \mb{x}_2 )   \Ddel( \mb{x}_1 - \mb{x}_3 ) \Ddel( \mb{x}_1 - \mb{x}_4 ) \Ddel( \mb{x}_1 - \mb{x}_5 ) \xi_{56} + 5 \cyc \nn \\
&+ \frac{ 1 }{ \bar{n}^4  }     \Ddel( \mb{x}_1 - \mb{x}_2 )   \Ddel( \mb{x}_1 - \mb{x}_3 ) \Ddel( \mb{x}_4 - \mb{x}_5 ) \Ddel( \mb{x}_4 - \mb{x}_6 ) \xi_{34} + 9 \cyc \nn \\
& + \frac{ 1 }{ \bar{n}^4  }     \Ddel( \mb{x}_1 - \mb{x}_2 )   \Ddel( \mb{x}_1 - \mb{x}_3 ) \Ddel( \mb{x}_1 - \mb{x}_4 ) \Ddel( \mb{x}_5 - \mb{x}_6 ) \xi_{45} + 14 \cyc \nn \\
&+ \frac{ 1 }{ \bar{n}^4  }     \Ddel( \mb{x}_1 - \mb{x}_2 )   \Ddel( \mb{x}_3 - \mb{x}_4 ) \Ddel( \mb{x}_3 - \mb{x}_5 ) \Ddel( \mb{x}_3 - \mb{x}_6 )  + 14 \cyc \nn \\
&+ \frac{ 1 }{ \bar{n}^4  }     \Ddel( \mb{x}_1 - \mb{x}_2 )   \Ddel( \mb{x}_1 - \mb{x}_3 ) \Ddel( \mb{x}_4 - \mb{x}_5 ) \Ddel( \mb{x}_4 - \mb{x}_6 ) + 9 \cyc \nn \\
&+  \frac{1  }{\bar{n}^3} \Ddel( \mb{x}_1 - \mb{x}_2 )   \Ddel( \mb{x}_3 - \mb{x}_4 ) \Ddel( \mb{x}_5 - \mb{x}_6 ) \zeta_{246} + 14 \cyc \nn \\
&+  \frac{1  }{\bar{n}^3} \Ddel( \mb{x}_1 - \mb{x}_2 )  \Ddel( \mb{x}_1 - \mb{x}_3 )  \Ddel( \mb{x}_4 - \mb{x}_5 ) \zeta_{346} + 59 \cyc \nn \\
&+ \frac{1  }{\bar{n}^3} \Ddel( \mb{x}_1 - \mb{x}_2 )   \Ddel( \mb{x}_1 - \mb{x}_3 )    \Ddel( \mb{x}_1 - \mb{x}_4 )    \zeta_{156} + 14 \cyc \nn \\
&+  \frac{1  }{\bar{n}^3} \Ddel( \mb{x}_1 - \mb{x}_2 )   \Ddel( \mb{x}_3 - \mb{x}_4 )    \Ddel( \mb{x}_5 - \mb{x}_6 )  + 14 \cyc \nn \\
&+  \frac{1  }{\bar{n}^3} \Ddel( \mb{x}_1 - \mb{x}_2 )   \Ddel( \mb{x}_3 - \mb{x}_4 )    \Ddel( \mb{x}_5 - \mb{x}_6 )    \xi_{24} + 44 \cyc \nn \\
&+  \frac{1  }{\bar{n}^3} \Ddel( \mb{x}_1 - \mb{x}_2 )   \Ddel( \mb{x}_1 - \mb{x}_3 )    \Ddel( \mb{x}_4 - \mb{x}_5 )    \xi_{16} + 59 \cyc \nn \\
&+  \frac{1  }{\bar{n}^3} \Ddel( \mb{x}_1 - \mb{x}_2 )   \Ddel( \mb{x}_1 - \mb{x}_3 )    \Ddel( \mb{x}_4 - \mb{x}_5 )    \xi_{56} + 59 \cyc \nn \\
&+  \frac{1  }{\bar{n}^3} \Ddel( \mb{x}_1 - \mb{x}_2 )   \Ddel( \mb{x}_1 - \mb{x}_3 )    \Ddel( \mb{x}_1 - \mb{x}_4 )    \xi_{56} + 14 \cyc \nn \\
&+  \frac{1  }{\bar{n}^2} \Ddel( \mb{x}_1 - \mb{x}_2 )   \Ddel( \mb{x}_5 - \mb{x}_6 )  \eta_{2345} + 44 \cyc \nn \\
&+  \frac{1  }{\bar{n}^2} \Ddel( \mb{x}_1 - \mb{x}_2 )   \Ddel( \mb{x}_1 - \mb{x}_3 )  \eta_{1456} + 19  \cyc \nn \\
&+  \frac{1  }{\bar{n}^2} \Ddel( \mb{x}_1 - \mb{x}_2 )   \Ddel( \mb{x}_5 - \mb{x}_6 )  \xi_{34} + 44  \cyc \nn \\
&+  \frac{1  }{\bar{n}^2} \Ddel( \mb{x}_1 - \mb{x}_2 )   \Ddel( \mb{x}_5 - \mb{x}_6 )  \zeta_{234} + 89  \cyc \nn \\
&+  \frac{1  }{\bar{n}^2} \Ddel( \mb{x}_1 - \mb{x}_2 )   \Ddel( \mb{x}_5 - \mb{x}_6 )  \xi_{25}   \xi_{34} + 44  \cyc \nn \\
&+  \frac{1  }{\bar{n}^2} \Ddel( \mb{x}_1 - \mb{x}_2 )   \Ddel( \mb{x}_5 - \mb{x}_6 )  \xi_{23}   \xi_{45} + 89  \cyc \nn \\
&+  \frac{1  }{\bar{n}^2} \Ddel( \mb{x}_1 - \mb{x}_2 )   \Ddel( \mb{x}_1 - \mb{x}_3 )  \zeta_{456}  + 19  \cyc \nn \\
&+  \frac{1  }{\bar{n}^2} \Ddel( \mb{x}_1 - \mb{x}_2 )   \Ddel( \mb{x}_1 - \mb{x}_3 )  \xi_{34}   \xi_{56}  + 59  \cyc \nn \\
&+  \frac{1  }{\bar{n} } \Ddel( \mb{x}_1 - \mb{x}_2 )   \chi_{23456}  + 14  \cyc \nn \\
&+  \frac{1  }{\bar{n} } \Ddel( \mb{x}_1 - \mb{x}_2 )   \xi_{34}   \xi_{56}  +   44  \cyc \nn \\
&+  \frac{1  }{\bar{n} } \Ddel( \mb{x}_1 - \mb{x}_2 )   \xi_{26}   \zeta_{345}  +  59  \cyc \nn \\
&+  \frac{1  }{\bar{n} } \Ddel( \mb{x}_1 - \mb{x}_2 )  \eta_{3456}  +  14  \cyc \nn \\
&+  \frac{1  }{\bar{n} } \Ddel( \mb{x}_1 - \mb{x}_2 )  \zeta_{234} \xi_{56} +  89 \cyc \nn \\
&+  \sigma_{123456} ,
\end{align}
where $\chi  $ and $\sigma$ are the continuous 5-point and 6-point functions. Note that some of the diagrams correspond to more than one term in Eq.~\ref{eq:nnnnnn_correlation_funct}. 

Then in Fourier space, the 6-point function reads
\begin{align}
\label{eq:Y_d}
& \quad  \mathcal{Y}_{\rm d}(k_1,k_2,k_3,k_4,k_5,k_6) \nn \\
&=  \frac{1}{ \tilde{n}^5 }  +  [  \frac{1}{ \tilde{n}^4 } P(k_6 ) + 5 \cyc   ] +    [  \frac{1}{ \tilde{n}^4 } P(k_{123} ) + 9  \cyc   ]  \nn  \\
& +   [  \frac{1}{ \tilde{n}^4 } P(k_{56} ) + 14  \cyc   ]   + \textcolor{green}{   [  \frac{1}{ \tilde{n}^4 } \Ddel(\mb{k}_{56} ) + 14  \cyc   ]  }
 + \textcolor{green}{  [  \frac{1}{ \tilde{n}^4 } \Ddel(\mb{k}_{123} ) + 9  \cyc   ] }  \nn \\
&+  [  \frac{1}{ \tilde{n}^3 } B( k_{12}, k_{34}, k_{56} ) + 14  \cyc   ]   +   [  \frac{1}{ \tilde{n}^3 } B( k_{123}, k_{45}, k_{6} ) + 59  \cyc   ]  +   [  \frac{1}{ \tilde{n}^3 } B( k_{1234}, k_{5}, k_{6} ) + 14  \cyc   ]  + \nn \\
&+\textcolor{red}{ [  \frac{1}{ \tilde{n}^3 } \Ddel(\mb{k}_{12} ) \Ddel(\mb{k}_{34} )   +  14  \cyc   ] }   + 
\textcolor{green}{  [  \frac{1}{ \tilde{n}^3 } \Ddel(\mb{k}_{56} )  P(k_{12} )   +  44  \cyc   ] }
+ \textcolor{green}{  [  \frac{1}{ \tilde{n}^3 } \Ddel(\mb{k}_{45} )  P(k_{6} )   +  59  \cyc   ] } \nn \\
&+ \textcolor{green}{  [  \frac{1}{ \tilde{n}^3 } \Ddel(\mb{k}_{123} )  P(k_{6} )   +  59  \cyc   ] }
 +  \textcolor{green}{ [  \frac{1}{ \tilde{n}^3 } \Ddel(\mb{k}_{56} )  P(k_{6} )   +  14  \cyc   ] } \nn \\
 & +  [  \frac{1}{ \tilde{n}^2 } \mathcal{T}( k_{12}, k_3, k_4, k_{56} )  + 44  \cyc   ]  + 
 [  \frac{1}{ \tilde{n}^2 } \mathcal{T}( k_{123}, k_4, k_5, k_{6} )  + 19  \cyc   ]   \nn \\
 &+ \textcolor{red}{[  \frac{1}{ \tilde{n}^2 } \Ddel(\mb{k}_{12}) \Ddel(\mb{k}_{34}) P(k_3)   + 44  \cyc   ]  }
+ \textcolor{green}{   [  \frac{1}{ \tilde{n}^2 } \Ddel(\mb{k}_{56}) B( k_{12}, k_3, k_4 )  +  89  \cyc   ]  }  \nn \\  
&+   \textcolor{green}{  [  \frac{1}{ \tilde{n}^2 } \Ddel(\mb{k}_{34}) P( k_{12} ) P(k_3)  +  44  \cyc   ] }
 +  \textcolor{green}{    [  \frac{1}{ \tilde{n}^2 } \Ddel(\mb{k}_{123}) P( k_{3} ) P(k_4)  +  89  \cyc   ]  }  \nn \\
&+  \textcolor{green}{   [  \frac{1}{ \tilde{n}^2 }  \Ddel(\mb{k}_{123})    B( k_{4} ,k_5, k_6 )  +  19  \cyc   ]  } 
+  \textcolor{green}{ [  \frac{1}{ \tilde{n}^2 }  \Ddel(\mb{k}_{56})    P( k_{4}) P(k_6 )  +  59  \cyc   ] }  \nn \\
&+ [ \frac{1 }{ \tilde{n} } \mathcal{X}( k_{12}, k_3, k_4, k_5, k_6  ) +   14  \cyc  ]    
+ \textcolor{red}{ [  \frac{1 }{ \tilde{n} }  \Ddel(\mb{k}_{12})   \Ddel(\mb{k}_{34})   P(k_3)   P(k_5)  +   44  \cyc ] }  \nn \\ 
&+    \textcolor{green}{    [  \frac{1 }{ \tilde{n} }  \Ddel(\mb{k}_{126})   P(k_6)   B(k_3, k_4,k_5)  +   59  \cyc ]  }
 +    \textcolor{green}{   [ \frac{1 }{ \tilde{n} }  \Ddel(\mb{k}_{12})   \mathcal{T}(k_3, k_4,k_5, k_6)  +   14  \cyc ]  } \nn \\ 
&+   \textcolor{green}{    [\frac{1 }{ \tilde{n} }  \Ddel(\mb{k}_{56})   B(k_{12}, k_3, k_4 ) P(k_5)  + 89 \cyc ] }  +  \mathcal{Y}_{123456} , 
\end{align}
where $\mathcal{X}$ and $\mathcal{ Y}$ are the Fourier transform of the continuous  5-point, and 6-point functions.   In Ref.~\cite{Matarrese:1997sk}, the 6-point function including the Poisson shot noise, but limited to connected terms only, was written down. They agree with the terms without $ \Ddel $ in the first three lines of Eq.~\ref{eq:Y_d}.

%\end{widetext}

We are going to classify the terms in $\mathcal{Y}_{\rm d}$  based on the number of Dirac delta functions. This is directly related to the correlator expansion we mentioned in Sec.~\ref{sec:Bcov_DMtheory}.   The terms with two Dirac delta functions in Eq.~\ref{eq:Y_d} (highlighted in red) are the Gaussian terms, and they are in the $PPP$ group. Diagrammatically,  they are represented by the disconnected diagrams with three disconnected parts, i.e.~the most disconnected diagrams in Fig.~\ref{fig:Poisson_shot_6_pt}. The terms with one Dirac delta function in  Eq.~\ref{eq:Y_d} (highlighted in green) belong to either the $PT$ or $BB$ group. They are the diagrams with two disconnected parts in Fig.~\ref{fig:Poisson_shot_6_pt}. Finally, the connected 6-point function contribution in Eq.~\ref{eq:Y_d} (in plain black) is represented by the connected diagrams  in Fig.~\ref{fig:Poisson_shot_6_pt}. In each group of terms, we can regard it as an expansion in  $ 1 / \tilde{n} $. Depending on the relative importance of  $ 1 / \tilde{n} $ and $P$, we can retain the terms with high or low  power of $ 1 / \tilde{n} $.   As  $\mathcal{Y}_{\rm d}$ contains connected 2, 3, 4, 5, and 6-point functions in general, it is formidable to evaluate in  the most general case. To proceed, in this paper, we will only explicitly evaluate the terms containing the continuous power spectrum or bispectrum.

Using the shorthand notation in Eq.~\ref{eq:int_D_notation}, the shot noise contribution to the covariance can be written as 
\beq
C  = \int \mathcal{D}  \, \mathrm{cov} \mathcal{Y}_{\rm d} (\mb{p},\mb{q},\mb{r},\mb{p}',\mb{q}',\mb{r}' ),  
\eeq
where $\mathrm{cov} \mathcal{Y}_{\rm d} $ is defined as 
\beqa
 \mathrm{cov} \mathcal{Y}_{\rm d}  (\mb{p},\mb{q},\mb{r},\mb{p}',\mb{q}',\mb{r}' )  
=      \mathcal{Y}_{\rm d}(\mb{p},\mb{q},\mb{r},\mb{p}',\mb{q}',\mb{r}') -\frac{1}{k_{\rm F}^3 } B_{\rm d}( \mb{p},\mb{q},\mb{r} )   B_{\rm d}( \mb{p}',\mb{q}',\mb{r}' )  . 
\eeqa

First the terms with two Dirac delta functions in Eq.~\ref{eq:Y_d} are the Gaussian terms.  They are non-vanishing only if triangle  $k_1k_2k_3$ is the same as $k_1'k_2'k_3' $. There are altogether three such terms in $ \mathcal{Y}_{\rm d} (\mb{p},\mb{q},\mb{r},\mb{p}',\mb{q}',\mb{r}')$,  and they are highlighted in red in Eq.~\ref{eq:Y_d}.   However, because of the triangle constraint, all the Dirac delta functions in these three terms must couple one of the vectors in $ k_1k_2k_3 $ with another one in $ k_1'k_2'k_3' $ to give non-vanishing contribution to the covariance. This leaves us with 
\beq
 \Big\{ \frac{1}{ \tilde{n}^3 } +  \frac{1}{ \tilde{n}^2 } [ P(p) + P(q) + P(r) ] 
+   \frac{1}{ \tilde{n} } [ P(p)P(q) + P(p)P(r) +  P(q)P(r)]      \Big\} 
  [ \Ddel(\mb{p} + \mb{p}' )\Ddel(\mb{q} + \mb{q}' ) \Ddel(\mb{r} + \mb{r}' )   +   5 \cyc ].  
\eeq
This term can be combined with the continuous Gaussian terms  $[ \Ddel(\mb{p} + \mb{p}' )\Ddel(\mb{q} + \mb{q}' ) \Ddel(\mb{r} + \mb{r}' )  +  5 \cyc  ] P(p) P(q) P(r) $.  Therefore, in the presence of Poisson shot noise,  Eq.~\ref{eq:BCov_Gaussian} is modified to 
\beq
\label{eq:C_G_bisp_shot}
 C_{\rm G}  =  \frac{ k_{\rm F}^3 }{V_{123} }  \delta_{k_1k_2k_3, k_1'k_2'k_3'  } s_{123} 
 \Big[  P(k_1) + \frac{1 }{ \tilde{n} } \Big]  \Big[ P(k_2) + \frac{1 }{ \tilde{n} } \Big]  \Big[  P(k_3) + \frac{1 }{ \tilde{n} } \Big]. 
\eeq
Eq.~\ref{eq:C_G_bisp_shot} agrees with the results in \cite{SCFFHM98}. Clearly, these terms would be classified as $PPP$ in the correlator expansion we mentioned in Sec.~\ref{sec:Bcov_DMtheory}.  Again, similar to the case of power spectrum, the Gaussian covariance of the halo bispectrum including the shot noise contribution can be obtained by replacing the continuous power spectrum with the halo power spectrum plus the shot noise contribution.  Similar to the comments for the Gaussian power spectrum covariance, they are called ``Gaussian'' here simply because they are on the same footing as the ``true'' Gaussian terms, not because  they arise from the Gaussian distribution.

We now look at the terms with one Dirac delta function in $\mathcal{Y}_{\rm d}$. There are altogether 14 such terms, and they are highlighted in green in  Eq.~\ref{eq:Y_d}.  Some of them can be computed analytically making use of Eq.~\ref{eq:Uintegral}. Among these terms, there are terms with Dirac delta function connecting three vectors, $\Ddel( \mb{p}_{ijk} )$. When three of the vectors are from the same bispectrum estimator, they are exactly cancelled by the corresponding terms in $\langle B\rangle  \langle B ' \rangle $. The net results due to  the terms with one Dirac delta function in  $\mathrm{cov} \mathcal{Y}_{\rm d} $ are
%% \beqa
%% \label{eq:C_NG_1_d}
%% C_{ \rm NG }^{(1)} & =& k_{\rm F}^3  \frac{  U(k_1, k_1') }{V_{123} V_{123}' } \Big\{  \frac{2 }{ \tilde{n}^4 } + \frac{ 1 }{  \tilde{n}^3 } [ P(k_2) + P(k_3) +P(k_2') + P(k_3')  ]  \nn \\
%% &+& \frac{ 1 }{  \tilde{n}^3 }[ P(k_1) +   P(k_2) + P(k_3) + P(k_1') + P(k_2') + P(k_3') ] + \frac{1  }{ \tilde{n}^3 } P(k_1) \nn \\
%% &+&  \frac{ 1 }{  \tilde{n}^2 }[ P(k_1) +   P(k_2') + P(k_3')][ P(k_1') + P(k_2) + P(k_3) ]   \nn \\
%% &+&   \frac{ 1 }{  \tilde{n}^2 } P(k_1) [ P(k_2) + P(k_3) +  P(k_2') + P(k_3')    ]      \Big\}  + 8 \cyc \nn \\
%% &+& \frac{ 1 }{\tilde{n}^2} \int \mathcal{D} \Ddel( \mb{p} +  \mb{p}' ) ( P(p) + \frac{1}{ \tilde{n} }   ) \Big[ P(| \mb{q} + \mb{r}| ) +     P(| \mb{q} + \mb{q}'| ) +  P(| \mb{q} + \mb{r}'| )        \Big]  + 8 \cyc
%% \eeqa
\beqa
\label{eq:C_NG_1_d}
C_{ \rm NG }^{(1)} & =& k_{\rm F}^3  \frac{  U(k_1, k_1') }{V_{123} V_{123}' } \Big\{  \frac{2 }{ \tilde{n}^4 } + \frac{ 1 }{  \tilde{n}^3 } [ P(k_2) + P(k_3) +P(k_2') + P(k_3')  ]  \nn \\
&+& \frac{ 1 }{  \tilde{n}^3 }[ P(k_1) +   P(k_2) + P(k_3) + P(k_1') + P(k_2') + P(k_3') ] + \frac{1  }{ \tilde{n}^3 } P(k_1) \nn \\
&+&  \frac{ 1 }{  \tilde{n}^2 }[ P(k_1) +   P(k_2') + P(k_3')][ P(k_1') + P(k_2) + P(k_3) ]   \nn \\
&+&   \frac{ 1 }{  \tilde{n}^2 } P(k_1) [ P(k_2) + P(k_3) +  P(k_2') + P(k_3')    ]      \Big\}  + 8 \cyc \nn \\
&+& \frac{ 1 }{\tilde{n}^2} \int \mathcal{D} \Ddel( \mb{p} +  \mb{p}' ) ( P(p) + \frac{1}{ \tilde{n} }   ) \Big[ P(| \mb{q} + \mb{r}| ) +     P(| \mb{q} + \mb{q}'| ) +  P(| \mb{q} + \mb{r}'| )        \Big]  + 8 \cyc  \nn \\           %%%%%%%%%%%%%%%%%%%%%%%%%%%%%%%%%%%%%%%%%%%%%%%%
&+&  \frac{ k_{\rm F}^3  U(k_1, k_1' ) }{ V V' } \Big \{  \frac{1}{ \tilde{n}^2 } \big[  B(k_1, k_2',k_3') +  B(k_1', k_2,k_3)  \big] + 8 \cyc \nn \\
&+& \frac{1}{ \tilde{n} } \big[\big( P(k_1') +  P(k_2) + P(k_3) \big) B( k_1, k_2',k_3' ) + \big( P(k_1) +  P(k_2') + P(k_3') \big) B( k_1', k_2 , k_3 )  \big] + 8 \cyc \Big\}  \nn \\
&+& \int \mathcal{D} \frac{1}{ \tilde{n}^2 } \Ddel( \mb{p} + \mb{p}' )  ]  \Big[  B(| \mb{q} + \mb{r} |, q' ,r' ) +  B(| \mb{q} + \mb{q}' |, r ,r' )      +  B(| \mb{q} + \mb{r}' |, r ,q' )    \nn \\  
& +&  B(| \mb{r} + \mb{q}' |, q ,r' ) +  B(| \mb{r} + \mb{r}' |, q ,q' )  +  B(| \mb{q}' + \mb{r}' |, q , r )  \Big]   + 8 \cyc   \nn \\ 
&+&  \int \mathcal{D} \frac{1}{ \tilde{n}  } \Ddel( \mb{p} + \mb{p}' )   P(p) \Big[  B(| \mb{q} + \mb{r} |, q' ,r' ) +  B(| \mb{q} + \mb{q}' |, r ,r' )  +    B(| \mb{q} + \mb{r}' |, r ,q' ) \nn \\
  & +&   B(| \mb{r} + \mb{q}' |, q ,r' )  +   B(| \mb{r} + \mb{r}' |, q ,q' )  +   B(| \mb{q}' + \mb{r}' |, q ,r )      \Big]    + 8 \cyc  \nn \\
&+& \dots, 
\eeqa
where the dots denote the term with the continuous 4-point function.

The terms without any Dirac delta function are the connected 6-point function and they are represented 
 by the connected diagrams in Fig.~\ref{fig:Poisson_shot_6_pt}.   These terms read
\beqa
\label{eq:C_NG_0_d}
C_{ \rm NG }^{ (0)}  &=&  \frac{ k_{\rm F}^3 }{\tilde{n}^5 } + \frac{ k_{\rm F}^3  }{\tilde{n}^4 } \big[ P(k_1) + 5 \cyc \big]  + \int \mathcal{D} \frac{1}{ \tilde{n}^4 }\Big\{   \big[P(| \mb{p} + \mb{p}'|) + 8 \cyc \big] + \big[ P(| \mb{p} + \mb{q}|) + 14 \cyc \big] \Big\}  \nn \\
&+& \int  \mathcal{D} \frac{1}{ \tilde{n}^3  }   B( | \mb{p} + \mb{q} |, | \mb{r} + \mb{p}' |, | \mb{q}' + \mb{r}' | )
 +  14 \cyc \nn \\
&+&  \int  \mathcal{D} \frac{1}{ \tilde{n}^3  }   B( | \mb{p} + \mb{q} + \mb{r} |, | \mb{p}' + \mb{q}' |, r' )
 +  59 \cyc \nn \\
 &+&  \int  \mathcal{D} \frac{1}{ \tilde{n}^3  }   B( | \mb{p} + \mb{q} + \mb{r} + \mb{p}'    |, q', r' ) + 14 \cyc    \nn \\
 &+& \dots, 
\eeqa
where the dots denotes the terms involving higher order connected correlators. 

\end{widetext}

\begin{figure}[!htb]
\centering
\includegraphics[width=0.9\linewidth]{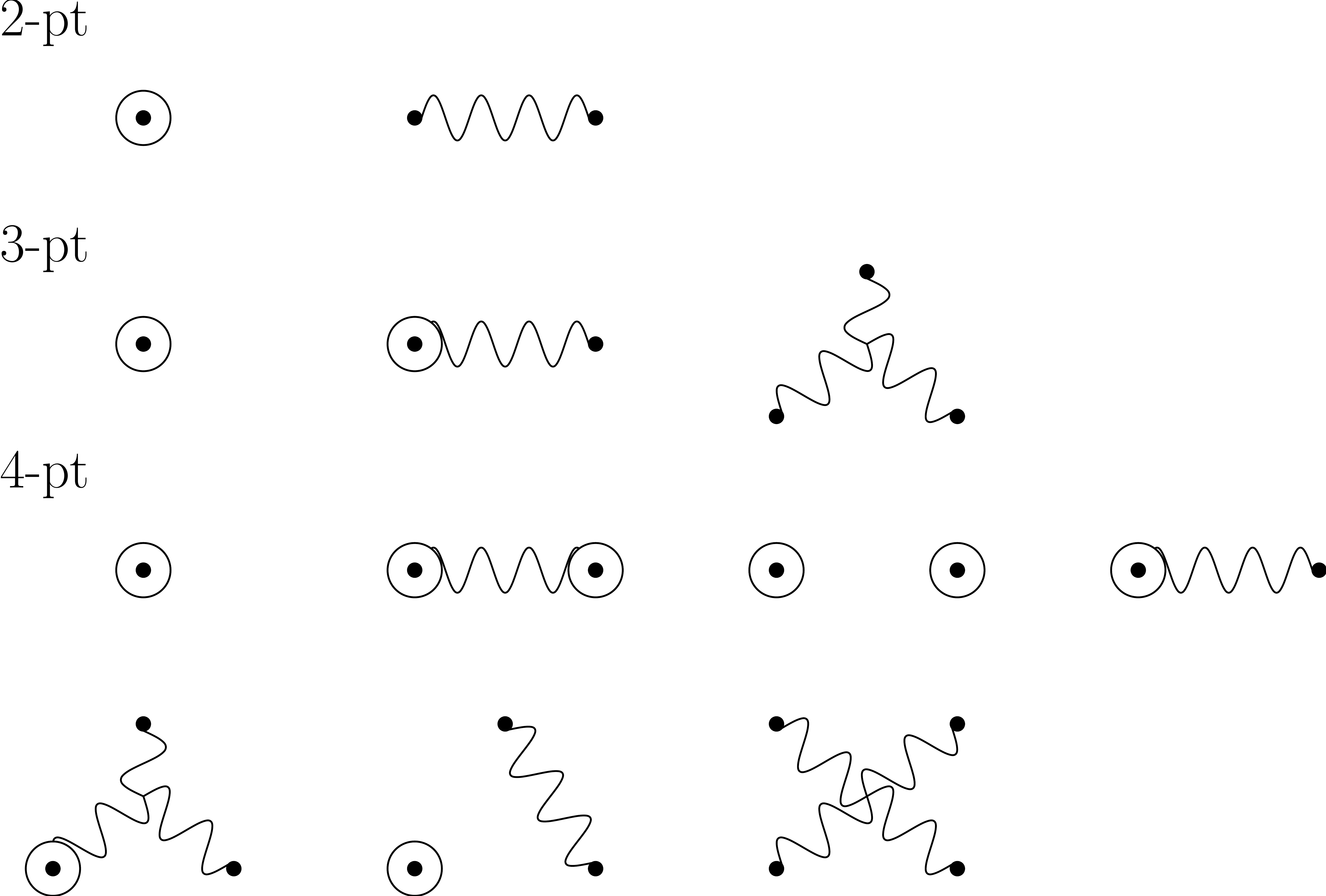}
\caption{ Diagrammatic representation of the 2-point, 3-point and 4-point correlation including the shot noise contributions.  }
\label{fig:Poisson_2_3_4_pt}
\end{figure}

\begin{figure}[!htb]
\centering
\includegraphics[width=0.9\linewidth]{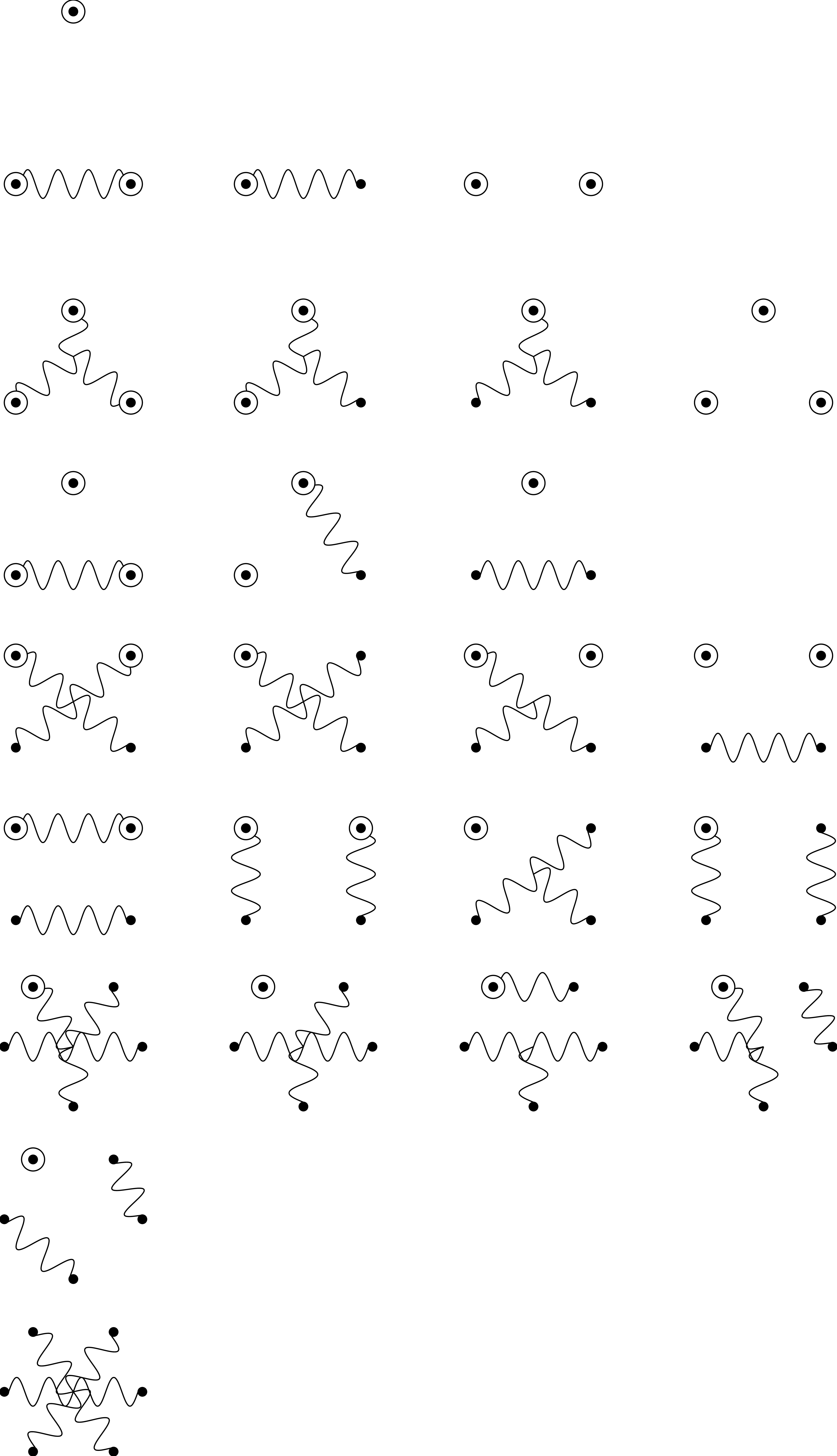}
\caption{ Diagrammatic representation of the 6-point correlation including the shot noise contributions.  }
\label{fig:Poisson_shot_6_pt}
\end{figure}

\section{ The independence of the signal-to-noise ratio on the binning  }
\label{sec:BinningDependence}
In this section, we discuss the possible dependence of the signal-to-noise ratio, S/N, on the binning width $\Delta k  $. We first consider the case of power spectrum and then move to the bispectrum. Of course small inaccuracies arise when a coarse binning is used as the field varies across the bin.  This is not the case we discuss here, instead we investigate whether the S/N explicitly depends on the binning to the lowest order.

For power spectrum, the Gaussian covariance scales with the binning as $ (\Delta k )^{-1} $, while the trispectrum contribution does not depend on $ \Delta k $. The latter case is true for both the matter power spectrum case Eq.~\ref{eq:CP_trisp_general} and the Poisson model result Eq.~\ref{eq:CNG_Poisson}. Suppose we change the binning in $\delta $ from  $\Delta k$ to  $ g \Delta k $. For  illustration purposes, let us take $ g=2$. When the binning is coarse grained by a factor of $g=2$, the coarse-grained data vector $S'$ is related to the original data vector $S$ as
\beq
S'_i = \frac{ S_{2i-1} +  S_{2i}}{ 2 } . 
\eeq
i.e.~an average over the neighbouring bins. For the sake of simplicity here we use a simple average instead of the phase-volume weighted one, but the discussion is still valid when a weighted mean is used.  Correspondingly, the covariance matrix of the coarse-grained data vector, $C'$, is given by
\beqa
\label{eq:CP_averaging}
 C'_{\, ij} &=& \langle S'_i S'_j \rangle  \nn \\
&=& \frac{1 }{4} \big( \langle S_{2i-1}S_{2j-1} \rangle  +  \langle S_{2i-1}S_{2j} \rangle   \nn \\
&& \quad \quad  + \langle S_{2i}S_{2j-1}\rangle + \langle S_{2i}S_{2j} \rangle \big). 
\eeqa 
Therefore the coarse-grained covariance matrix is obtained by locally averaging the square block in the original matrix. In fact, the scalings of the Gaussian and non-Gaussian parts with respect to $\Delta k $ respect the averaging prescription Eq.~\ref{eq:CP_averaging}. Clearly the non-Gaussian matrix elements are invariant with respect to binning up to the accuracy of the field represented by the binned value. When the Gaussian covariance matrix is coarse grained, the diagonal element of the coarse-grained one is obtained by averaging the $g$ diagonal elements and $g(g-1)$ off-diagonal ones, which are zeroes. This is equivalent to simply averaging over the diagonal elements, and we get the  $1/g$ scaling for the diagonal element.

We can expand the inverse of the covariance matrix perturbatively as 
\beqa
\label{eq:Cinv_perturbative}
C^{-1} &=&  ( C_{\rm G} + C_{\rm NG} )^{-1} \nn \\ 
      &=&  ( I + C_{\rm G}^{-1} C_{\rm NG} )^{-1} C_{\rm G}^{-1}   \nn \\
      &=& C_{\rm G}^{-1} - C_{\rm G}^{-1} C_{\rm NG} C_{\rm G}^{-1} +( C_{\rm G}^{-1} C_{\rm NG})^2 C_{\rm G}^{-1} + \dots . \nn \\ 
\eeqa
This series expansion is valid when the non-Gaussian part is smaller than the Gaussian one in some appropriate sense. 

  Plugging Eq.~\ref{eq:Cinv_perturbative} into the signal-to-noise ratio $ S^{T} C^{-1} S $, we can check the binning dependence term by term. Clearly the first term  $S^T C_{\rm G}^{-1} S$ is invariant because although the number of bins is reduced by a factor of  $g$, the Gaussian precision matrix is enhanced by $g$.  Because the enhancement by a factor of $g$ in  $  C_{\rm G}^{-1} $ compensates the reduction in the number of rows in  $ C_{\rm NG} $, the end result of $S^T C_{\rm G}^{-1} C_{\rm NG} C_{\rm G}^{-1} S$ is also invariant.  By inspecting the perturbation series term by term, we conclude that the signal-to-noise for the power spectrum does not depend on the binning to the lowest order. This result was also stated in Ref.~\cite{Takahashi:2009ty}.

We now consider the case of bispectrum. When the binning is rescaled by a factor of $g$, say $g=2$, the triangles in the bins $[2i-1, 2i][ 2j-1, 2j][2k-1,2k]$ are mapped into the triangle $[ i]'[j]'[k]'$ in the coarse-grained case. Here the triangle sides are in units of the fundamental mode $k_{\rm F}$.  For example, when the binning is changed from $\Delta k = k_{\rm F}$ to $ 2 k_{\rm F}$, the scalene triangles, [7][5][3], [7][5][4], [7][6][3], [7][6][4], [8][5][3], [8][5][4], [8][6][3], and [8][6][4]  are mapped to $[4]'[3]'[2]'$. For triangles with some symmetries, i.e.~the isosceles and equilateral triangles, the counting is slightly different. For example, for the equilateral triangle set [7,8][7,8][7,8], there are four distinct triangle sets [7][7][7], [7][7][8], [7][8][8], and [8][8][8] and they are mapped to $[4]'[4]'[4]'$. We should not include triangles such as [8][7][7] as it is identical to  [7][7][8] and its inclusion  would cause the covariance matrix to be singular.  The symmetry factor is important for the consistency of counting.  We comment that using a coarse binning for bispectrum is not very accurate for the bins with the smallest sides as triangles of many different shapes are mapped to a certain coarse-grained one. Yet for triangles of larger lengths, triangles are mapped to a triangle of similar shape by the coarse-graining transformation, and hence we expect that the coarse-grained field  reflects the original one accurately.

The Gaussian bispectrum covariance scales with the binning as $(\Delta k )^{-3}$. The leading disconnected non-Gaussian contribution to the dark matter bispectrum covariance  and also the non-Gaussian terms $ C_{\rm NG}^{(1)} $ (Eq.~\ref{eq:C_NG_1_d}) scale as $(\Delta k)^{-1}$ and couple only triangles with at least one side equal to each other.  On the other hand  $C_{\rm NG}^{(0)} $ (Eq.~\ref{eq:C_NG_0_d}) does not depend on  $\Delta k $ and it couples all triangles.  Again using Eq.~\ref{eq:Cinv_perturbative}, we can check if the S/N changes when the binning is rescaled by a factor of $g$, such as $g=2$.  For scalene triangles,  the number of bins is reduced by a factor of 8, while $C_{\rm G}^{-1}  $ is enhanced by a factor of 8, and hence it is invariant with respect to binning.  For other shapes, such as the equilateral triangles, taking the symmetry factor $s_{123}$ into account, we can also show that $S^T C_{\rm G}^{-1} S$ is invariant. As the non-Gaussian term $C_{\rm NG}^{(0)} $ does not scale with $\Delta k $, by reasoning similar to the case of power spectrum, we also deduce that this term is invariant under bin width rescaling.  For the  terms that scale as  $(\Delta k)^{-1}$,  the coupling is non-trivial in the covariance matrix, and it is hard to make an analytical argument.  By considering some explicit examples for scalene, isosceles and equilateral triangles, we check that this particular scaling and coupling also result in the S/N  invariant with respect to the binning width.  Thus we have verified that the bispectrum S/N is  invariant with respect to $\Delta k$ to the leading order.

\bibliography{references}

\end{document}